\documentclass[aps,prd,superscriptaddress,preprintnumbers,reprint,nofootinbib,showpacs]{revtex4-1}
\pdfoutput=1

\usepackage{graphicx}
\usepackage{adjustbox}
\usepackage{xcolor}
\usepackage{epsfig}
\usepackage{setspace}
\usepackage{amssymb}
\usepackage{amsmath,amsfonts,mathtools}
\usepackage{amsthm}
\usepackage{fixmath}
\usepackage{bm}
\usepackage{braket}
\usepackage{hyperref}
\usepackage{lineno}

\def\lMSbt{\Lambda_{\rm \overline{MS}}^{N_f=3}}
\def\lMSb{\Lambda_{\rm \overline{MS}}}
\def\als{\alpha_s} 
\newcommand{\qbq}{{Q\bar Q}}
\def\cdf{\chi^2/{\rm d.o.f.}}
\def\MS{\overline{\rm MS}}
\newcommand{\be}{\begin{equation}}
\newcommand{\ee}{\end{equation}}
\newcommand{\bea}{\begin{eqnarray}}
\newcommand{\eea}{\end{eqnarray}}
\newcommand{\nn}{\nonumber}
\newcommand{\al}[1]{\vskip-3ex\begin{align}#1\end{align}}
\renewcommand{\bm}{\mathbold}
\newcommand{\capstretch}{\setstretch{1.1}}

\definecolor{red}      {rgb}{0.8,0.0,0.0}
\definecolor{green}    {rgb}{0.0,0.6,0.0}
\definecolor{darkblue} {rgb}{0.0,0.1,0.7}
\definecolor{brown}    {rgb}{0.6,0.1,0.0}
\definecolor{gray}     {rgb}{0.6,0.6,0.6}
\definecolor{darkgreen}{rgb}{0.0, 0.545098, 0.0}
\definecolor{orange}   {RGB}{238,80,25}
\definecolor{purple}   {rgb}{0.5,0.0,0.5}
\definecolor{babypink} {rgb}{0.64, 0.44, 0.44}

\begin{document}

\title{Determination of the QCD coupling from the
static energy and the free energy}

\author{Alexei Bazavov}
\affiliation{Department of Computational Mathematics, Science and Engineering and Department of Physics and Astronomy,
Michigan State University, East Lansing, MI 48824, USA}

\author{Nora Brambilla}
\affiliation{Physik Department, Technische Universit\"at M\"unchen,
James-Franck-Strasse 1, D-85748 Garching, Germany}
\affiliation{
Institute for Advanced Study, Technische Universit\"at M\"unchen, 
Lichtenbergstrasse 2a, 
D-85748 Garching, Germany}

\author{Xavier \surname{Garcia i Tormo}}
\affiliation{
Institut de Ci\`encies del Cosmos, Universitat de Barcelona, Mart\'\i$\,$ i Franqu\`es 1, E-08028 Barcelona, Catalonia, Spain}

\author{P\'eter Petreczky}
\affiliation{Physics Department, Brookhaven National Laboratory,
  Upton, NY 11973, USA}

\author{Joan Soto}
\affiliation{
Institut de Ci\`encies del Cosmos, Universitat de Barcelona, Mart\'\i$\,$ i Franqu\`es 1, E-08028 Barcelona, Catalonia, Spain}
\affiliation{Departament de F\'isica Qu\`antica i
Astrof\'isica, 
Universitat de Barcelona, Mart\'\i$\,$ i
Franqu\`es 1, E-08028 Barcelona, Catalonia, Spain}

\author{Antonio Vairo}
\affiliation{Physik Department, Technische Universit\"at M\"unchen,
James-Franck-Strasse 1, D-85748 Garching, Germany}

\author{Johannes Heinrich Weber}
\affiliation{Department of Computational Mathematics, Science and Engineering and Department of Physics and Astronomy,
Michigan State University, East Lansing, MI 48824, USA}
\affiliation{
Excellence Cluster ORIGINS,
Boltzmannstrasse 2, D-85748 Garching, Germany
}

\date{\today}
\preprint{TUM-EFT 111/18; INT-PUB-19-028;}

\begin{abstract}
We present two determinations of the strong coupling \(\als\). 
The first one is from the static energy at three-loop accuracy, 
and may be considered an update of earlier determinations by some of us. 
The new analysis includes new lattice data at smaller lattice spacings, 
and reaches distances as short as \(0.0237\,{\rm fm}\). 
We present a comprehensive and detailed estimate of the error sources that 
contribute to the uncertainty of the final result, $\als(M_Z)=0.11660^{+0.00110}_{-0.00056}$. 
The second determination is based on lattice data for the singlet free 
energy at finite temperature up to distances as small as \(0.0081\,{\rm fm}\), 
from which we obtain $\als(M_Z)=0.11638^{+0.0009 5}_{-0.00087}$.
\end{abstract}


\maketitle

\section{Introduction}\label{sec:intr}

A precise determination of the strong coupling $\als$ is of key importance 
both for the theory of strong interactions, which is determined by such 
fundamental parameter, and for investigations of new physics beyond the 
Standard Model.
Cross section calculations at the Large Hadron Collider, for example 
suffer from the uncertainties related to our limited knowledge of $\als$.
The last decades have witnessed an impressive effort in the $\als$ extraction 
from an ample range of physical observables with a broad sweep of different methods.
Notwithstanding all these efforts, the Particle Data Book (PDG)   world average value for
$\als$ in 2018, $\als(M_Z) =0.1181 \pm 0.0011$, has an overall uncertainty
that has almost doubled with respect to the PDG average in 2013, 
$\als(M_Z)=0.1185 \pm 0.0006 $~\cite{Tanabashi:2018oca}. 
This is due to the fact that the uncertainty of several determinations that 
enter the average is dominated by errors of theoretical origin, 
which are often difficult to precisely assess.
It is important therefore to use a variety of different ways to extract $\als$
and to validate each extraction at the best of the state of the art.

In this paper we aim at an improved determination of \(\als\) both by 
updating earlier extractions from the static energy and by proposing a new 
method that involves the singlet free energy.

The static energy is an observable up to an additive constant, and it is a 
function of the distance $r$ between the static quark and the static antiquark. 
In the limit of massless dynamical quarks, it depends only on $\als$. 
It can be calculated on the lattice for any distance 
\(r = \sqrt{x^2+y^2+z^2}\), where \(x\), \(y\) and \(z\) are 
integer multiples of the lattice spacing.
For short distance $r$ it can be calculated in perturbation theory in QCD 
as a function of $\als$ using nonrelativistic QCD effective field theories (EFT). 
Noteworthy, perturbation theory is accurate for this quantity at three loops, 
and the tree-level result is already sensitive to \(\als\). 
The comparison between the perturbative expression and lattice data at distances small 
enough to be accessible to 
perturbation theory is a good way to obtain a precise determination of $\als$. 
For this endeavor it is crucial to have lattice data covering an 
interval of sufficiently small distances in order to be sensitive to the 
minute details of the curvature of the static energy, namely, with 
sufficiently fine lattice spacing. 
At the present day, these lattices still pose a major challenge. 
In this paper we will use lattices~\cite{Bazavov:2017dsy} with an 
extraordinarily fine lattice spacing \(a = 0.0249\,{\rm fm}\) to achieve 
a systematically improved extraction of $\als$.

Additionally we exploit a new idea. 
One reason for which it is challenging to reach such fine lattice 
spacings in lattice QCD simulations with dynamical quarks is that one 
has to simultaneously maintain the control over finite volume effects arising 
from the propagation of the lightest hadronic modes, namely, 
\mbox{the Goldstone bosons}, at the pion scale.
A lattice simulation at high enough temperature avoids this infrared problem, and 
thus enables reaching much finer lattice spacings using smaller volumes. 
We use finite temperature lattices with unprecedentedly fine lattice 
spacing \(a = 0.00848\,{\rm fm}\)~\cite{Bazavov:2018wmo}. 
The singlet static free energy is again a function of the static quark-antiquark 
distance and has been calculated on the lattice~\cite{Bazavov:2018wmo} and 
perturbatively~\cite{Berwein:2017thy}. 
The comparison between the two offers a novel and independent method to extract 
a precise determination of $\als$.

The rest of the paper is organized as follows. 
In Section~\ref{sec:lattice} we give an account of the gauge ensembles and lattice 
correlators that we use and discuss the relevant systematic effects in the lattice data. 
In Section~\ref{sec:static} after briefly recalling the procedure used in 
\mbox{Ref.}~\cite{Bazavov:2014soa} to extract $\als$ from the static energy, 
we compare between the lattice data and the weak-coupling result, discuss our 
estimates of systematic uncertainties, and finally obtain an updated 
extraction of \(\als\). 
In Section~\ref{sec:finite} we discuss the singlet free energy at finite temperature, 
and outline the relation with the static potential and static energy at zero temperature. 
We discuss the relevant scale hierarchies, compare
with the lattice, and proceed to extract $\als$ at even shorter distances. 
Section~\ref{sec:disc}  contains some  discussion of our results, as well as 
a comparison with previous related works. 
Finally, in Sec.~\ref{sec:concl} we present a short summary of the main results and conclude. 
In Appendix~\ref{app:ensembles} we discuss the gauge ensembles used in this study. 
In Appendix~\ref{app:discretization} we discuss discretization artifacts at short 
distances in detail.
In Appendix~\ref{app:pert} we list the coefficients appearing in the perturbative results 
used in this paper.

\section{Lattice setup}\label{sec:lattice}

\begin{table} 
\parbox{.98\linewidth}{
  \begin{tabular}{|c|c|c|c|c|c|c|c|}
    \hline
    \multicolumn{8}{|c|}{\(m_l=m_s/20\):} \\
    \hline
    $ \beta $ & a\,(\rm{fm}) & \(N_\sigma,N_\tau\) & $am_s$ & $m_\pi L$ & \#TUs & \#MEAS & Ref. \\
    \hline
    7.373 & 0.060 & \(64^3\times 48\) & 0.0250 & 2.3 & 4623 & 1000 & \cite{Bazavov:2014pvz} \\
    7.596 & 0.049 & \(64^4\) & 0.0202  & 2.6 & 4757 & 1000& \cite{Bazavov:2014pvz} \\
    7.825 & 0.040 & \(64^4\) & 0.0164  & 2.0 & 4768 & 1000 & \cite{Bazavov:2014pvz} \\
  \hline
    \hline
    \multicolumn{8}{|c|}{\(m_l=m_s/5\):} \\
    \hline
    $ \beta $ & a\,[\rm{fm}] & \(N_\sigma,N_\tau\) & $am_s$ & \(m_\pi L\) & \#TUs & \#MEAS & Ref. \\
    \hline
    8.000 & 0.035 & \(64^4\) & 0.01299 & 3.6 & 4616 & 1000 & \cite{Bazavov:2017dsy} \\
    8.200 & 0.029 & \(64^4\) & 0.01071 & 3.1 & 4616 & 1000 & \cite{Bazavov:2017dsy} \\
    8.400 & 0.025 & \(64^4\) & 0.00887 & 2.6 & 4616 & 1000 & \cite{Bazavov:2017dsy} \\
  \hline
  \end{tabular}
  \caption{\capstretch
\label{tab:t0 fine}
  Parameters for the fine $T=0$ ensembles. 
  In the seventh column we indicate the number of correlator 
  measurements performed (MEAS).
  }
}
\end{table}

We require the lattice result of the static energy at the smallest available 
distances in order to compare with the weak-coupling calculations. 
For this reason we employ the 2+1 flavor ensembles using the highly improved 
staggered quark (HISQ) action~\cite{Follana:2006rc} that have been generated 
for studies of the QCD Equation of State~\cite{Bazavov:2014pvz, Bazavov:2017dsy}. 
The spatial volume is given by \(V_3=(a N_\sigma)^3\), and  
the physical length of the Euclidean time direction is \(aN_\tau\). 
We summarize the zero temperature ensembles\footnote{
The lattice observables (gluon action density, and sea quark condensates) 
contributing to the Equation of State have divergent, additive contributions, 
which depend only on the lattice spacing. 
Renormalization group invariant observables are obtained by subtracting the 
\(T=0\) observables from the \(T>0\) observables at each lattice spacing. 
} 
with \(N_\tau=64\) 
in \mbox{Tab.}~\ref{tab:t0 fine}, which correspond to two different pion masses in the continuum limit. 
Namely, these ensembles have different light sea quark masses, \(m_l=m_s/20\) or \(m_l=m_s/5\). 
The strange quark mass \(m_s\) is at its physical value, namely, such 
that the mass of the hypothetical \(\eta_{s\bar s}\) meson is reproduced as 
\(m_{\eta_{s\bar s}}  = 686\,{\rm MeV}\). 
The lattice spacing $a$ has been fixed by the \(r_1\) scale, which is defined by the equation 
\al{
 r^2 \left . \frac{d E(r)}{d r} \right|_{r=r_1}=1,
}
see \mbox{Refs.}~\cite{Bazavov:2014pvz, Bazavov:2017dsy} for the details. 
Here, and in the following \(E(r)\) denotes the QCD static energy calculated 
on the lattice. 
We use the value \(r_1= 0.3106(17)\,{\rm fm}\) determined from the pion decay 
constant.  
Since the quark mass dependence of \(r_1/a\) is small, the \(m_l=m_s/20\) and 
\(m_l=m_s/5\) results can be combined to produce a parametrization of 
\(r_1/a\) as a function of the bare lattice gauge coupling 
\(\beta\)~\cite{Bazavov:2017dsy}. 
The tunneling between different topological sectors is suppressed 
for these gauge ensembles.
Nevertheless, no statistically significant difference has been observed 
for bulk observables other than the light quark condensate~\cite{Bazavov:2017dsy} 
in different topological sectors. 
For the distances we consider in this study the static energy~\cite{Weber:2018bam} 
does not vary significantly between the different topological sectors.

In addition, we use the finite temperature results of the singlet free 
energy with $N_{\sigma}/N_{\tau}=4$ and $N_\tau=12$, or $16$. 
These ensembles correspond to the thermal QCD medium at temperatures 
\(T=1/(aN_\tau)\). 
The finite temperature ensembles have been generated using lattice 
parameters that would correspond to the same two pion masses (at zero 
temperature) in the continuum limit as the zero temperature 
ensembles~\cite{Bazavov:2018wmo}.
For these we use the same parametrization of \(r_1/a\). 
We give an account of these ensembles in \mbox{Tables}~\ref{tab:nt 12} 
and~\ref{tab:nt 16} in Appendix~\ref{app:ensembles}.

\begin{figure*}
\centering
\hskip-1.em
\includegraphics[width=6.2cm]{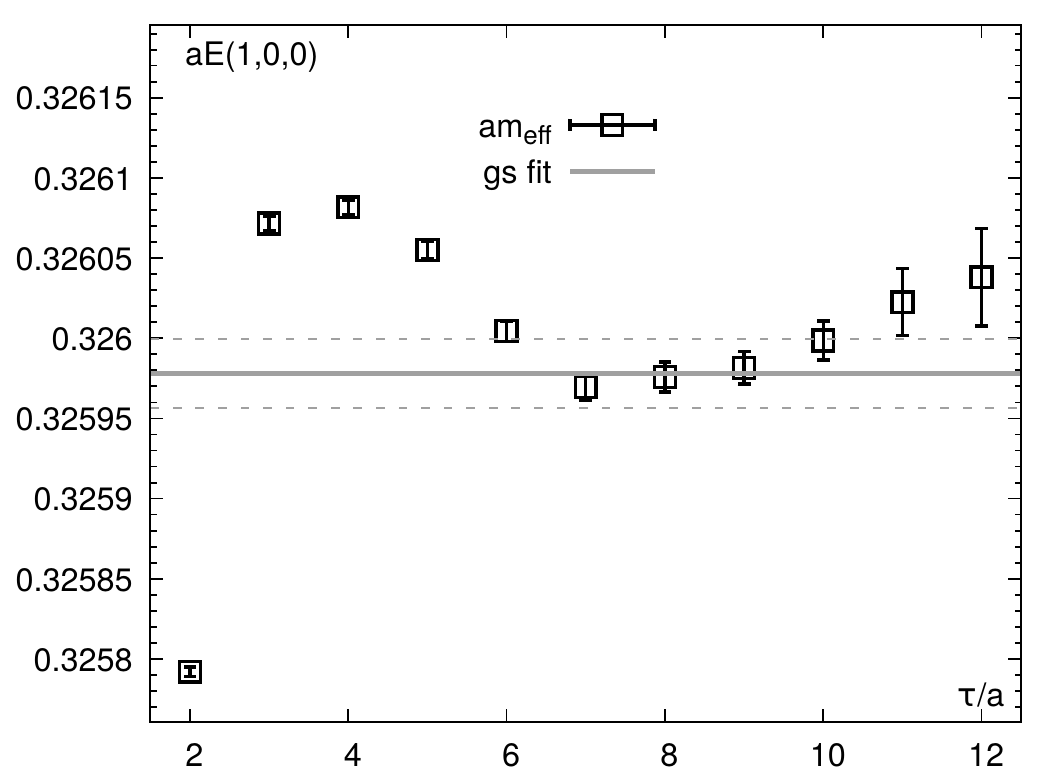}
\hskip-1.em
\includegraphics[width=6.2cm]{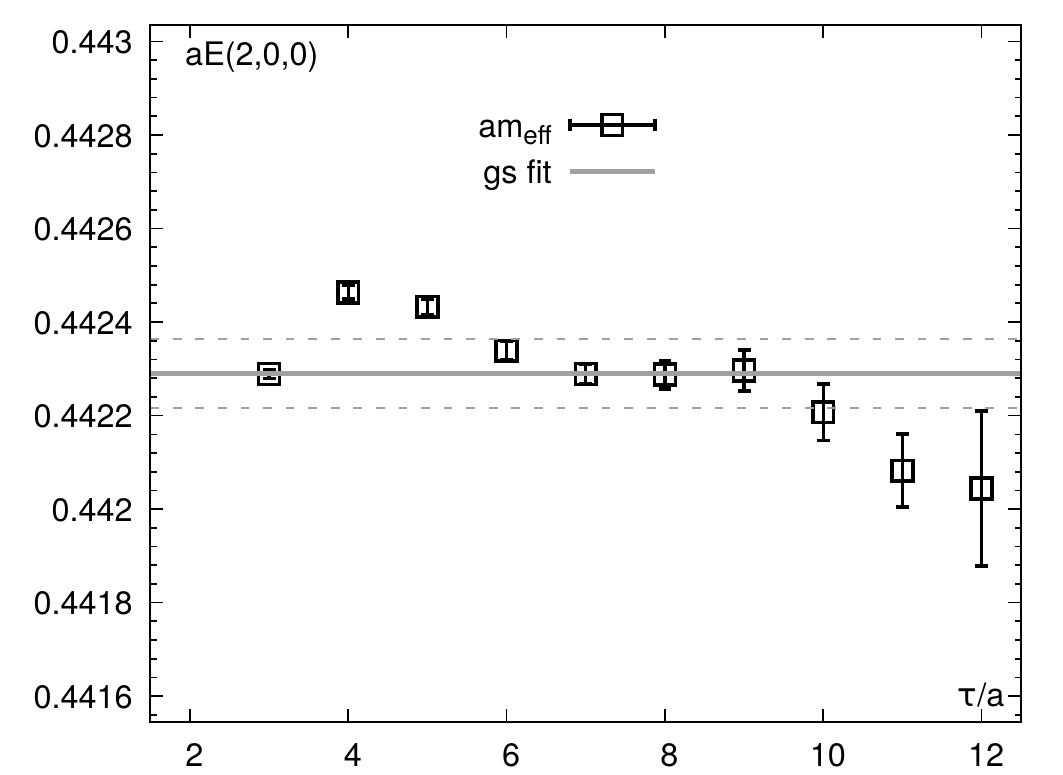}
\hskip-1.em
\includegraphics[width=6.2cm]{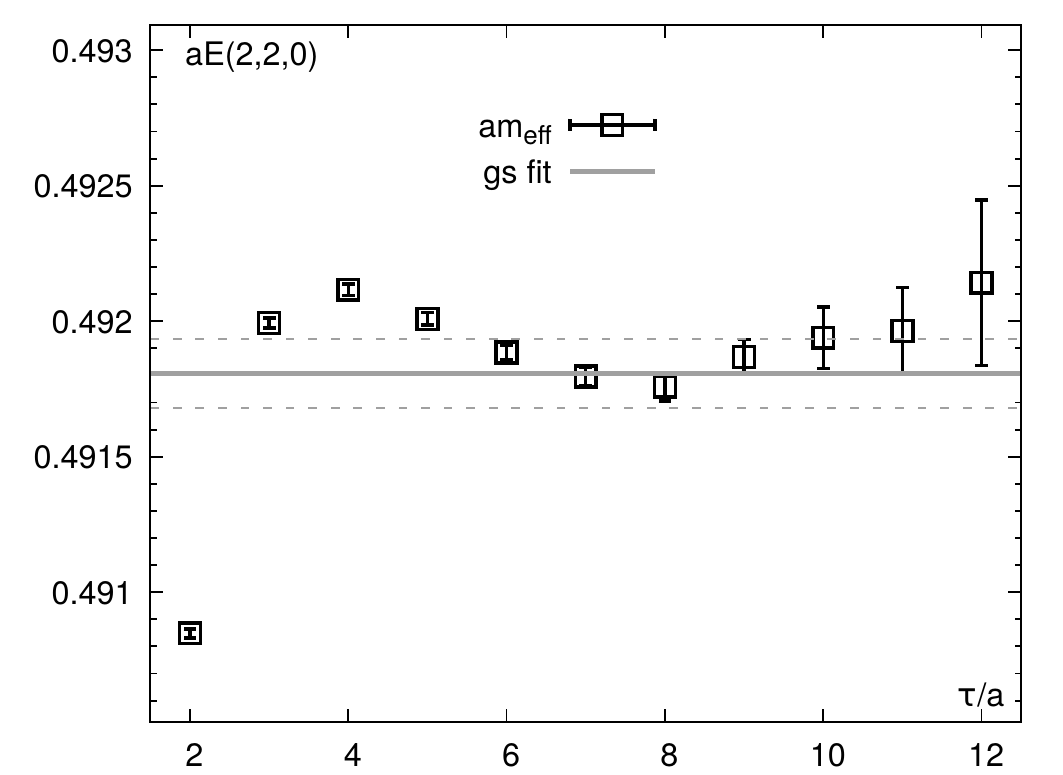}

\hskip-1.em
\includegraphics[width=6.2cm]{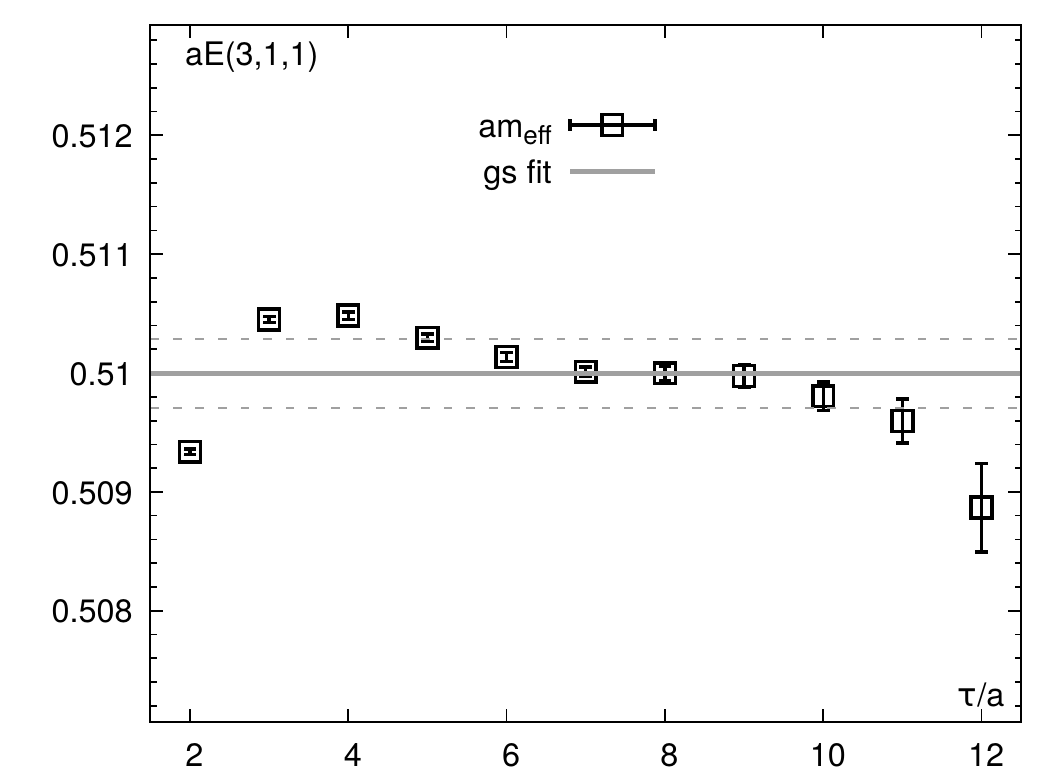}
\hskip-1.em
\includegraphics[width=6.2cm]{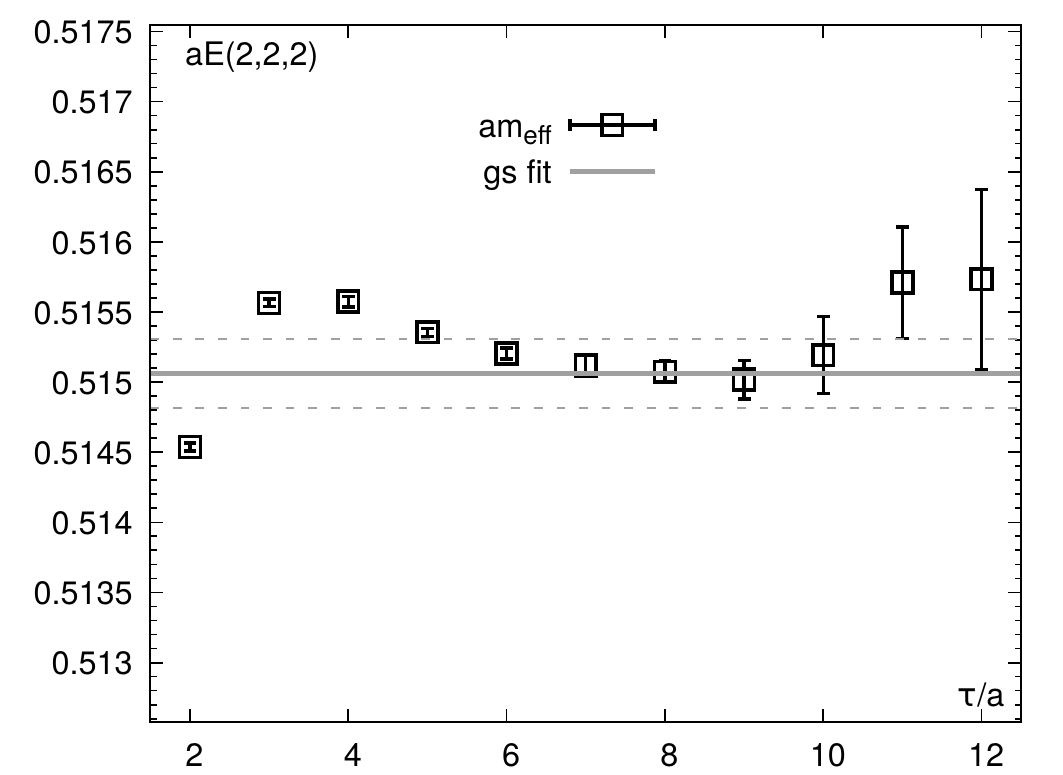}
\hskip-1.em
\includegraphics[width=6.2cm]{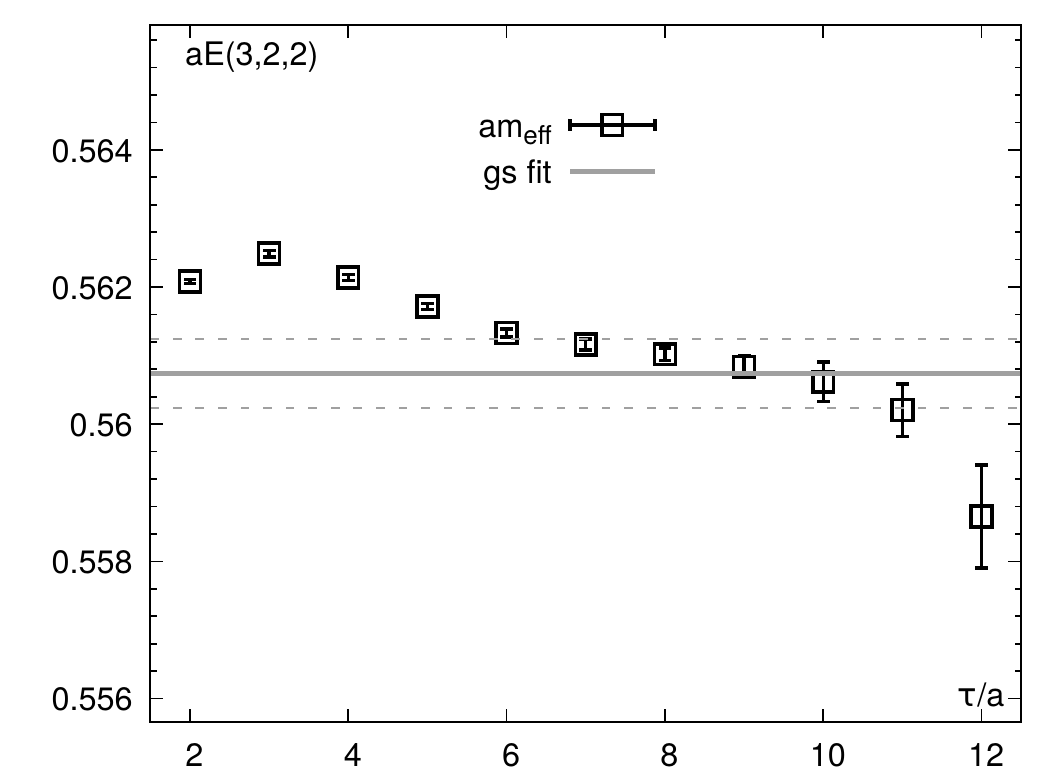}
\caption{\capstretch
Effective mass of the static energy in units of the inverse lattice spacing 
using Wilson line correlators in Coulomb gauge. 
We show \(\beta=8.2,\ r/a=\sqrt{1}\),  \(\beta=8.4,\ r/a=\sqrt{4}\), and 
\(\beta=8.4,\ r/a=\sqrt{8}\) (top, from left to right), and 
\(\beta=8.4,\ r/a=\sqrt{11}\), \(\beta=8.4,\ \sqrt{12}\), and 
\(\beta=8.2,\ r/a=\sqrt{17}\) (bottom, from left to right) as function of 
the Euclidean time \(\tau/a\).
Black open squares indicate the mean and statistical error of the effective 
mass. 
Although, we can usually identify an unambiguous plateau in the window 
\(7 \le \tau/a \le 11\), we have to restrict the fit to narrower windows 
in multiple cases that are similar to those shown above. 
There is no evident pattern regarding \(\beta\), \(r/a\) or \(r\) for the 
appearance of these unstable plateaus. 
The gray solid line indicates the central value of the ground state 
fit, the dashed gray lines indicate the one sigma interval for the 
total error that includes the systematic errors due to the dependence 
on the fit interval. 
}
\label{fig:effm}
\end{figure*}

\begin{figure*}
\centering
\hskip-1.em
\includegraphics[width=6.2cm]{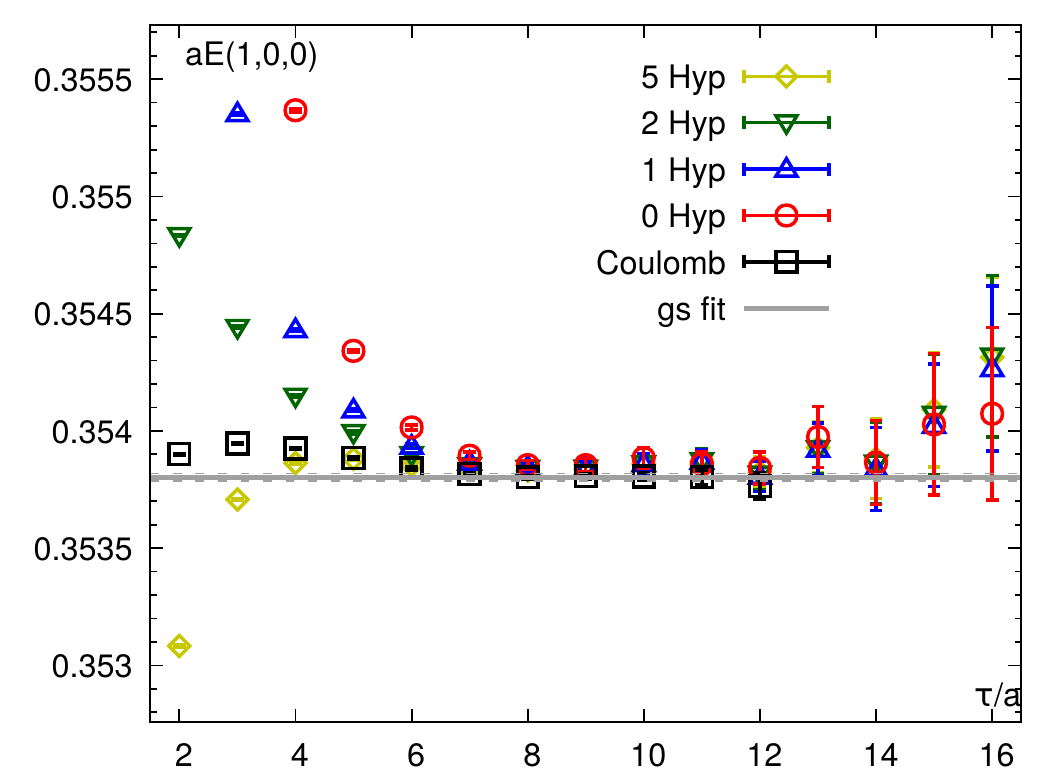}
\hskip-1.em
\includegraphics[width=6.2cm]{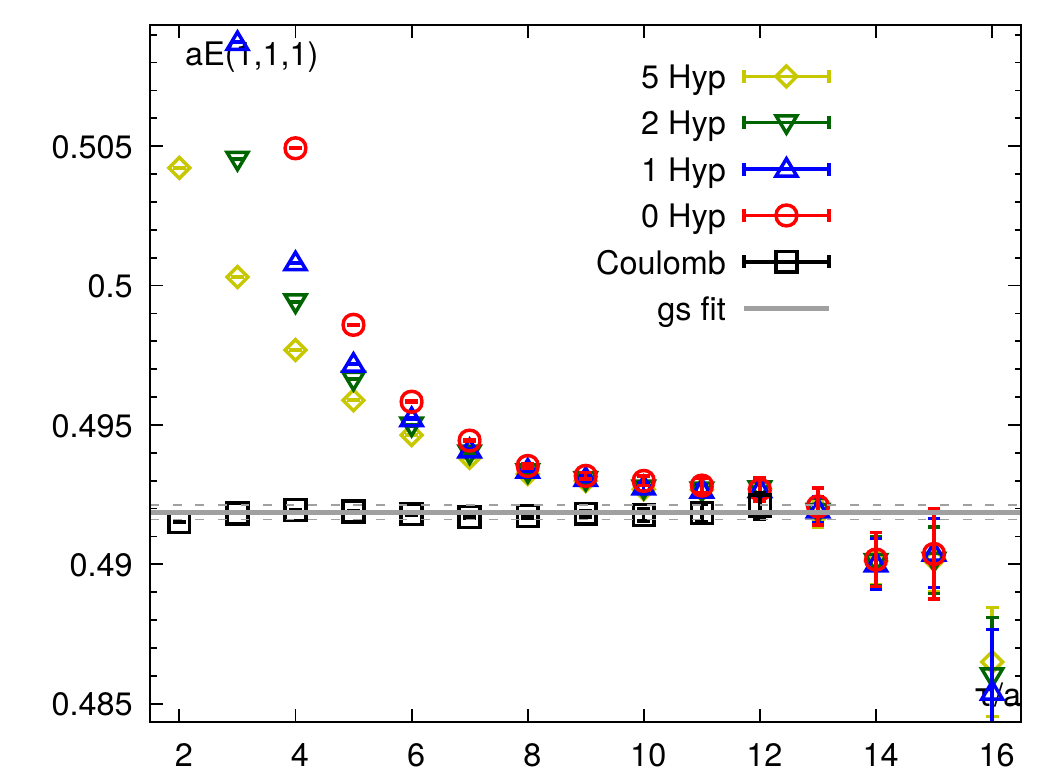}
\hskip-1.em
\includegraphics[width=6.2cm]{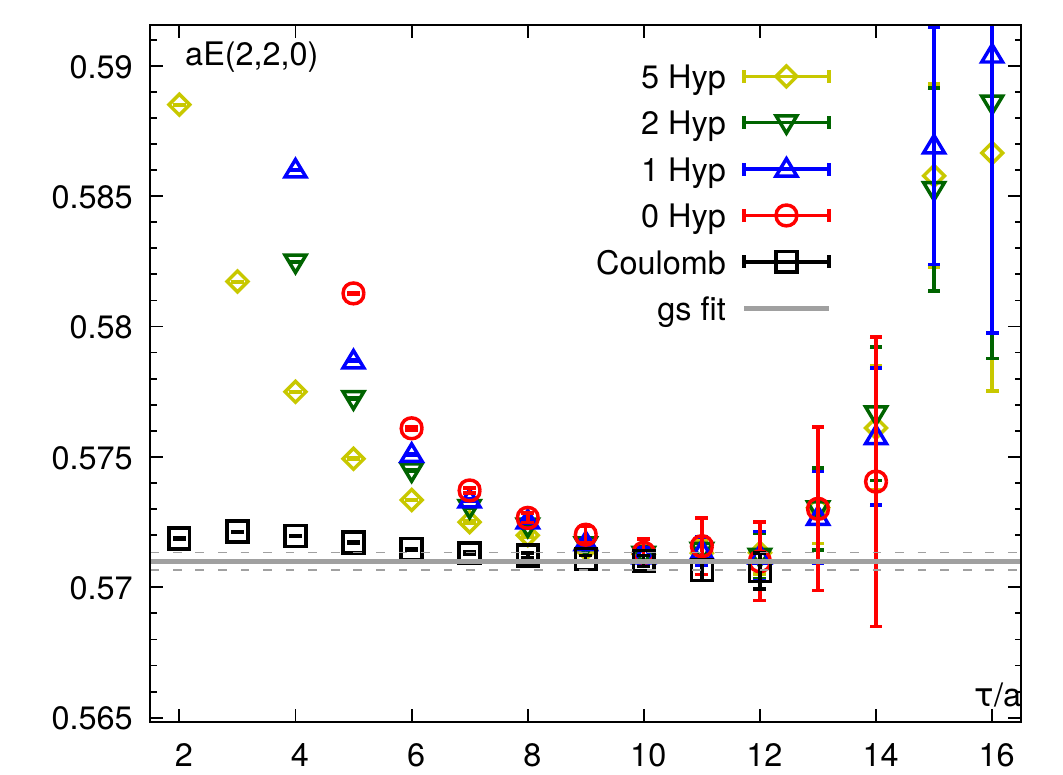}
\caption{\capstretch
Effective mass of the static energy in units of the inverse lattice spacing 
using Wilson loops with or without spatial smearing. 
We show \(\beta=7.825\) with \(r/a=\sqrt{1}\), \(\sqrt{3}\), and 
\(\sqrt{8}\) (from left to right) as function of the Euclidean time \(\tau/a\).
Black open squares indicate the mean and statistical error of the effective 
mass for Wilson lines in Coulomb gauge, colored symbols indicate mean and 
statistical errors of the static energy for Wilson loops. 
The gray solid line indicates the central value of the ground state 
fit, the dashed gray lines indicate the one sigma interval. 
The Wilson loops have the more severe excited state contamination, which 
is increasingly suppressed by the smearing.
More smearing is required for larger \(r/a\), and the signal may drown into 
statistical noise before the masses of Wilson loops reach the same plateau 
value as the Wilson line correlators. 
}
\label{fig:wloop}
\end{figure*}

The static energy can be calculated using Wilson line correlators 
in Coulomb gauge or Wilson loops. 
We use the former because it is more convenient for practical reasons. 
For additional crosschecks we also studied Wilson loops for 
$\beta=7.825$. 
When extracting the static energy from
Wilson loops the spatial lines have to be smeared in order
to obtain a reasonable signal-to-noise ratio and suppress excited
states. 
We used one, two and five levels of HYP smearings \cite{Hasenfratz:2001hp} 
in our calculations. 
In the studies of the Equation of State~\cite{Bazavov:2014pvz, Bazavov:2017dsy}, 
the static energy was extracted from two-exponential fits to the Wilson line 
correlators using a fixed range of Euclidean time \(\tau/a\) that was chosen 
independently from \(r/a\).
This previous calculation is not adequate for our analysis.
Here we improve the extraction of the static energy in the following
way.
\begin{itemize}
\item 
We explicitly include the correlation matrix of the 
static energy in the fits to determine \(\als\). 
For this reason we have to recalculate the static energy and preserve 
its statistical correlations. 
\item 
We explicitly distinguish at short distances between data with distinct 
path geometries instead of averaging over them. 
This increases the number of data for each \(\beta\) compared 
to the earlier analysis~\cite{Bazavov:2014soa}. 
\item 
At small $\tau$ the correlators of Wilson lines are contaminated by
excited states, while at large $\tau$ the statistical fluctuations are large,
and the effective masses sometimes do not show a clear plateau, see 
\mbox{Fig.}~\ref{fig:effm} for some of the worst-case scenarios.
Therefore a careful choice of the lower end, $\tau_{\rm min}$, and upper end, 
$\tau_{\rm max}$, of the fit interval is needed to obtain reliable results 
of the static energy. 
The corresponding fit interval $[\tau_{\rm min}:\tau_{\rm max}]$ will
depend on the distance $r$. 
We estimated the systematic errors due to the choice of the fit interval. 
The procedure of extracting the static energy is demonstrated in 
\mbox{Fig.}~\ref{fig:effm}, where the solid lines show the static energy and 
the dashed lines the corresponding total uncertainty.
We summarize the fit windows used for obtaining the ground state 
in \mbox{Tab.}~\ref{tab:gsfit} in Appendix~\ref{app:ensembles}.
\end{itemize}
At short distances, namely, for \(r \lesssim 0.14\,{\rm fm}\), we observe the 
non-monotonic behavior of the effective mass at small \(\tau/a\), typically, 
within \(\tau \lesssim 0.14\,{\rm fm}\). 
This non-monotonic behavior is not restricted to the Wilson line correlator 
in Coulomb gauge, but also appears in Wilson loops with spatially-smeared 
spatial Wilson lines, see \mbox{Fig.}~\ref{fig:wloop}. 
For this reason, these non-monotonicities are due to the gauge 
ensembles with improved gauge action instead of being due to the details 
of the interpolating operator.
The effective masses obtained from correlators of Wilson lines in Coulomb gauge
and from Wilson loops with HYP smeared spatial lines converge to the same plateau
at large $\tau$.

The current and previous results for the static energy are consistent within 
errors.
For the given reasons the new result is more precise at short 
distances and 
has a better estimate of the errors. 
It supersedes the previous calculation~\cite{Bazavov:2014pvz, Bazavov:2017dsy}. 

The last major source of systematic uncertainties of the lattice data is 
due to discretization artifacts at short distances, which are caused by 
the breaking of rotational symmetry from O(3) in the continuum to the 
cubic group \(W_3\) on the lattice. 
Even for an improved gauge action these artifacts are 
significantly larger than any of the other uncertainties of the lattice 
result. 
The artifacts express themselves in the form of a variation 
of the lattice static energy around the continuum static energy at any given 
distance \(r\). 
The sign and the size of this variation depends on the geometry of the 
shortest paths connecting the two sites at distance \(r\) on the lattice. 
This variation is typically large for on-axis distances, \mbox{i.e.}, 
\(\bm r/a=(n,0,0)\), while being smaller for off-axis distances, and shrinks to 
the point of numerical irrelevance as the distances become larger 
than \(r/a \gtrsim 5\). 
These lattice artifacts could be understood qualitatively in leading-order 
perturbation theory, \mbox{i.e.} at tree level. 
The analysis presented in Appendix~\ref{app:discretization} shows that the 
large artifacts at $r/a=1$ are about $8\%$ for the Wilson action and $4\%$ 
for the L\"uscher--Weisz action that we use, see 
\mbox{Fig.}~\ref{fig:tree-level}. 
The tree-level discretization effects decrease rapidly at larger distances,
and become very small for $r/a\ge \sqrt{8}$. 
One can use the tree-level results to reduce lattice artifacts also in the 
interacting theory.
This procedure is called the tree-level improvement and discussed in 
Appendix~\ref{app:discretization}.

The statistical errors of the static energy obtained on the lattice
increase when increasing the separation $r$. 
For $r/a\ge \sqrt{8}$ the statistical errors in the static energy
are typically larger than the discretiation errors after tree-level improvement.
Thus for these distances discretization errors can be neglected. 
For $\sqrt{5} \le r/a \le \sqrt{8}$ the statistical and discretization errors are comparable. 
In this region one can deal with the discretization errors by estimating 
them and combining with the statistical errors. 
For very short distances $r/a<\sqrt{5}$ the discretization effects are much 
larger that the statistical errors and should be estimated carefully and 
corrected for, before the lattice results can be compared to the weak-coupling 
calculations. 
The corresponding procedure is described in Appendix~\ref{app:discretization}.

\section{Extracting \(\als\) from the static energy}\label{sec:static}

The static energy of a quark-antiquark pair is directly accessible in lattice calculations. 
Our lattice calculation of the static energy for \(N_f=2+1\) light flavors has been discussed in \mbox{Sec.}~\ref{sec:lattice}.
In the weak-coupling regime the static energy is known up to three loops,
\mbox{i.e.}, up to order $\als^4$  in the small $\als$ expansion. 
The resummation of $\ln \als$ terms is also known at subleading order.
Hence, since realistic lattice calculations for $N_f=2+1$ are available at 
short enough distances where the small $\als$ expansion is reliable, $\als$ can be 
extracted by comparing the lattice data for the static energy with its 
perturbative expression for the case of three (approximately) massless quarks. 

This program was initially carried out  in \mbox{Ref.}~\cite{Bazavov:2012ka} and 
considerably improved in \mbox{Ref.}~\cite{Bazavov:2014soa}, to a large extent 
due to the incorporation of shorter distance data on finer lattices. 
Here, we further improve the $\als$ extraction by including three even finer 
lattices. 
The abundance of data at short distances allows us to perform the analysis 
omitting the data with $r/a < \sqrt{5}$ that are susceptible to significant discretization artifacts. 
It also permits us to quantify the impact of using such data with or without 
appropriate corrections. 
Rather than making a separate analysis for each lattice spacing and averaging 
the final results, here we make a global fit to all lattice data leaving as a 
free parameter a normalization constant for each lattice spacing. 
This approach is well-motivated from the earlier observation that no 
remaining lattice spacing dependence can be resolved from the three lattice 
spacings in \mbox{Ref.}~\cite{Bazavov:2014soa} and that the scale uncertainties for 
all considered lattice spacings are similar. 
We also take into account the statistical correlations between the fluctuations 
of data at different distances on the lattice.  
Eventually the effect of these correlations turns out to be much smaller 
than the statistical uncertainty, even though it reduces the uncertainty of 
the extracted value of \(\als\).

\begin{figure}
\centering
\includegraphics[width=9.0cm]{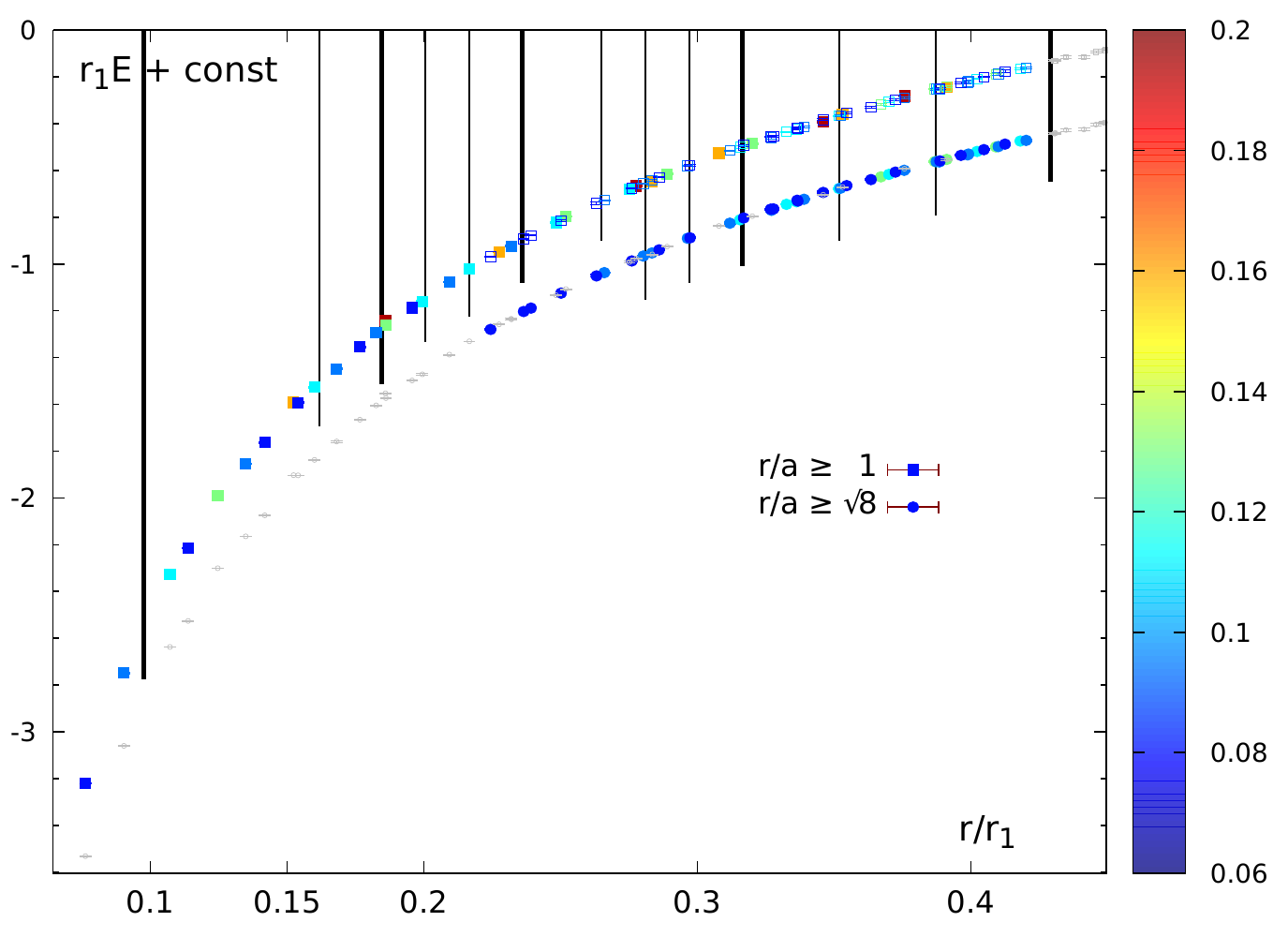}
\caption{\capstretch
Lattice data for the static energy in units of \(r_1\).
The data are shown in vertically displaced sets corresponding to different 
minimal distances in units of the lattice spacing \(r/a\) that were used in 
the fits. 
We restricted the data included in each fit by the maximal distances in 
physical units \(r/r_1\) (indicated by vertical lines).
The thicker vertical lines roughly indicate the maximal distances listed in 
\mbox{Tabs.}~\ref{tab:ra ge 8}, \ref{tab:summary T=0}, \ref{tab:beta 8.4}, and \ref{tab:ra le 4}. 
The color indicates the lattice spacing in units of \(r_1\). 
Filled colored symbols in the set of points labeled as \(r/a \ge 1\) are not 
accessible in the fits that use only \(r/a \ge \sqrt{8}\). 
Gray symbols either indicate data that were not used in the fits
with a given minimal \(r/a\), but are below the respective maximal \(r/r_1\), or that are above the maximal \(r/r_1\) that we included in the fits.  
}
\label{fig:tzerofitranges}
\end{figure}

We focus on data for $r\lesssim 0.14\,{\rm fm}$ or \(r \lesssim 0.45\,r_1\), 
see Fig.~\ref{fig:tzerofitranges}. 
The shortest distance that we can access, namely, due to a single lattice 
spacing\footnote{
Actually, this is 4\% smaller than the lattice spacing itself due to the tree-level improvement. 
} 
on our finest lattice (at zero temperature) is \(0.0237\,{\rm fm}\). 
In \mbox{Ref.}~\cite{Bazavov:2014soa} only data with 
$0.057\,{\rm fm} \lesssim r \lesssim 0.16\,{\rm fm}$ was used for the extraction. 
For that range it was shown that the small $\als$ expansion was reliable. 
Since we use only shorter distances in this paper, the perturbative expansion 
remains reliable.
In \mbox{Ref.}~\cite{Bazavov:2014soa}, separate analyses with data for each 
gauge ensemble were performed in the range \(r/a \ge \sqrt{2}\). 
In this work, thanks to the finer lattices and the combined analysis, we 
can afford to vary the minimal distance \(r/a\) and quantify the corresponding 
uncertainty. 
Data on coarser lattices in the same \(r\) range as well as data on finer lattices 
may be excluded due to restriction of the \(r/a\) range in the analysis. 
This is shown with more detail in \mbox{Fig.}~\ref{fig:tzerofitranges}. 
It was also shown in \mbox{Ref.}~\cite{Bazavov:2014soa} that there is no 
sensitivity to nonperturbative contributions giving rise to power-like 
corrections. 
This is important, since sea quark mass effects, finite volume effects, or 
effects due to an incorrect expectation value of the topological charge 
would make themselves noticeable in such contributions\footnote{
These contributions are also related to the second renormalon of the 
singlet potential, which needed to be accounted for explicitly in 
\mbox{Refs.}~\cite{Takaura:2018lpw, Takaura:2018vcy} because the range of 
data used in those references extends to much larger distances. 
}, if they were not 
negligibly small. 

To obtain the perturbative result for the static energy we follow 
\mbox{Ref.}~\cite{Bazavov:2014soa}. 
We use the perturbative result for the force, \(F(r) = d E/dr\), 
and integrate it to obtain the energy. 
The constant contribution obtained this way is irrelevant and subsumed into 
the normalization constants that we have to fix anyway when comparing the 
weak-coupling results to the lattice data at each \(\beta\).
A choice of the renormalization scale $\nu$ (also called the soft scale, 
see below) proportional to $1/r$ avoids large logarithms of the 
form \(\ln(\nu r)\)~\cite{Bazavov:2014soa}. 
We call the choice $\nu = 1/r$ our standard choice for the 
renormalization scale. 
For this choice we have~\cite{Bazavov:2014soa}
\al{\label{eq:F3L}
F(r) 
& = \frac{C_F}{r^2}\als(1/r)\Bigg[1\nonumber\\
&+\frac{\als(1/r)}{4\pi}\Big(\tilde{a}_1-2\beta_0\Big)\nonumber\\
&+\frac{\als^2(1/r)}{(4\pi)^2}\Big(\tilde{a}_2-4\tilde{a}_1\beta_0-2\beta_1\Big)\nonumber\\
&+\frac{\als^3(1/r)}{(4\pi)^3}\Big(\tilde{a}_3-6\tilde{a}_2\beta_0-4\tilde{a}_1\beta_1-2\beta_2\nonumber\\
&+a_3^L\ln\frac{C_A\als(1/r)}{2}\Big)+\mathcal{O}(\als^4,\als^4\ln^2\als)\Bigg],
}
where the first three terms correspond to tree-level, one-loop, and 
two-loop order, respectively, the fourth term corresponds to three-loop order. 
The strong coupling, $\als$, is understood, here and in the following, 
as renormalized in the $\MS$ scheme. 
At any time the strong coupling can be traded for the corresponding QCD scale. 
At three flavors in the $\MS$ scheme the QCD scale is $\lMSbt$. 
The coefficients in \mbox{Eq.}~\eqref{eq:F3L} are defined in 
\mbox{Ref.}~\cite{Tormo:2013tha} and reproduced in Appendix~\ref{app:pert}. 
At three loops there is a contribution proportional to $\ln \als$
as well as a non-logarithmic term. 
The origin of these terms can be understood using the effective field theory 
approach, namely, potential non-relativistic QCD 
(pNRQCD)~\cite{Brambilla:1999xf,Brambilla:1999qa}. 
The term, pointed out in \mbox{Ref.}~\cite{Appelquist:1977tw}, proportional 
to $\ln \als$ comes from the energy scale $\sim \als/r$, which is called the 
ultra-soft scale to distinguish it from the energy scale $1/r$, which is 
referred to as the soft scale.
Equation~\eqref{eq:F3L} is accurate at next-to-next-to-next-to-leading order (N$^3$LO). 

\begin{figure}
\centering
\includegraphics[width=8.6cm]{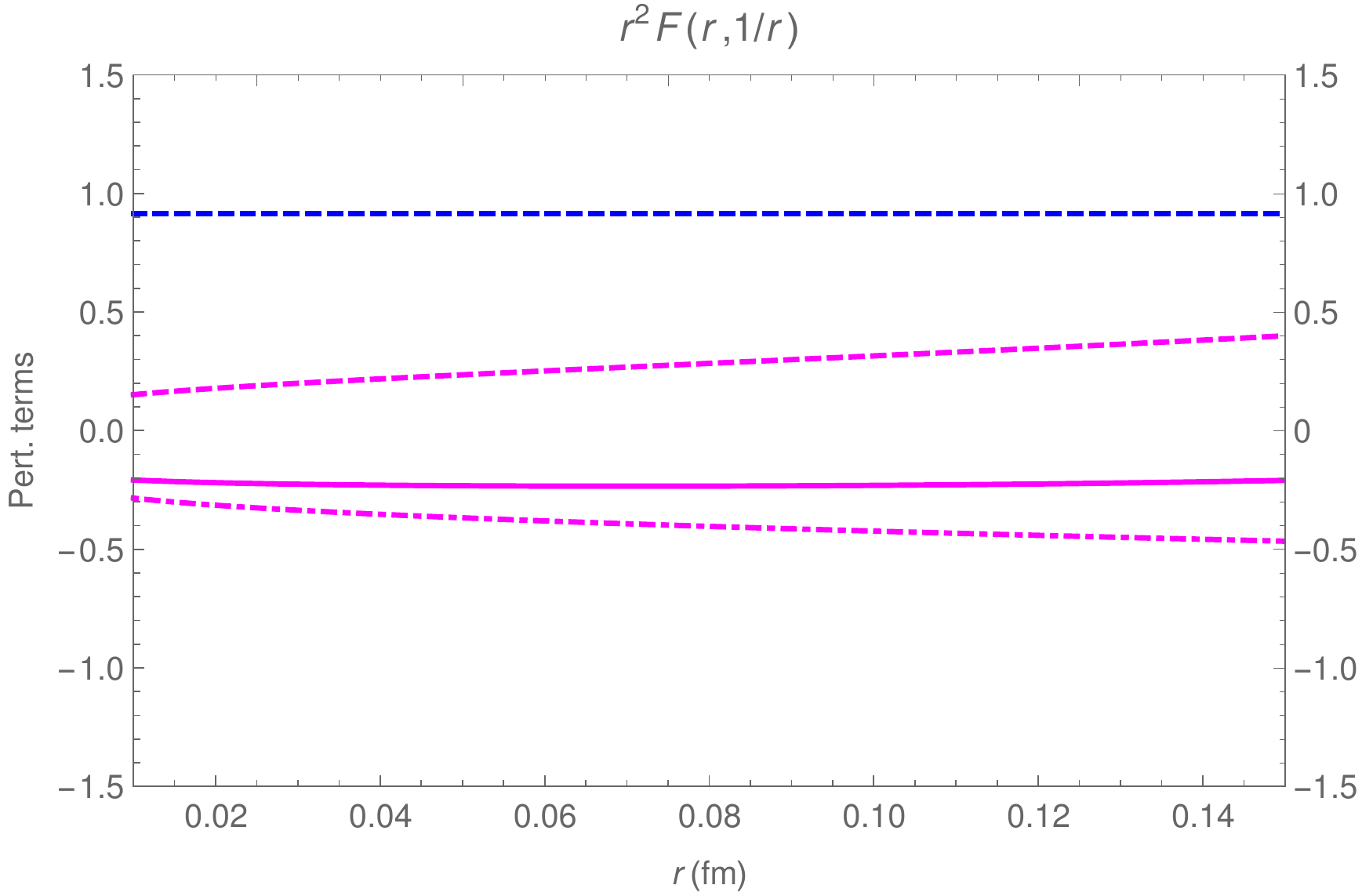}
\caption{\capstretch
The magenta lines stand for three-loop contributions:
dashed (soft, fourth line of \mbox{Eq.}~\eqref{eq:F3L}), 
solid (ultra-soft, last line of \mbox{Eq.}~\eqref{eq:F3L}),
and dot-dashed (resummed ultra-soft, sum of fourth and fifth 
lines of \mbox{Eq.}~\eqref{eq:FN3LL}).
The blue dashed line stands for the two-loop contribution 
(third line of \mbox{Eq.}~\eqref{eq:F3L}).
A factor \(\als(1/r)^3\) has been dropped from all terms. 
We use the standard value $\mu_{us}=C_A \als(1/r)/(2r)$ for the ultra-soft scale. 
We have taken $\lMSbt=315$ MeV.
}
\label{counting}
\end{figure}

For sufficiently small values of the coupling, \mbox{i.e.}, at small enough 
distances the ultra-soft logarithm \(\ln(C_A\als(1/r)/2)\) 
eventually becomes large and can be resummed.    
This can be done using the renormalization group equations in 
pNRQCD~\cite{Pineda:2000gza, Brambilla:2004jw, 
Brambilla:2009bi}.
The expression for the force after performing the resummation of the 
ultra-soft logarithms reads~\cite{Bazavov:2014soa} 
\al{\label{eq:FN3LL}
F(r,\nu=1/r) 
& =  \frac{C_F}{r^2}\als(1/r)\Bigg[1\nonumber\\
&+\frac{\als(1/r)}{4\pi}\Big(\tilde{a}_1-2\beta_0\Big)\nonumber\\
&+\frac{\als^2(1/r)}{(4\pi)^2}\Big(\tilde{a}_2-4\tilde{a}_1\beta_0-2\beta_1\Big)\nonumber\\
& -\frac{\als^2(1/r)}{(4\pi)^2}\frac{a_3^L}{2\beta_0}\ln\frac{\als(\mu_{us})}{\als(1/r)}\nonumber\\
& +\frac{\als^2(1/r)\als(\mu_{us})}{(4\pi)^3}a_3^L\ln\frac{C_A\als(1/r)}{2r\mu_{us}}\nonumber\\
&+\frac{\als^3(1/r)}{(4\pi)^3}\Big(\tilde{a}_3-6\tilde{a}_2\beta_0-4\tilde{a}_1\beta_1-2\beta_2\Big)\nonumber\\
& +\mathcal{O}(\als^4)\Bigg].
}
We choose as a standard value for the ultra-soft scale \(\mu_{us}=C_A \als(1/r)/(2r)\) where \(C_A=N_c\). 
The coefficients in \mbox{Eq.}~\eqref{eq:FN3LL} are defined in 
\mbox{Ref.}~\cite{Tormo:2013tha} and reproduced in Appendix~\ref{app:pert}. 

At leading order in $\ln \als$ we can set $\mu_{us}$ to $1/r$ and 
\mbox{Eq.}~\eqref{eq:FN3LL} reduces to \mbox{Eq.}~\eqref{eq:F3L}. 
The first four terms of \mbox{Eq.}~\eqref{eq:FN3LL} include all terms of the 
form $\als^{3+n}\ln^n\als$, \mbox{i.e.} it is accurate at 
next-to-next-to-leading logarithmic order (N$^2$LL). 
Keeping all terms in \mbox{Eq.}~\eqref{eq:FN3LL}, we obtain an expression 
that is accurate at \emph{three-loop order with leading ultra-soft resummation}. 

In \mbox{Fig.}~\ref{counting}, we show the two-loop contribution to the force, 
\mbox{i.e.}, 
the third term of \mbox{Eq.}~\eqref{eq:F3L}, along with 
different $\als^4$ contributions as function of the distance $r$. 
Here and in the rest of the paper we use, if not differently specified, 
the standard scales \(\nu=1/r\) and \(\mu_{us}=C_A \als(1/r)/(2r)\). 
One can see that the $\als^4 \ln \als$ contribution is never larger than the 
non-logarithmic $\als^4$ contribution in the entire distance range used in our 
study, and both contributions are smaller than the two-loop contribution. 
If we resum the ultra-soft logarithms (\mbox{i.e.} by replacing the last line 
of \mbox{Eq.}~\eqref{eq:F3L} by the fourth and fifth lines of \mbox{Eq.}~\eqref{eq:FN3LL}), 
the corresponding term becomes slightly larger in absolute value, see the 
dot-dashed line in \mbox{Fig.}~\ref{counting}, but it is still significantly 
smaller than the two-loop contribution. 
Therefore, as already noticed in~\cite{Bazavov:2014soa}, the leading ultrasoft 
logarithms can be counted as being of the same order as the other 
non-logarithmic three-loop terms, while subleading logarithms may be included 
in the uncertainty due to the missing four-loop contributions. 
In \mbox{Fig.}~\ref{counting}, we see also that there is a partial 
cancellation between the soft and ultra-soft contributions at order $\als^4$. 
This cancellation, however, may be accidental and may not happen at higher 
orders. 
Our final result will come from comparing lattice data with perturbation 
theory at \emph{three-loop order with leading ultra-soft resummation}, and we 
will use the difference between the N$^3$LO expression of 
\mbox{Eq.}~\eqref{eq:F3L} and the three-loop expression with leading 
ultra-soft resummation from \mbox{Eq.}~\eqref{eq:FN3LL} to estimate the size of 
unknown higher-loop contributions. 
Other ways to estimate the higher-loop contributions will be discussed in the 
following and will enter eventually in our final error assessment.

We proceed as follows. 
On the lattice data side, we put all six data sets (see 
\mbox{Tab.}~\ref{tab:t0 fine}) together by adding an arbitrary 
subtraction constant to each set that is determined by the fit. 
On the continuum side, we generate a grid of $\als (M_Z, N_f=5)$ input 
values ranging from $0.1140$ to $0.1200$ in steps of $0.0001$. 
We use the perturbative result for the force\footnote{
We use perturbative running and decoupling to convert each grid value of 
\(\als(M_Z,N_f=5)\) to the corresponding \(\lMSbt\) value, which we use in 
the analytical expressions of the force, \mbox{Eqs.}~\eqref{eq:F3L} 
and~\eqref{eq:FN3LL}.  
The details of the running and decoupling are the same as in 
\mbox{Ref.}~\cite{Bazavov:2014soa}. 
Namely, we use four-loop running, with the charm quark mass equal to 
\(1.6\,{\rm GeV}\) and the bottom quark mass equal to \(4.7\,{\rm  GeV}\). The effects of higher-order terms in the running are negligible with the current accuracies. 
}, given in 
\mbox{Eq.}~\eqref{eq:FN3LL}, restricting 
ourselves to the \emph{three-loop with leading ultra-soft resummation} 
expression, for the reasons discussed above. 
For each of the \(\als\) input values we calculate the static energy $E(r)$ by 
numerically integrating \mbox{Eq.}~\eqref{eq:FN3LL} and fit the lattice data 
to it, thus determining the subtraction constants. 
Hence, the subtraction constants for different sets are obtained in 
independent fits, since the different ensembles are statistically independent. 
Of course the subtraction constants depend on the input value of \(\als\). 
For the determination of the subtraction constants we consider fits with or 
without the full correlation matrix, which does not lead to statistically 
distinguishable results. 
The smallest $\cdf$ determines the preferred value of $\als (M_Z)$ 
by the lattice data. 
We calculate the \(\cdf\) with or without the full correlation matrix. 
Typically, the minimum of the \(\cdf\) varies by at most 
\(\delta^{\rm corr}=\pm 0.0001\) upon inclusion of the full correlation 
matrix, which is always smaller than the statistical error. 
We consider this difference as part of the statistical error that is already 
accounted for. 
The subtraction constants in the correlated fit are also consistent with the 
subtraction constants in the uncorrelated fits within fractions of the 
statistical error.
We repeat this procedure for several values of $\max(r)$, vary the 
smallest distances \(\min(r/a)\) in units of the lattice spacing 
considered in the analysis, and vary the treatment of discretization 
artifacts at the shortest distances. 
Eventually we repeat the entire procedure for expressions with different 
values of the soft scale,  and for the two-loop and three-loop (without resumming the ultra-soft contribution) expressions. 

Discretization artifacts are significant at distances $r/a \lesssim 3$, 
with increasing importance at smaller and smaller distances as detailed 
in \mbox{Sec.}~\ref{sec:lattice}.  
We use different strategies to handle the residual discretization 
artifacts, see Appendix~\ref{app:discretization} for details: 
\begin{enumerate}
  \item[(i)] 
  ignoring them, \mbox{i.e.}, we engage in no further corrections beyond 
  the tree-level improvement. 
  In this case we restrict the fit to \(r/a \ge \sqrt{8}\) and omit \(r/a = \sqrt{12}\); 
  \item[(ii)] 
  correcting for them, by following a strategy similar to \mbox{Ref.}~\cite{Bazavov:2014soa}, \mbox{i.e.}, nonperturbative improvement (see Appendix~\ref{app:discretization});
  \item[(iii)] 
  accounting for them, by enlarging the statistical errors to fully cover the 
  discretization artifacts, \mbox{i.e.}, tree-level improvement and systematic 
  error estimates. 
  This approach is suitable for \(r/a \ge \sqrt{5}\). 
\end{enumerate}

\begin{table*}
\begin{adjustbox}{angle=0}
\begin{tabular}{|c|c|c|c|c|c|c|c|c|c|}
\hline
Artifacts & \(\max(r)\) [fm] & d.o.f. & uncorr. & corr. & 
\(\als\) & \(\delta^{\rm stat}\) & \(\delta^{\rm lat}\) & 
\(\delta^{\rm pert}\) & \(\als^{2L}\)  
\\
\hline
 nonperturbative improvement & 0.097 & 10 & 0.33 & 0.47 & 
 0.11658 & 0.00072 & 0.00021 &
 \(^{+0.00152}_{-0.00045}\) &  0.11672
\\
 tree-level improvement & 0.097 & 9 & 0.39 & 0.54 & 
 0.11679 & 0.00050 & 0.00021 &
 \(^{+0.00156}_{-0.00045}\) &  0.11696
\\
\hline
 nonperturbative improvement & 0.131 & 42 & 0.33 & 0.40 &
 0.11668 & 0.00051 & 0.00014 &
 \(^{+0.00193}_{-0.00064}\) &  0.11684
\\
 tree-level improvement & 0.131 & 39 & 0.40 & 0.48 &
 0.11673 & 0.00037 & 0.00014 &
 \(^{+0.00190}_{-0.00062}\) &  0.11697
\\
\hline

\end{tabular}
\end{adjustbox}
\caption{\capstretch
Fits with \(r/a \ge \sqrt{8}\) and different treatment of discretization 
artifacts, here, nonperturbative improvement or only tree-level improvement.  
We list uncorrelated and correlated \(\cdf\) in columns 4 and 5. 
In column 9 we list the perturbative error for the corresponding fit window. 
The last column displays the outcome for \(\als\) at two-loop order.
The tree-level improved calculations have fewer degrees of freedom, 
since the data with \(r/a=\sqrt{12}\) have been excluded, see text.
}
\label{tab:ra ge 8}   
\end{table*}

First, we analyze the data for \(r/a \ge \sqrt{8}\) without considering the 
discretization artifacts beyond the tree-level improvement. 
In these fits we omit the data with \(r/a =\sqrt{12}\) due to their exceptionally 
large discretization artifacts. 
We generally obtain \(\cdf\) between 0.3 and 0.6 for the best fits, which 
rises steeply outside a narrow window around the central value of \(\als\). 
We estimate the systematic error due to discretization artifacts from the 
difference to fits with \(r/a \ge \sqrt{8}\) with tree-level improvement or 
with nonperturbative improvement as being typically about \(\delta^{\rm lat}=\pm 0.00021\).
\mbox{Tab.}~\ref{tab:ra ge 8} clearly shows that any errors due to discretization 
artifacts are safely contained in the statistical uncertainty. 
We point out that the analysis for \(\max(r) \le 0.12\,{\rm fm}\) and 
\(r/a \ge \sqrt{8}\) does not depend on gauge ensembles with the smaller sea 
quark mass, \mbox{i.e.}, uses exclusively ensembles with \(m_l=m_s/5\). 

\begin{figure}
\includegraphics[width=9.0cm]{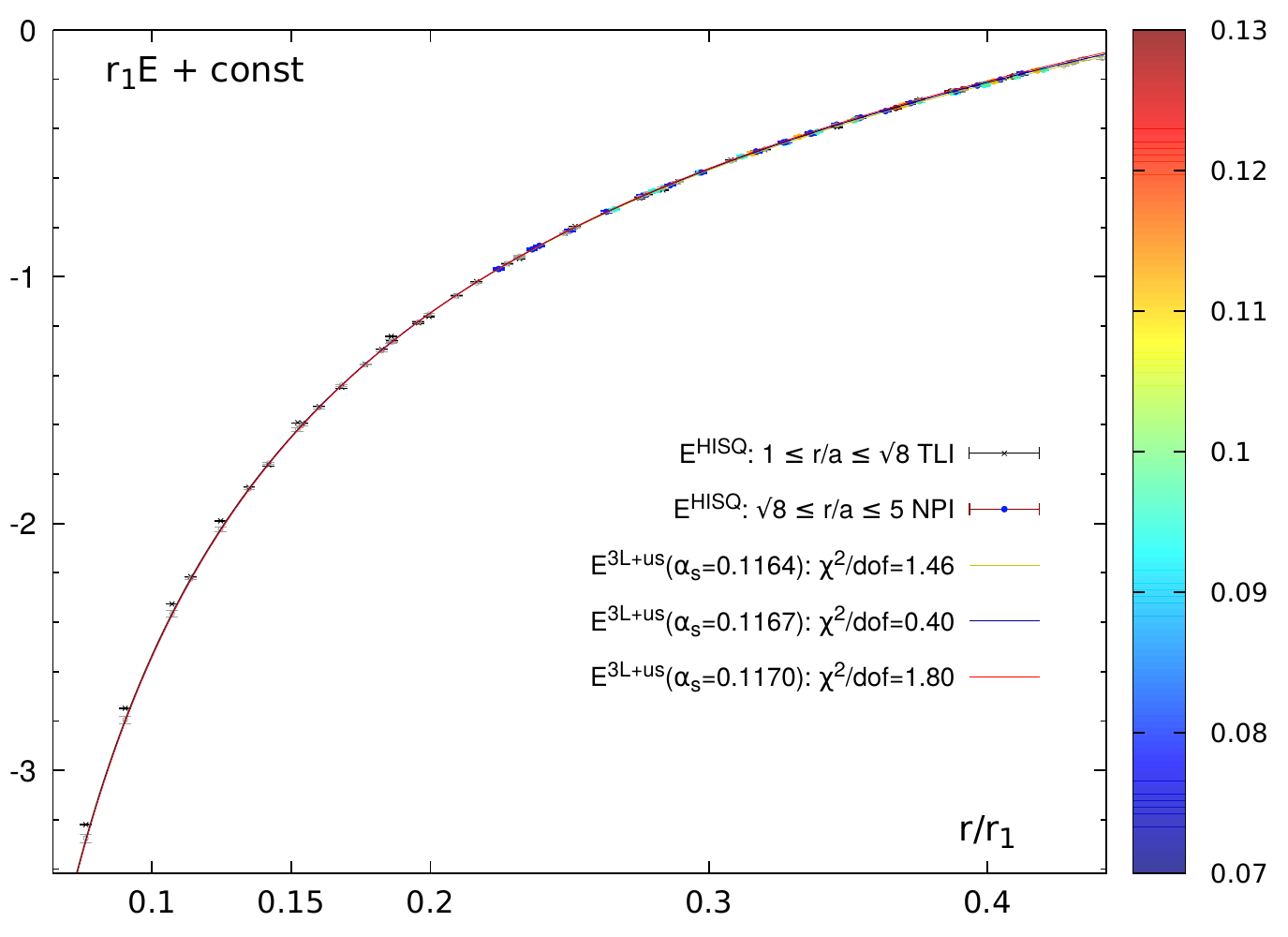}
\caption{\capstretch
Normalized lattice data and weak-coupling result for the static energy 
in units of \(r_1\).
The colored or gray bullets show the nonperturbatively improved (NPI) lattice data, 
the black crosses show the tree-level improved (TLI) lattice data for \(r/a < \sqrt{8}\).
Only the colored symbols are used in the fit with \(r/a \ge \sqrt{8}\). 
The \emph{three-loop with leading ultra-soft resummation} result with standard scales is shown 
for the \(\als(M_Z)\) grid values corresponding to the best fit, and one 
standard deviation (statistical) lower or higher \(\als(M_Z)\) grid 
values. 
The \(\cdf\) correspond to NPI lattice data with \(\min(r/a)=1\) and \(\max(r)=0.131\,{\rm fm}\) (fifth row of \mbox{Tab.}~\ref{tab:summary T=0}).
}
\label{fig:main}
\end{figure}

Second, for $1 < r/a < \sqrt{8}$ the discretization artifacts are too large to 
be ignored.
The data can still be used after correcting for the discretization artifacts 
through the nonperturbative improvement. 
The \(\cdf\) is about the same, unless data at large distances are discarded. 
In this case, the \(\cdf\) becomes even smaller, due to the conservative error 
estimates for the corrections.
We do not observe any larger deviation from the results for 
\(r/a \ge \sqrt{8}\), and, hence, estimate that the uncertainty due to the 
corrections is generally about the same as for \(r/a \ge \sqrt{8}\), 
namely, \(\delta^{\rm lat}=\pm 0.00021\). 
The discretization errors are safely contained in the statistical uncertainty.  
The results for fits in four representative fit intervals\footnote{
We considered many more, but restrict the discussion to a representative 
set of fits.} using nonperturbatively corrected discretization errors are 
summarized in \mbox{Tab.}~\ref{tab:summary T=0}. 
In each fit interval all results are consistent. 
We find slightly lower central values for fits with only 10 data or less, 
which, however, are never statistically significant.  
We show the tree-level or nonperturbatively improved lattice data together 
with the \emph{three-loop with leading ultra-soft resummation} result with 
standard scales in \mbox{Fig.}~\ref{fig:main}. 
Given the large variation over the \(r\) range fine details are 
difficult to resolve.
It can be seen that the fit to the data with large \(r/a\) misses the data 
with \(r/a=1\), \mbox{i.e.}, the first data for each \(\beta\),  unless the 
discretization artifacts are taken care of with an approach beyond the 
tree-level improvement. 

\begin{figure}
\centering
\includegraphics[width=8.6cm]{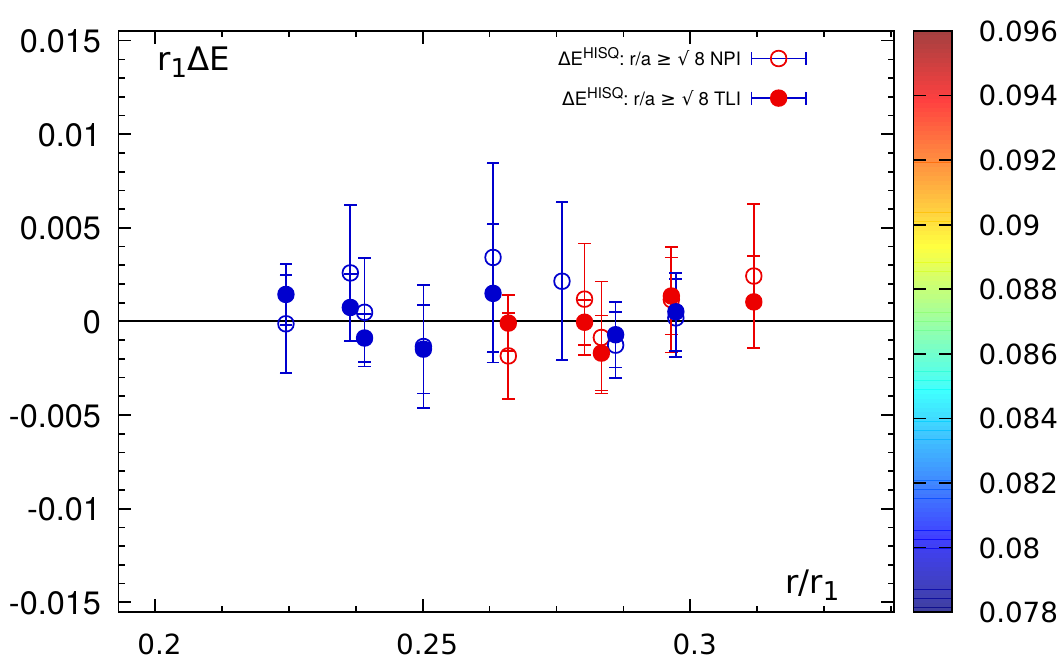}
\includegraphics[width=8.6cm]{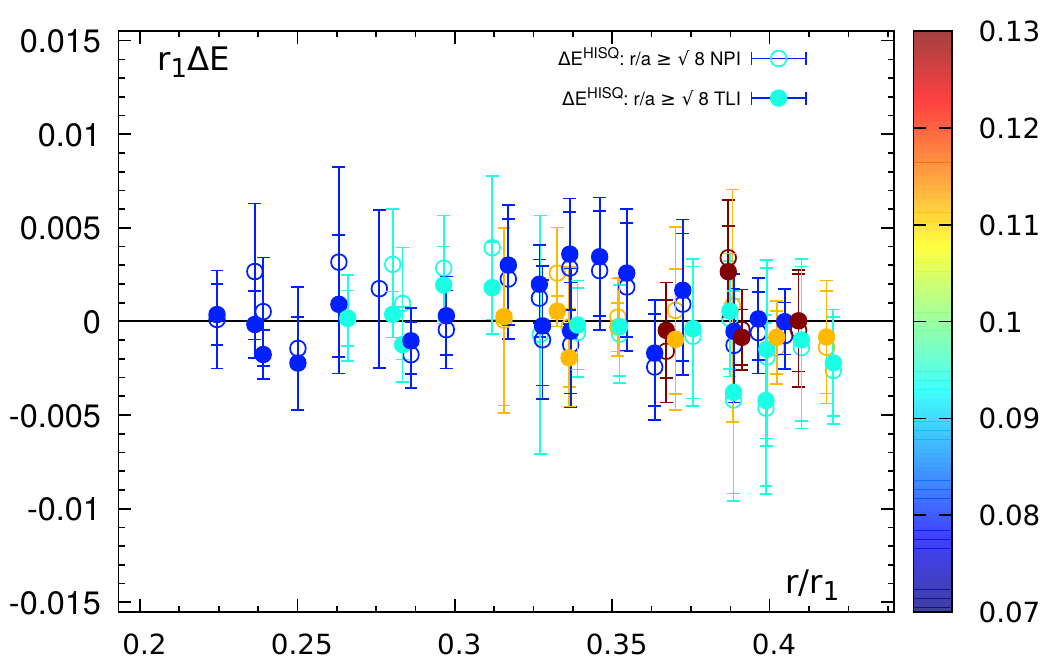}
\caption{\capstretch
Difference between nonperturbatively improved (NPI)  lattice data, or the only 
tree-level improved (TLI) lattice data, and the weak-coupling result for the 
static energy in units of \(r_1\) (after removal of the additive constant).
The open circles show the nonperturbatively improved lattice data, the 
filled bullets the only tree-level improved lattice data. 
For the fits we have used \(r/a \ge \sqrt{8}\). 
Differences between the central values of \(\als\) are \(0.00021\) 
or smaller for the fits using different treatment of discretization 
artifacts.  
The two panels correspond to \(\max(r) = 0.098\), and 
\(0.131\,{\rm fm}\). 
}
\label{fig:diff0}
\end{figure}

We show in \mbox{Fig.}~\ref{fig:diff0} the differences 
between the normalized lattice data and the \emph{three-loop with leading 
ultra-soft resummation} result with standard scales, either using 
nonperturbatively improved data, or data with only tree-level improvement. 
In the considered range the nonperturbative improvement is irrelevant.

\begin{figure}
\centering
\includegraphics[width=8.6cm]{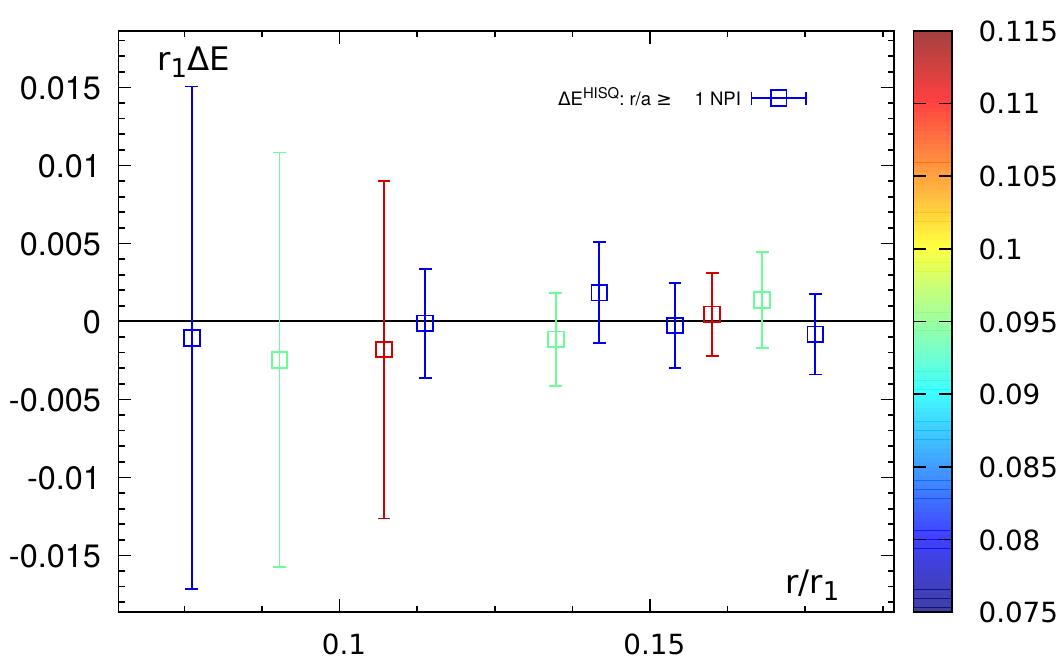}
\includegraphics[width=8.6cm]{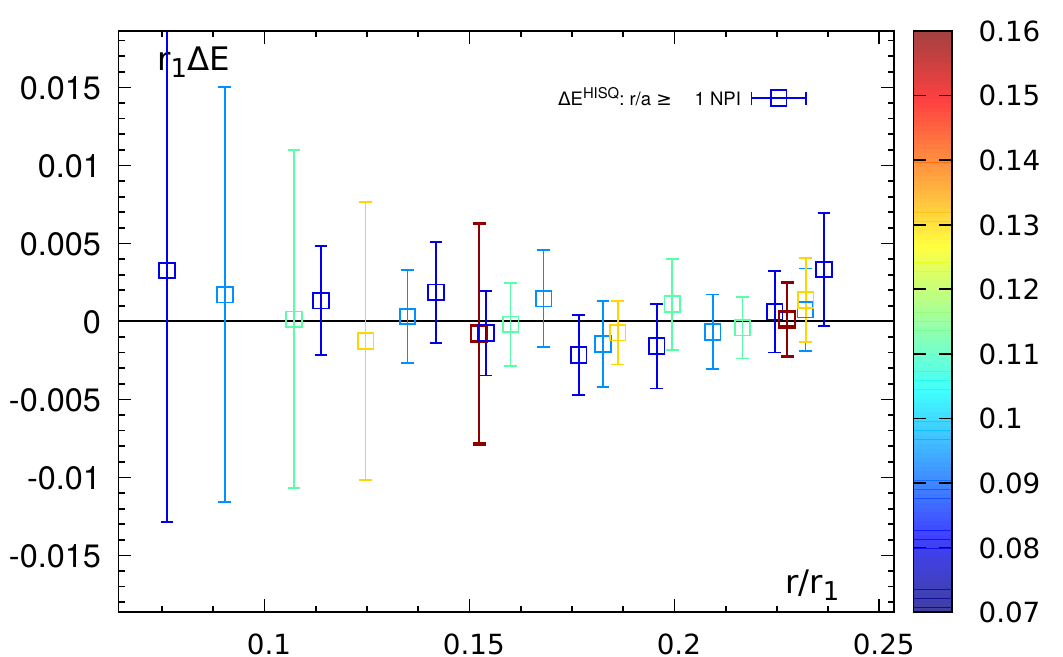}
\includegraphics[width=8.6cm]{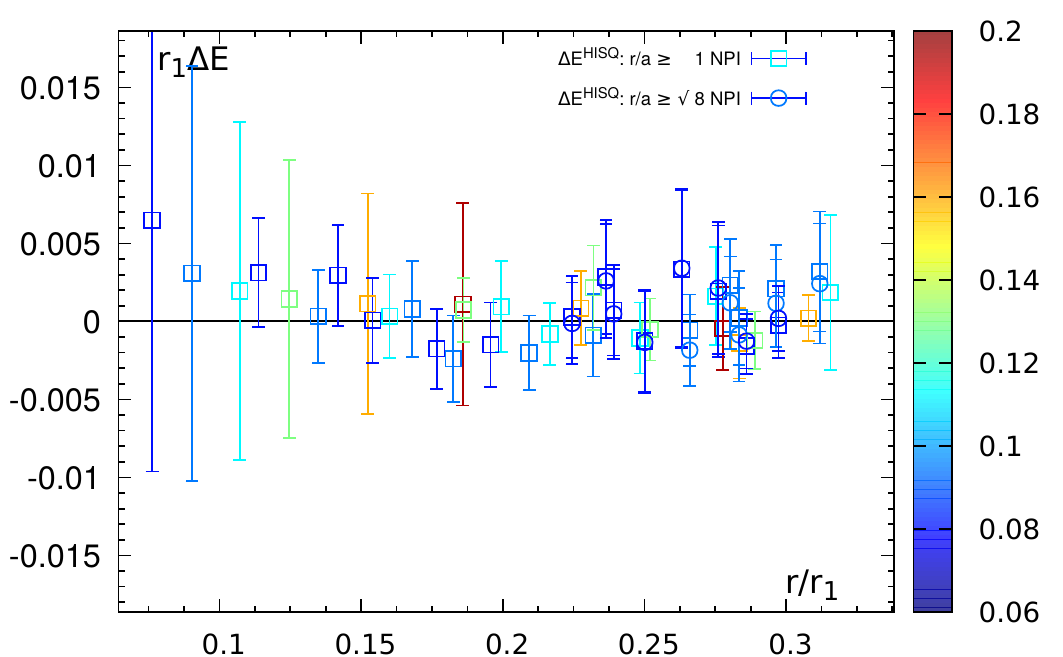}
\includegraphics[width=8.6cm]{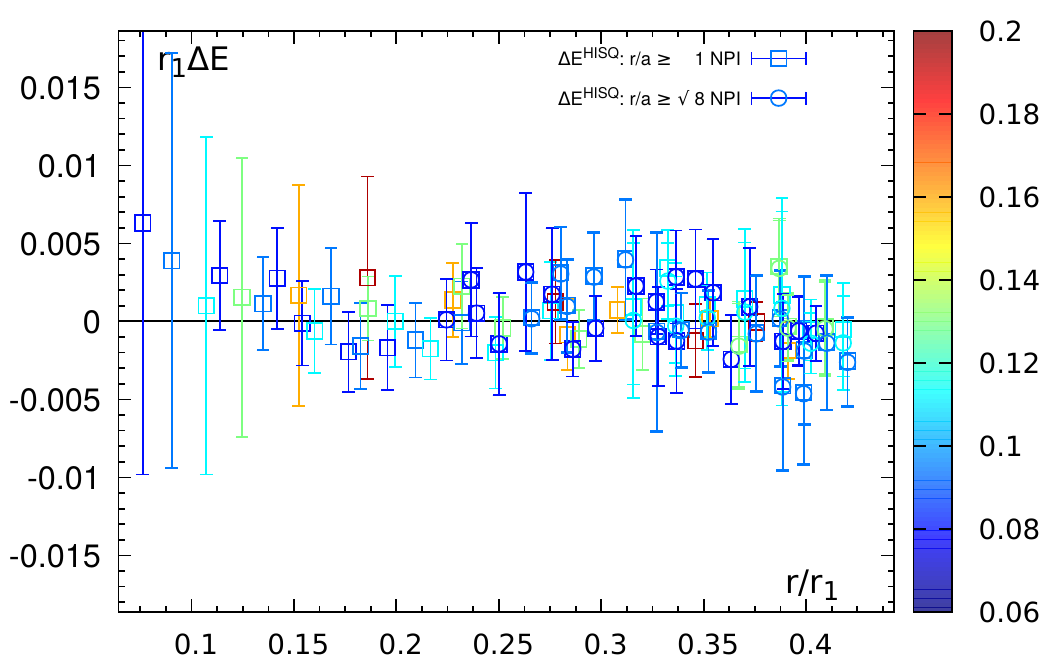}
\caption{\capstretch
Difference between normalized, nonperturbatively improved (NPI) lattice data 
and the weak-coupling result for the static energy in units of \(r_1\). 
For \(\max(r) <0.056\,{\rm fm}\), inaccuracy of the nonperturbative 
improvement appear to be responsible for a slight decrease of the 
central value of \(\als\). 
The four panels correspond to \(\max(r) = 0.055\), \(0.073\), 
\(0.098\), and \(0.131\,{\rm fm}\) (from top to bottom). 
}
\label{fig:diff1}
\end{figure}

We show the differences between the normalized, nonperturbatively improved 
lattice data and the \emph{three-loop with leading ultra-soft resummation} result with standard 
scales in \mbox{Fig.}~\ref{fig:diff1} for fits with \(r/a \ge 1\). 
The difference is generally smaller than \(0.007/r_1\) (about 4 MeV) and 
fluctuates around zero with no apparent trend. 
At distances smaller than \(0.15\,r_1\), the difference tends to be marginally 
below zero for \(\max(r) = 0.055\,{\rm fm}\) and slightly above zero otherwise. 
The central value of \(\als\) is \(0.00027\) lower for fits restricted to the 
shortest distance interval (\(\max(r) = 0.055\,{\rm fm}\)) than for 
\(\max(r) = 0.131\,{\rm fm}\).  
This difference may be indicative of the imperfect removal of the 
discretization artifacts, but is still compatible with the estimate of the 
discretization error using \(\min(r/a)=\sqrt{8}\), and is more than covered 
by statistical uncertainties. 

\begin{table*}
\begin{adjustbox}{angle=00}
\begin{tabular}{|c|c|c|c|c|c|c|c|c|
c|}
\hline
\(\min(\tfrac{r}{a})\) & \(\max(r)\) [fm] & d.o.f. & uncorr. & corr. & 
\(\als\) & \(\delta^{\rm stat}\) & \(\delta^{\rm lat}\) & 
\(\delta^{\rm pert}\) & \(\als^{2L}\)  
\\
\hline
\(\sqrt{1}\) & 0.055 &  6 & 0.14 & 0.18 & 
 0.11641 & 0.00051 & 0.00021 & 
 \(^{+0.00081}_{-0.00026}\) & 0.11638 
\\
\hline
\(\sqrt{1}\) & 0.073 & 17 & 0.22 & 0.27 &
 0.11660 & 0.00041 & 0.00021 & 
 \(^{+0.00099}_{-0.00031}\) & 0.11661 
\\
\hline
\(\sqrt{1}\) & 0.098 & 36 & 0.31 & 0.34 &
 0.11666 & 0.00034 & 0.00021 & 
 \(^{+0.00121}_{-0.00036}\) & 0.11672
\\
\(\sqrt{8}\) & 0.097 & 10 & 0.33 & 0.47 & 
 0.11658 & 0.00072 & 0.00021 &
 \(^{+0.00152}_{-0.00045}\) & 0.11672
\\
\hline
\(\sqrt{1}\) & 0.131 & 74 & 0.35 & 0.40 &
 0.11668 & 0.00027 & 0.00021 & 
 \(^{+0.00154}_{-0.00049}\) & 0.11683
\\
\(\sqrt{8}\) & 0.131 & 42 & 0.33 & 0.40 &
 0.11668 & 0.00051 & 0.00021 &
 \(^{+0.00193}_{-0.00064}\) & 0.11684
\\
\hline

\end{tabular}
\end{adjustbox}
\caption{\capstretch
Representative fits using nonperturbatively improved data 
in different intervals. 
We list uncorrelated and correlated \(\cdf\) in columns 4 and 5. 
In column 9 we list the aggregate perturbative error for the corresponding 
fit window. 
The last column displays the outcome for \(\als\) at two-loop order.
}
\label{tab:summary T=0}   
\end{table*}

We perform another set of tests to investigate the systematic uncertainties 
of the lattice result.
Namely, we test for a bias due to combining multiple lattice spacings 
into a single analysis that assumes that the data are within uncertainties 
consistent with the continuum limit.  
We separately analyze the data at the finest lattice spacing, \(\beta=8.4\). 
Here we also explore whether using data with \(r/a \ge \sqrt{5}\) or 
\(\sqrt{8}\) with errors that are inflated by estimates of the discretization 
artifacts has a significant impact on our results. 
We use the same uncertainty estimate for the influence of the discretization 
artifacts that we obtained previously, since it is larger than any estimates 
that we would obtain by comparing the results in this set of tests. 
Table~\ref{tab:beta 8.4} shows that the separate analysis for the finest 
lattice spacing is consistent within statistical errors or within the even 
smaller estimate of the uncertainties due to the treatment of discretization artifacts. 
While the preferred central value in this analysis appears to be marginally 
higher than in the analysis that combines multiple spacings, this is not 
relevant for our final result. 

\begin{table*}
\begin{adjustbox}{angle=0}
\begin{tabular}{|c|c|c|c|c|c|c|c|c|c|c|c|c|c|c|c|c|c|c|}
\hline
Artifacts & \(\min(r/a)\) & \(\max(r)\) [fm] & d.o.f. & uncorr. & corr. & 
\(\als\) & \(\delta^{\rm stat}\) & \(\delta^{\rm lat}\) & 
\(\delta^{\rm pert}\) 
\\
\hline
 tree-level improvement with syst. errors & \(\sqrt{5}\) & 0.098 & 9 & 0.29 & 0.36 & 
 0.11687 &  0.00040 & 0.00021 &
 \(^{+0.00146}_{-0.00043}\) 
\\
 tree-level improvement with syst. errors & \(\sqrt{8}\) & 0.098 &  7 & 0.28 & 0.35 & 
 0.11685 &  0.00049 & 0.00021 &
 \(^{+0.00157}_{-0.00046}\) 
\\
 tree-level improvement & \(\sqrt{8}\) &  0.098 & 6 & 0.50 & 0.60 & 
 0.11683 &  0.00035 & 0.00021 &
 \(^{+0.00155}_{-0.00043}\) 
\\
\hline
 tree-level improvement with syst. errors & \(\sqrt{5}\) & 0.121 & 18 & 0.37 & 0.45 &
 0.11687 & 0.00038 & 0.00021 &
 \(^{+0.00164}_{-0.00051}\) 
\\
 tree-level improvement with syst. errors & \(\sqrt{8}\) & 0.121 & 16 & 0.38 & 0.46 &
 0.11687 & 0.00044 & 0.00021 &
 \(^{+0.00176}_{-0.00055}\) 
\\
 tree-level improvement & \(\sqrt{8}\) & 0.121 & 15 & 0.48 & 0.58 &
 0.11687 & 0.00033 & 0.00021 &
 \(^{+0.00172}_{-0.00053}\) 
\\
\hline

\end{tabular}
\end{adjustbox}
\caption{\capstretch
Fits using only the finest lattice spacing, \(\beta=8.4\), and different 
treatment of discretization artifacts, \mbox{i.e.}, with or without systematic 
error estimates and the tree-level improvement.
We list uncorrelated and correlated \(\cdf\) in columns 5 and 6. 
In column 10 we list the perturbative error for the corresponding fit window. 
For the lattice error due to discretization artifacts we use the 
results of the analysis with multiple lattice spacings. 
The tree-level improved calculations have fewer degrees of freedom, 
since the data with \(r/a=\sqrt{12}\) have been excluded, see text.
}
\label{tab:beta 8.4}   
\end{table*}

\begin{figure}
\centering
\includegraphics[width=8.6cm]{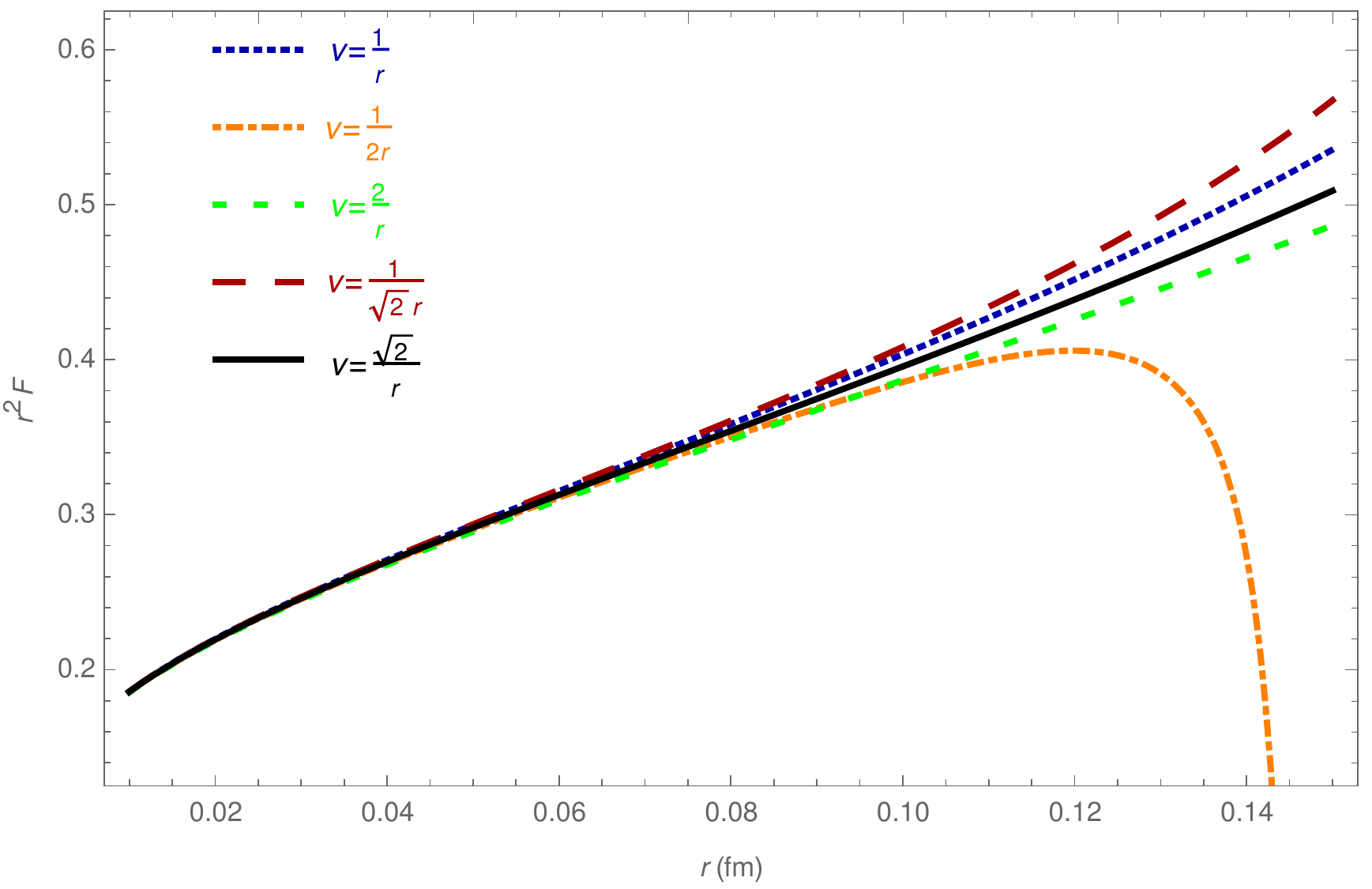}
\caption{\capstretch
The force for different values of the soft scale $\nu$. 
The ultra-soft scale is kept fixed to $\mu_{us} =C_A \als (1/r)/(2r)$. 
We have taken $\lMSbt=315$ MeV.
}
\label{nonmon}
\end{figure}

In order to estimate the perturbative error of the small $\als$ expansion, 
i.e., the error made by neglecting contributions beyond three loops, we 
proceed similarly to \mbox{Ref.}~\cite{Bazavov:2014soa} using three methods 
to quantify the uncertainty of the perturbative calculation. 
Namely, we add and subtract a term with the typical size of the soft four-loop 
contribution, we analyze the dependence on the soft scale $\nu$, and we 
use the difference between the \emph{three-loop order with leading 
ultra-soft resummation} and the three-loop result. 
We estimate the perturbative error in each method as the difference 
between \(\als(M_Z)\) calculated in that way and the \(\als(M_Z)\) value 
obtained from the \emph{three-loop with leading ultra-soft resummation} 
expression at standard scales.
In \mbox{Ref.}~\cite{Bazavov:2014soa}, the error due to the soft scale
variation was estimated by taking $\nu=\sqrt{2}/r$ and $\nu=1/\sqrt{2}r$. 
In this study, we enlarge the range of scale variation by considering 
$\nu=2/r$ and $\nu=1/(2r)$.
In doing so, we realize that the $\nu$ dependence of the extracted 
$\als$ is not monotonous: 
it has a minimum close to $\nu=1/\sqrt{2}r$, see \mbox{Fig.}~\ref{nonmon}. 
For the lowest soft scale \(\nu=1/(2r)\) the use of perturbation theory 
becomes inadequate much earlier than for other choices. 
Namely, for \(\max(r) \gtrsim 0.1\,{\rm fm}\), the fits do not describe 
the data in a satisfying manner, and \(\cdf\) increases much more rapidly 
for \(\nu=1/(2r)\) than for the other choices of the soft scale. 
This is expected on the grounds of the observation that the 2014 analysis 
showed that the scale \(\nu=1/(\sqrt{2}r)\) was adequate up to about 
\(r=r_1/2\). 
If we exclude those larger distances the upper perturbative error is 
always given in terms of the soft scale variation to $\nu=2/r$.
We observe that for \(\max(r) \gtrsim 0.09\,{\rm fm}\) the soft scale 
variation to \(\nu=1/(\sqrt{2}r)\) contributes to a larger estimate of the 
lower perturbative error than the soft higher-order terms, 
in line with the 2014 results.
However, for smaller distances \(\max(r) \lesssim 0.09\,{\rm fm}\) the 
estimate of the lower perturbative error due to inclusion of soft higher-order 
terms exceeds the estimate of the lower perturbative error due to soft scale 
variation. 
We take the most extreme among the error estimates from the soft contribution 
as the upper and lower perturbative errors due to the soft contribution. 
We also estimate the perturbative error by evaluating the difference between 
the \emph{three-loop with leading ultra-soft resummation} result and the 
fixed-order three-loop result.
The fixed-order three-loop result always yields a lower central value than the 
\emph{three-loop with leading ultra-soft resummation} result. 
In order to be conservative, we apply the difference as a symmetric error. 
This has an even larger effect on the lower perturbative error than the 
soft scale variation or the soft higher-order terms. 
Moreover, the soft contribution to the lower perturbative error decreases for 
smaller values of \(\max(r)\) more rapidly than the ultra-soft contribution. 
We stress that the lower perturbative error is always smaller than the 
statistical error for \(r \lesssim 0.1\,{\rm fm}\).  
We add this symmetric perturbative error due to (inclusion or not of) 
ultra-soft resummation to the perturbative error due to the soft contribution 
in quadrature. 
The reasoning for adding these errors in quadrature is that it is unknown 
whether there is a similar partial compensation between soft and ultra-soft 
contributions at higher orders as it happens at three loops, and as such, 
we have to treat the corresponding uncertainties as independent. 
Lastly, we compare the \emph{three-loop with leading ultra-soft resummation} and the two-loop results (first three lines of \mbox{Eq.}~\eqref{eq:F3L}).
As can be read off from \mbox{Tabs.}~\ref{tab:ra ge 8} 
and~\ref{tab:summary T=0}, the difference to the two-loop result never exceeds 
\(+0.00025\) and decreases for smaller values of \(\max(r)\), \mbox{i.e.}, it 
is smaller than the statistical errors and smaller than the other effects due 
variation of soft scale, soft higher order terms, or variation of the 
ultra-soft resummation. 
The \(\cdf\) does not change significantly between using the two-loop or 
\emph{three-loop with leading ultra-soft resummation} results. 
Hence, we confirm the criterion for having the lattice data in the 
perturbative regime. 
We observe that the smaller $\max(r)$ is, the smaller the variation of the 
central value of \(\als\) between fits with different forms of the 
weak-coupling results becomes. 
\mbox{Tab.}~\ref{tab:summary T=0} shows clearly that, for a given 
\(\min(r/a)\), the perturbative errors are dramatically reduced at smaller 
distances, as expected, while the statistical error increases as less data 
are used to constrain the fits.

Let us summarize the considerations of the preceding paragraphs. 
We have to use \(\max(r) \lesssim 0.1\,{\rm fm}\) to perform the full scale 
variation and keep the perturbative uncertainties fully under control. 
We should ideally use significantly more than 10 data point to limit 
the impact of the imperfectly treated discretization artifacts.  
Given the considerations of the preceding paragraphs, we take the result for 
\(1 \le r/a \le 5\) and \(\max(r) =0.073\,{\rm fm}\), namely, \(\als(M_Z)=0.11660\) 
as our final result, which corresponds to \(r_1\lMSbt=0.4943\). 
The uncertainty of the scale \(r_1\) is \(\pm 0.0017\,{\rm fm}\), 
which yields an error of 
\(\delta^{\rm scale}= \pm 1.7\,{\rm MeV}\) for \(\lMSbt\), and 
\(\delta^{\rm scale} = \pm 0.00010\) for \(\als(M_Z, N_f=5)\). 
Therefore, the final result and full error budget of our zero temperature lattice 
calculation are given as 
\al{
 &\phantom{\delta}\als(M_Z, N_f=5) = 0.11660^{+0.00110}_{-0.00056}, 
 \label{eq:as t=0} \\
 &\delta \als(M_Z, N_f=5) = 
 (41)^{\rm stat} (21)^{\rm lat} (10)^{r_1} 
 (^{+95}_{-13})^{\rm soft} (28)^{\rm us} ,
 \label{eq:as t=0 error} 
} 
or in terms of \(\lMSbt\) as 
\al{
 &\phantom{\delta}\lMSbt =  314.0^{+15.5}_{-8.0}\,{\rm MeV}, 
 \label{eq:lq t=0} \\
 &\delta \lMSbt = 
 (5.8)^{\rm stat} (3.0)^{\rm lat} (1.7)^{r_1} 
 (^{+13.4}_{-1.8})^{\rm soft} (4.0)^{\rm us}
 \,{\rm MeV}.
\label{eq:lq t=0 error}
}
We have added the statistical error and the lattice discretization error of 
the static energy, the total error of the \(r_1\) scale, and the perturbative 
error in quadrature. 

In order to compare the current analysis to the previous 
analysis~\cite{Bazavov:2014soa}, we use the smaller window $[1/(\sqrt{2}r),\sqrt{2}/r]$
for the variation of the soft scale $\nu$, and do not account 
for the uncertainty arising from the difference between resumming or not the 
leading ultra-soft logarithms to obtain
\al{
 &\delta^{\sqrt{2}} \als(M_Z, N_f=5) = (41)^{\rm stat} (21)^{\rm lat} (10)^{r_1} (^{+37}_{-13})^{\rm pert}, \\
 &\delta^{\sqrt{2}} \lMSbt = (5.8)^{\rm stat} (3.0)^{\rm lat} (1.7)^{r_1} (^{+5.2}_{-1.8})^{\rm pert}\,{\rm MeV}.
}
In this case, the perturbative error is not dominant anymore. 
Thus, the presented analysis has approximately halved the uncertainties
of \mbox{Ref.}~\cite{Bazavov:2014soa}. 
Nevertheless our final errors are only 10\% (upper error) and 30\% (lower 
error) smaller than the ones in~\cite{Bazavov:2014soa}, since we have 
accounted for the other possible sources of uncertainty listed above.
The central values of Eq.~\eqref{eq:as t=0} and of the final result in
\mbox{Ref.}~\cite{Bazavov:2014soa} coincide.

\section{Extracting \(\als\) from the singlet free  energy}\label{sec:finite}

In this section, we consider the extraction of the strong coupling 
from the singlet free energy at non-zero temperature, as it is expected
that at small distances medium effects are small.
We define the singlet free energy in terms of the correlation function 
of two thermal Wilson lines in Coulomb gauge
\begin{equation}
F_S(r,T)=-T\ln \left(\frac{1}{N_c}\langle {\rm Tr}\left[ W(r) W^{\dagger}(0) \right]\rangle\right).
\end{equation}
At distances much smaller than the inverse temperature 
$r T \ll 1$, we can write using pNRQCD~\cite{Berwein:2017thy}
\begin{equation}
F_S(r,T)=V_s(r,\mu_{us})+\delta F_S(r,T,\mu_{us}),
\end{equation}
where $\mu_{us}$ is the ultra-soft scale.
The form of the thermal correction depends on the scale hierarchy. 
One could consider the case
$1/r \gg \als/r \gg T \gg m_D \sim g T$ or the case
$1/r \gg T \gg m_D \sim g T \gg \als/r $. 
In the former case 
$\mu_{us} \sim \als/r$ and 
$\delta F_S(r,T,\mu_{us})=\delta E_{US}(\mu_{us})+\Delta F_S(r,T)$ 
with $E_{US}(\mu_{us})$ being the ultra-soft contribution to
the static energy in the vacuum. 
In the latter case $\mu_{us} \sim T$ and 
$\delta F_s(r,T,\mu_{us})$ has been calculated to order $g^5$, \mbox{i.e.}, 
see \mbox{Eqs.}~(16)~--~(19) in \mbox{Ref.}~\cite{Bazavov:2018wmo}.
In the latter case the cancellation of the ultra-soft factorization scale 
dependence cannot be verified because of the unknown $g^6$ contribution 
to $\delta F_s(r,T,\mu_{us})$. 
Since $V_s(r,\mu_{us})$ has a term $\sim \als^3 \ln(\mu_{us} r)/r$, however, the 
difference between the $T=0$ static energy and singlet free energy, 
$E(r)-F_S(r,T)$ should have a term $\sim \als^3 \ln (r T)/r$. 
This complicates
the extraction of the strong coupling from the singlet free energy. 
The matching between NRQCD and pNRQCD also induces a term $\sim g^6 T$ 
in $F_S(r,T)$ for both scale hierarchies~\cite{Berwein:2017thy}, 
which also needs to be considered. 

\begin{figure}
\includegraphics[width=9cm]{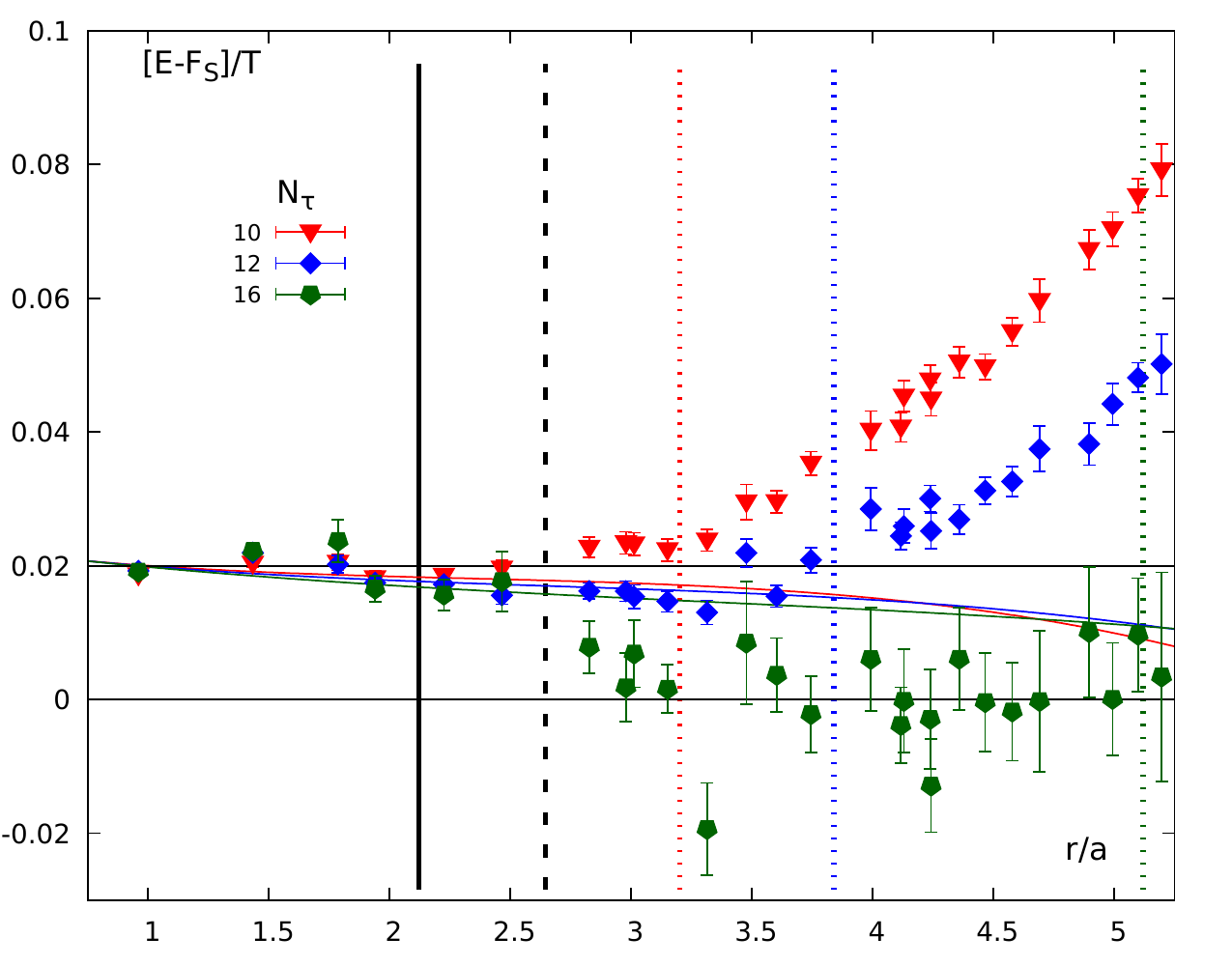}
\caption{\capstretch
The difference between the static energy at $T=0$ and the singlet 
free energy for $\beta=8.4$ calculated with $N_{\tau}=10,~12$ and $16$ in 
units of the temperature as function of distance in units of the lattice spacing. 
The lines correspond to the $\delta F(r,T)$ calculated at order $g^5$. 
The dotted vertical lines show the boundary $r T=0.3$ for different $N_{\tau}$. 
The solid or dashed vertical lines show the boundaries between 
\(r/a < \sqrt{5}\) and \(r/a \ge \sqrt{5}\), or between 
\(r/a < \sqrt{8}\) and \(r/a \ge \sqrt{8}\), respectively.   
}
\label{fig:esmfs}
\end{figure}

The singlet free energy has been studied on the lattice in 
\mbox{Ref.}~\cite{Bazavov:2018wmo} using a wide temperature range and 
several lattice spacings, \mbox{i.e.}, several temporal extents $N_{\tau}$. 
The shortest distance that we can access, due to a single lattice 
spacing on our finest lattice at \(T>0\), is \(0.00814\,{\rm fm}\).
For our analysis the relevant data correspond to $N_{\tau}=10,~12$ and $16$, 
since $rT$ has to be small. 
From the analysis of \mbox{Ref.}~\cite{Bazavov:2018wmo} we know that thermal 
effects are small for $rT \lesssim 0.3$.
To understand the temperature dependence of the singlet free energy in more 
detail we show the lattice results for the difference\footnote{
The discretization artifacts between both quantities cancel exactly at tree 
level. 
Nonetheless, we still use the tree-level improved distance in 
\mbox{Fig.}~\ref{fig:esmfs}, in order to permit the visual distinction between 
data that are inequivalent in the full QCD result.
Moreover, we matched the perturbative result to the lattice data at \(r/a=1\) 
using a constant that mimics the effect of the unknown \(g^6 T\) term. 
} 
between the singlet free energy at \(T>0\) and the static energy at \(T=0\) for $\beta=8.4$ 
(corresponding to the finest zero temperature lattice) in 
\mbox{Fig.}~\ref{fig:esmfs}. 
For other $\beta$ values the results are similar.

From the figure we see that for $r/a \le \sqrt{6}$ the difference approaches a 
constant proportional to the temperature according to the above expectations 
and no temperature effects beyond this constant can be seen in this range 
within the errors of the lattice results. 
Therefore, we conclude that in this regime the scale hierarchy 
$1/r \gg \als/r \gg T$ is appropriate. 
We treat the finite temperature data in this range as if at zero 
temperature and fit them with the \emph{three-loop with leading ultra-soft resummation} result 
for the static energy with standard scales. 
Alternatively one could use the $N_{\tau}=12$ data in the range where thermal 
effects appear to be small together with the two-loop result for the static 
energy, where the problem of the US scale dependence does not enter at all.  
This, however, would lead to a larger theoretical uncertainty.

For distances $r/a>\sqrt{6}$ we see some temperature dependence. 
For the lattice data with $N_{\tau}=12$ this temperature dependence is 
to some extent captured by the known $g^5$ result for $\delta F_S(r,T)$, 
but for $N_{\tau}=10$ and $16$ it is off. 
On the one hand, for $N_{\tau}=16$, \mbox{i.e.}, for a lower temperature, the 
temperature effects are larger than what is expected based on the $g^5$ result. 
Since the deviation of the temperature effects from a constant or from the 
\(g^5\) result is quite similar to the size of the typical discretization 
artifacts or the statistical errors of the data for \(r/a > \sqrt{6}\), it 
is unclear whether this is truly an effect due to the finite temperature. 
On the other hand, for \(N_\tau=10\), where the same temperature 
corresponds to a coarser lattice, the temperature effects are in the 
opposite direction and may be caused by temperature-dependent discretization 
errors~\cite{Bazavov:2018wmo}.

The known \(g^5\) result consists of two contributions with opposite sign but 
similar magnitude for the temperatures under consideration, and, therefore, 
the overall temperature effect is small due to cancellations.
As these two contributions are due to either non-static or static Matsubara 
modes at orders \(g^4\) or \(g^5\), respectively, it is clear that this is 
an accidental cancellation in the temperature window of our lattice simulations. 
Furthermore, there is no reason to expect that discretization artifacts in 
both contributions are similar enough to achieve a cancellation between them. 
In fact, it has been shown that the \(\mathcal{O}(a^2)\) discretization errors 
appear to be more pronounced in the difference \(E-F_S\) than in \(E\) or 
\(F_S\) individually~\cite{Bazavov:2018wmo}. 

For these reasons, we conclude that for $r/a>\sqrt{6}$ we do not have sufficient 
understanding of thermal effects to use $F_S$ for extraction of $\als$ with 
fully controlled uncertainties. 
Nonetheless, the cancellations that appear to be at work both in the \(g^5\) 
result and in the lattice data suggest that a fit of the lattice data with 
the zero temperature result may still be possible as a cross-check. 
Hence, we again treat the finite temperature data in this range as if at 
zero temperature and fit them with the \emph{three-loop with leading ultra-soft resummation} result with standard scales.

In order to determine \(\als\) from the singlet free energy at \(T>0\) 
we proceed as follows. 
We analyze the \(T=0\) static energy result (\(N_\tau=64\)) in the same 
\(r/a\) intervals for which we expect that temperature  effects in the 
singlet free energy are under control or are small due to accidental 
cancellations. 
We use the same weak-coupling result, namely, the 
\emph{three-loop with leading ultra-soft resummation} result with standard scales to obtain 
\(\als\) from the singlet free energy. 
We estimate the uncertainty due to discretization artifacts to be the same 
as in the zero temperature analysis, \mbox{i.e.}, 
\(\delta^{\rm lat}=\pm 0.00021\). 
Similarly, we estimate the uncertainty due to temperature effects from 
the difference of the central values of \(\als\) for any combination of 
\(N_\tau\) values in each fit interval. 

First, we perform fits in the range where temperature effects are expected to 
be constant and the scale hierarchy \(1/r \gg \als/r \gg T \gg m_D \sim g T\) 
applies. 
We report the result of representative fits with \(r/a \le 2\) in 
\mbox{Tab}.~\ref{tab:ra le 4}, which corresponds to \(rT \lesssim 0.17\) 
or \(0.13\) for \(N_\tau=12\) or \(16\), respectively. 
The central values of \(\als\) in fits to the singlet free energy with 
\(N_\tau=12\) or \(16\) bracket the result for the static energy at zero 
temperature, with deviations being smaller than any other uncertainties.  
Hence, we conclude that we do not resolve nonconstant \(T>0\) effects as 
expected.  
As for the \(T=0\) analysis, we see a marginal decrease of the 
central value of \(\als\) upon restriction to shorter distances. 
This decrease is about the same magnitude as in the \(T=0\) analysis, 
but is delayed to significantly shorter distances, which supports our 
interpretation that it is due to imperfections of the nonperturbative 
improvement procedure.
Within the smallest statistical uncertainties or within the even smaller 
estimates of the lattice discretization uncertainties, all of these results 
are consistent with the zero temperature analysis of the previous section, 
while the individual estimates of perturbative errors are systematically 
smaller, compare with 
\mbox{Tab.}~\ref{tab:summary T=0}. 

Next, we perform fits in the range where temperature effects beyond a constant 
appear to be small due to accidental cancellations. 
We assume the same scale hierarchy \(1/r \gg \als/r \gg T \gg m_D \sim g T\), 
noting that it is at present not verifiable if the ultra-soft factorization 
scale dependence cancels in this case.   
Namely, we perform the fits with \(r/a \le 3\) and report the results in 
\mbox{Tab}.~\ref{tab:ra le 4} as well\footnote{
Using an even larger fit range up to \(r/a \le \sqrt{12}\) leads to the same conclusions, \mbox{i.e.}, 
the nonconstant thermal effects are still numerically irrelevant.
}. 
This corresponds to \(rT \lesssim 0.25\) or \(0.19\)
for \(N_\tau=12\) or \(16\) respectively.
\(\cdf\) is practically unchanged for \(N_\tau=12\), but increases slightly for \(N_\tau=16\). 
Remarkably, the agreement with the previous \(T=0\) analysis, using up to 
\(r/a \le 5\), does not change significantly as \(\max(r)\) becomes larger. 
This suggests that the differences between the results are not caused by 
effects of the finite temperature, but rather by the more severe influence 
of the imperfections of the nonperturbative improvement procedure in 
the \(T=0\) analysis with \(r/a \le 2\) or \(r/a \le 3\). 
Hence, the estimate of the \(\delta^{T>0}\) errors for \(r>0.05\,{\rm fm}\) 
cannot be separated from the \(\delta^{\rm lat}\) errors due to lattice 
discretization uncertainties. 

\begin{table*}
\begin{adjustbox}{angle=0}
\begin{tabular}{|c|c|c|c|c|c|c|c|c|c|c|c|c|c|}
\hline
 \(\min(\tfrac{r}{a})\) & 
 \(\max(\tfrac{r}{a})\) & \(N_\tau\) & \(\max(rT)\) & \(\max(r)\) [fm] & d.o.f. & uncorr. & corr. & 
 \(\als\) & \(\delta^{stat}\) & \(\delta^{syst}\) & \(\delta^{T>0}\) & 
 \(\delta^{pert}\) & \(\als^{2L}\) 
\\
\hline
 1 & 2 & 16 & 0.125 & 0.030 & 17 & 0.12 & 0.14 & 
 0.11621 & 0.00098 & 0.00021 & 0.00017 & -- & --
\\
 1 & 2 & 12 & 0.167 & 0.030 & 17 & 0.18 & 0.22 & 
 0.11638 & 0.00080 & 0.00021 & 0.00017 &
 \(^{+0.00043}_{-0.00016}\) & 0.11629 
\\
\hline
 1 & 2 & 64 & n/a & 0.057 & 6 & 0.12 & 0.13 & 
 0.11646 & 0.00057 & 0.00021 & n/a & 
 \(^{+0.00081}_{-0.00026}\) & 0.11643 
\\
 1 & 2 & 16 & 0.125 & 0.057 & 30 & 0.19 & 0.20 & 
 0.11640 & 0.00089 & 0.00021 & 0.00006 & -- & --
\\
 1 & 2 & 12 & 0.167 & 0.057 & 30 & 0.16 & 0.18 & 
 0.11651 & 0.00074 & 0.00021 & 0.00005 &
 \(^{+0.00060}_{-0.00021}\) & 0.11644 
\\
\hline
 1 & 2 & 64 & n/a & 0.078 & 12 & 0.15 & 0.16 & 
 0.11657 & 0.00051 & 0.00021 & n/a &
 \(^{+0.00100}_{-0.00030}\) & 0.11658 
\\
 1 & 2 & 16 & 0.125 & 0.078 & 35 & 0.22 & 0.24 & 
 0.11651 & 0.00082 & 0.00021 & 0.00006 & -- & --
\\
 1 & 2 & 12 & 0.167 & 0.078 & 37 & 0.19 & 0.22 & 
 0.11663 & 0.00063 & 0.00021 & 0.00006 &
 \(^{+0.00081}_{-0.00025}\) & 0.11661  
\\
\hline
 1 & 2 & 64 & n/a & 0.096 & 15 & 0.15 & 0.16 & 
 0.11659 & 0.00048 & 0.00021 & n/a &
 \(^{+0.00112}_{-0.00031}\) & 0.11663 
\\
 1 & 2 & 12 & 0.167 & 0.091 & 41 & 0.22 & 0.25 & 
 0.11667 & 0.00063 & 0.00021 & 0.00008 &
 \(^{+0.00088}_{-0.00027}\) & 0.11667 
\\
\hline
\hline
 1 & 3 & 16 & 0.1875 & 0.030 & 30 & 0.34 & 0.39 & 
 0.11658 & 0.00073 & 0.00021 & 0.00003 & -- & --
\\
 1 & 3 & 12 & 0.25 & 0.030 & 30 & 0.20 & 0.28 & 
 0.11661 & 0.00058 & 0.00021 & 0.00003 &
 \(^{+0.00046}_{-0.00017}\) & 0.11652 
\\
\hline
 1 & 3 & 64 & n/a & 0.055 & 6 & 0.14 & 0.18 & 
 0.11641 & 0.00051 & 0.00021 & n/a & 
 \(^{+0.00081}_{-0.00026}\) & 0.11638 
\\
 1 & 3 & 16 & 0.1875 & 0.058 & 69 & 0.42 & 0.46 & 
 0.11672 & 0.00068 & 0.00021 & 0.00031 & -- & -- 
\\
 1 & 3 & 12 & 0.25 & 0.057 & 68 & 0.20 & 0.26 & 
 0.11671 & 0.00054 & 0.00021 & 0.00030 &
 \(^{+0.00062}_{-0.00021}\) & 0.11665 
\\
\hline
1 & 3 & 64 & n/a & 0.073 &  17 & 0.22 & 0.27 & 
 0.11660 & 0.00041 & 0.00021 & n/a &
 \(^{+0.00099}_{-0.00031}\) & 0.11661 
\\
 1 & 3 & 16 & 0.1875 & 0.077 & 82 & 0.56 & 0.61 & 
 0.11682 & 0.00064 & 0.00021 & 0.00022 & -- & --
\\
 1 & 3 & 12 & 0.25 & 0.077 & 84 & 0.24 & 0.30 & 
 0.11680 & 0.00047 & 0.00021 & 0.00020 &
 \(^{+0.00078}_{-0.00026}\) & 0.11677 
\\
\hline
1 & 3 & 64 & n/a & 0.096 &  28 & 0.28 & 0.30 & 
 0.11665 & 0.00035 & 0.00021 & n/a &
 \(^{+0.00115}_{-0.00035}\) & 0.11670 
\\
 1 & 3 & 16 & 0.1875 & 0.098 & 89 & 0.66 & 0.71 & 
 0.11686 & 0.00063 & 0.00021 & 0.00021 & -- & --
\\
 1 & 3 & 12 & 0.25 & 0.096 & 95 & 0.14 & 0.17 & 
 0.11682 & 0.00045 & 0.00021 & 0.00017 &
 \(^{+0.00090}_{-0.00028}\) & 0.11682 
\\
\hline
1 & 3 & 64 & n/a & 0.134 & 40 & 0.39 & 0.44 & 
 0.11668 & 0.00031 & 0.00021 & n/a &
 \(^{+0.00142}_{-0.00045}\) & 0.11680 
\\
 1 & 3 & 12 & 0.25 & 0.133 & 109 & 0.29 & 0.32 & 
 0.11684 & 0.00040 & 0.00021 & 0.00016 &
 \(^{+0.00115}_{-0.00037}\) & 0.11690 
\\
\hline
\end{tabular}
\end{adjustbox}
\caption{\capstretch
Fits with \(r/a \le 2\), or \(r/a \le 3\).
We list uncorrelated and correlated \(\cdf\) in columns 7 and 8. 
In column 13 we list the perturbative error for the corresponding fit window. 
The last column displays the outcome for \(\als\) at two-loop order.
\(\delta^{\rm pert}\) or \(\als^{2L}\) for \(N_\tau=16\) were not computed 
because the \(N_{\tau}=16\) data are not used for the final $\alpha_s$ result. 
Note that the \(\max(rT)\) and \(\delta^{T>0}\) columns are not 
applicable for the $N_{\tau}=64$ results.
}
\label{tab:ra le 4}   
\end{table*}

\begin{figure}
\centering
\includegraphics[width=8.6cm]{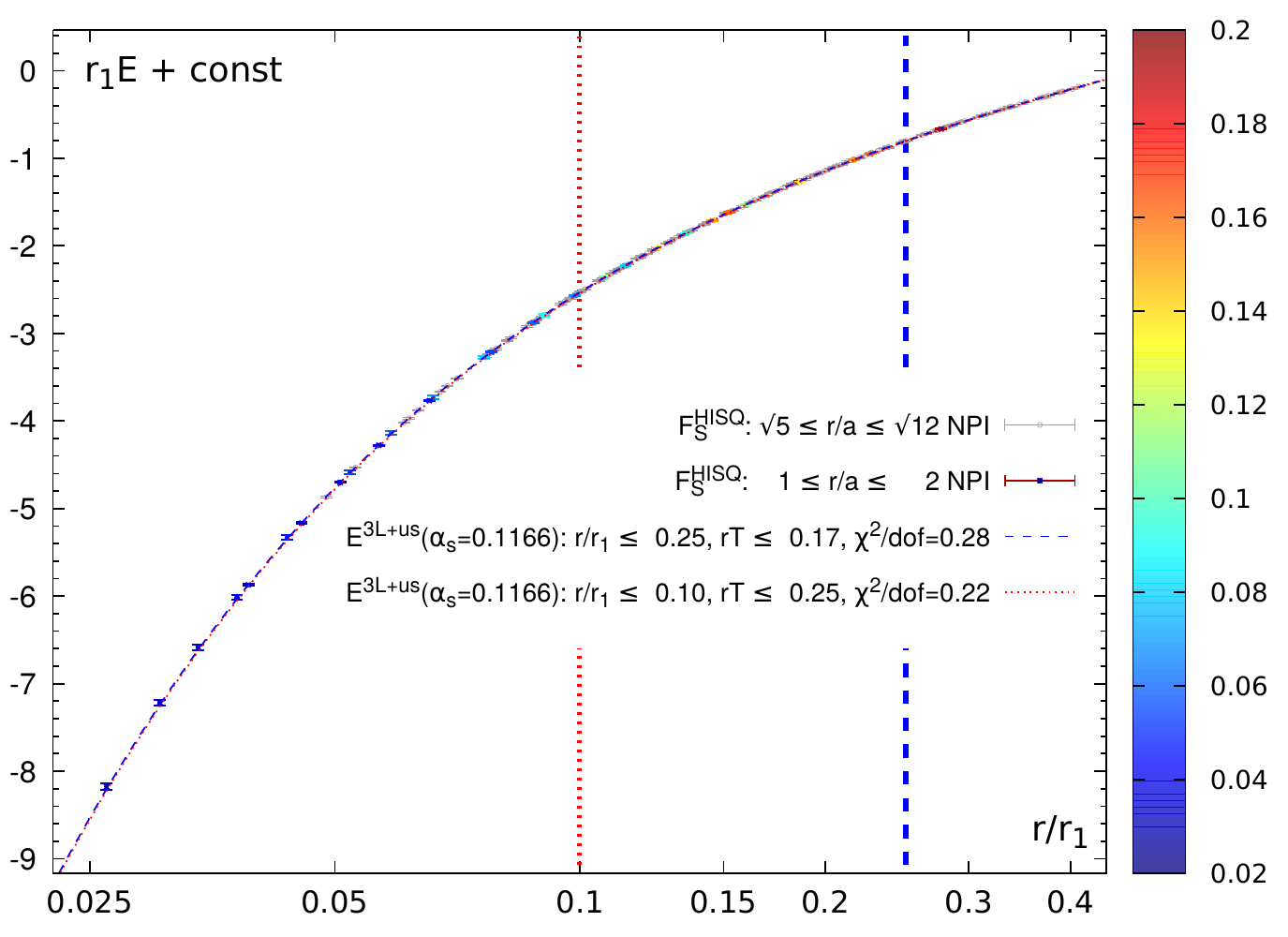}
\caption{\capstretch
Normalized lattice data and weak-coupling result for the static energy 
in units of \(r_1\).
We use a logarithmic scale for the coordinate axis. 
The colored or gray bullets show the nonperturbatively improved (NPI) lattice data for \(r/a \le 2\) or \(2 < r/a \le \sqrt{12}\).
The \emph{three-loop with leading ultra-soft resummation} with standard scales is shown for the 
\(\als(M_Z)\) grid values corresponding to the best fits for the \(r\) and 
\(rT\) intervals as indicated. 
The vertical lines of the same color indicate \(\max(r)\) of the fits. 
}
\label{fig:fintemp}
\end{figure}

Let us summarize the considerations of the preceding paragraphs. 
For \(T>0\) we have to use \(r/a \lesssim 2\) to guarantee the cancellation of 
ultra-soft factorization scale dependence, although we do not see any 
indication that it does not work for somewhat larger distances, \mbox{i.e.}, 
\(r/a \le 3\).  
The weak-coupling picture suggests that cancellation between different 
medium effects is responsible for the small thermal modification of the 
singlet free energy. 
While the medium effects according to the weak-coupling picture are not 
unambiguously resolved in the data, the data seem to be affected by an accidental 
cancellation in the temperature window under consideration. 
Perturbative uncertainties are dramatically reduced at smaller distances \(\max(r)\). 
In \mbox{Fig.}~\ref{fig:fintemp} we show our final \(T=0\) result with the 
reported \(\cdf\) for two different fits to the singlet free energy data. 
The data show no nontrivial thermal effects in the ranges considered and 
are consistent with the central value of the \(T=0\) result. 

Given the considerations of the preceding paragraphs, we take the 
\(N_\tau=12\) result for \(1 \le r/a \le 2\) and \(\max(r)=0.030\,{\rm fm}\), 
namely, \(\als(M_Z)=0.11638\) as our final result, 
which corresponds to \(r_1\lMSbt=0.4900\).
Furthermore, it has a largely independent error budget.
We note that the finite temperature result for these distances does not depend 
on gauge ensembles corresponding to the larger sea quark mass, \mbox{i.e.}, uses exclusively ensembles with \(m_l=m_s/20\). 
The scale uncertainty is the same as in the \(T=0\) analysis, 
\(\delta^{\rm scale}= \pm 1.7\,{\rm MeV}\) for \(\lMSbt\), and 
\(\delta^{\rm scale} = \pm 0.00010\) for \(\als(M_Z, N_f=5)\). 
Therefore, the final result and full error budget of our finite temperature 
lattice calculation are given as 
\al{
 &\phantom{\delta}\als(M_Z, N_f=5) =0.11638^{+0.00095}_{-0.00087}, 
 \label{eq:as t>0} \\
 &\delta \als(M_Z, N_f=5) = 
 \nonumber\\&\phantom{==}
 (80)^{\rm stat} (21)^{\rm lat} (17)^{T>0} (10)^{r_1} 
 (^{+40}_{-06})^{\rm soft} (15)^{\rm us},
 \label{eq:as t>0 error} 
} 
or in terms of \(\lMSbt\) as 
\al{
 &\lMSbt =  310.9^{+13.5}_{-12.3}\,{\rm MeV}, 
 \label{eq:lq t>0} \\
 &\delta \lMSbt =
 \nonumber\\&\phantom{==} 
 (11.3)^{\rm stat} (3.0)^{\rm lat} (2.4)^{T>0} (1.7)^{r_1} 
 (^{+5.6}_{-0.8})^{\rm soft} (2.1)^{\rm us}\,{\rm MeV}.
 \label{eq:lq t>0 error} 
}
We have added the statistical error, the lattice discretization error of the 
static energy, the finite temperature error due to using the singlet free 
energy, the total error of the \(r_1\) scale, and the perturbative 
error in quadrature. 

\section{Discussion}\label{sec:disc}

In the sections~\ref{sec:static} and \ref{sec:finite}, we have 
presented two extractions of the strong coupling constant from two different 
observables, the static energy on zero temperature lattices, and the singlet 
free energy on finite temperature lattices. 

Both results, \mbox{i.e.}, \(\als(M_Z,N_f=5)=0.11660^{+0.00110}_{-0.00056}\), 
and \(\als(M_Z,N_f=5)=0.11638^{+0.00095}_{-0.00087}\), are in excellent 
agreement.
This is despite the facts that they are based on several statistically 
independent sets of gauge ensembles corresponding to the QCD vacuum  or the 
quark-gluon plasma and have been obtained using two different sea quark masses. 
The systematic uncertainties due to the lattice discretization and due to the 
lattice scale are not independent, but do not play a significant role 
in either of the two error budgets. 
Moreover, the two results are obtained in different \(r\) ranges, 
\([0.0237\,{\rm fm},0.0734\,{\rm fm}]\), or \([0.0081\,{\rm fm},0.0301\,{\rm fm}]\), 
which  overlap only in a narrow window. 
As such, we may even consider the perturbative uncertainties of the two 
analyses as practically independent. 
We stress that every single fit in our analysis is contained already 
within the purely statistical errors of either the final \(T=0\) or the final 
\(T>0\) results as long as the soft scale is at \(\nu=1/r\). 

\begin{figure}
\centering
\includegraphics[width=8.6cm]{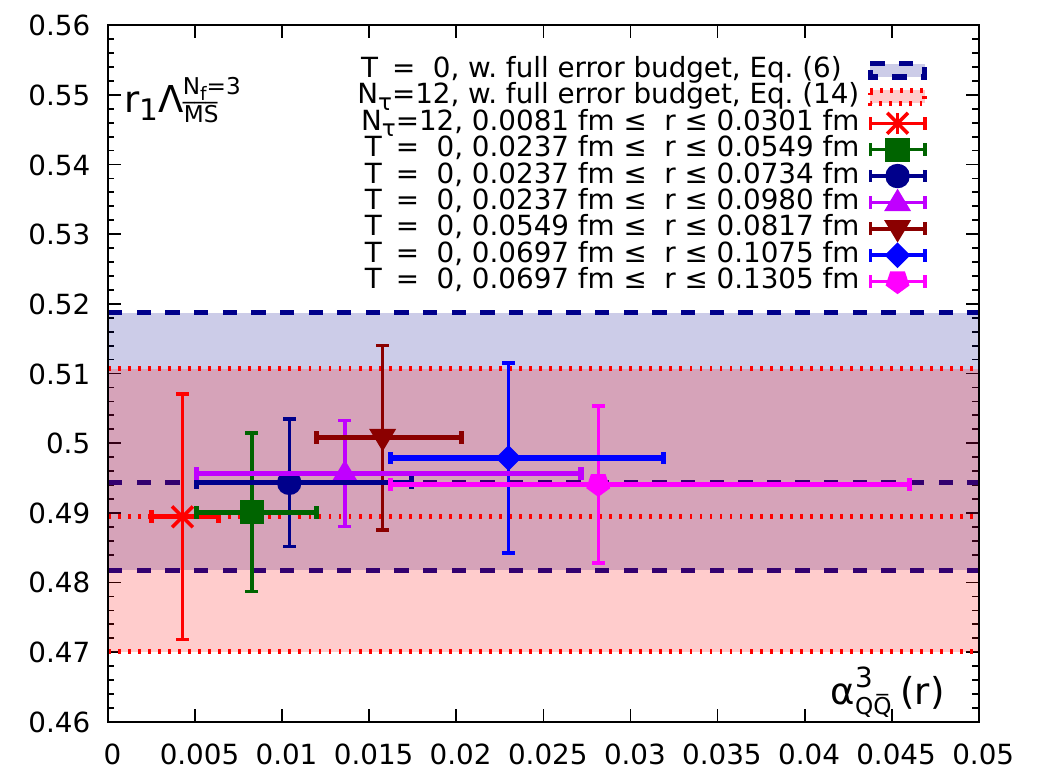}
\caption{\capstretch
The QCD scale \(r_1 \lMSbt\) as a function of \(\alpha_{\qbq}^3(r)\). 
The error bars of \(r_1 \lMSbt\) correspond to statistical errors only. 
For \(\alpha_{\qbq}^3(r)\) we use the value at \(r=(\min(r)+\max(r))/2\) as the central value, and the error bars of \(\alpha_{\qbq}^3\) 
correspond to varying \(r\) from \(\min(r)\) to \(\max(r)\) (statistical errors are negligible).  
The data with \(0.0549\,{\rm fm} \le r\) is selected from our \(T=0\) 
analysis with \(r/a \ge \sqrt{5}\) using only \(\beta=8.4\), compare with 
\mbox{Tab.}~\ref{tab:beta 8.4}.
The data with \(0.0697\,{\rm fm} \le r\) are selected from our \(T=0\) 
analysis with \(r/a \ge \sqrt{8}\), compare with 
\mbox{Tab.}~\ref{tab:ra ge 8}. 
All other data sets are with \(r/a \ge 1\). 
All individual results are in agreement with our final results, 
\mbox{Eqs.}~\eqref{eq:lq t=0} and~\eqref{eq:lq t>0}, which are displayed 
as overlapping bands with the corresponding full error budgets.  
}
\label{fig:aqq3}
\end{figure}

There has been an observation in pure Yang--Mills theory that the approach to 
the continuum limit of \(r_0\lMSb^{N_f=0}\) extracted in the \(\qbq\)-scheme\footnote{
The gauge coupling in the \(\qbq\)-scheme is defined as  
\(\alpha_{\qbq}(1/r)=r^2 F(r)/C_F\).} 
appears to change if data with \(\alpha_{\qbq}^3 \lesssim 0.01\) are 
considered~\cite{Husung:2017qjz}. 
Namely, shorter distances corresponding to smaller 
\(\alpha_{\qbq}\) seemed to be related to smaller \(r_0\lMSb^{N_f=0}\) 
in the continuum limit, with about 10\% lower continuum value in the pure 
Yang--Mills theory analysis. 
We show such a comparison for our data in \mbox{Fig.}~\ref{fig:aqq3}. 
The fits with \(\max(r) \lesssim 0.05\,{\rm fm}\) are safely in the 
\(\alpha_{\qbq}^3 \lesssim 0.01\) range, and are consistent with either a 
stable value or a modest drop. To account for possible artifacts in our 
analysis due to the use of small \(r/a\), we have included the 
systematic error estimate \(\delta^{\rm lat}= \pm 0.00021\).

Our result from the static energy supersedes the result of the previous 
analysis~\cite{Bazavov:2014soa}, \(\als(M_Z,N_f=5) = 0.1166^{+0.0012}_{-0.0008}\), 
obtained using three gauge ensembles with light quark masses \(m_l=m_s/20\). 
Although it reproduces the previous central value, it includes more 
conservative estimates of the known sources of systematic error. 
Nevertheless it achieves to reduce the uncertainties through the 
use of finer lattice spacings.

In \mbox{Ref.}~\cite{Takaura:2018vcy} the strong coupling was calculated 
from the static energy using two different analyses of spatially-smeared 
Wilson loop data on three gauge ensembles with a pion mass of 
about \(300\,{\rm MeV}\), which are both consistent with our results. 
The presented two-step analysis (obtaining the continuum limit first 
and then extracting the strong coupling) yielded 
\(\als(M_Z,N_f=5) = 0.1166^{+0.0021}_{-0.0020}\), while the second analysis using a single global fit yielded a higher value 
\(\als(M_Z,N_f=5) = 0.1179^{+0.0015}_{-0.0014}\). 

Our result is slightly smaller than the PDG average~\cite{Tanabashi:2018oca} 
\(\als(M_Z,N_f=5) = 0.1181(11)\) as well as the recent FLAG average~\cite{Aoki:2019cca} \(\als(M_Z,N_f=5) = 0.11823(81)\). 
It is also smaller than the result of the ALPHA 
collaboration~\cite{Bruno:2017gxd} \(\als(M_Z, N_f=5)=0.11852(84)\), 
which uses step scaling and the Schr\"odinger functional method. These three values are, however, correlated. 
Our result is smaller than the results of the HPQCD collaboration using 
pseudoscalar quarkonium correlators or small Wilson 
loops~\cite{McNeile:2010ji,Chakraborty:2014aca} with three or four sea quark 
flavors, namely, \(\als(M_Z,N_f=5) =0.1183(7)\), or 
\(\als(M_Z,N_f=5) =0.11822(74)\), respectively. 
Our result agrees with the very recent lattice determination in terms of 
pseudoscalar quarkonium correlators~\cite{Petreczky:2019ozv}, 
\(\als(M_Z,N_f=5) = 0.1159(12)\), or the preceding analysis of charmonium 
correlators~\cite{Maezawa:2016vgv}, \(\als(M_Z,N_f=5) = 0.11622(83)\), 
or the analyses of higher quarkonium moments~\cite{Nakayama:2016atf} 
\(\als(M_Z,N_f=5)=0.1176(26)\). 
Moreover, our result is consistent within errors with the recent lattice 
determinations from the hadronic vacuum polarization~\cite{Hudspith:2018bpz} 
\(\als(M_Z,N_f=5)=0.1181(27)^{+0.0008}_{-0.0022}\), or from the gauge-fixed gluon 
propagator in Landau gauge~\cite{Zafeiropoulos:2019flq}, 
\(\als(M_Z,N_f=5)=0.1172(11)\). 

Our result is also consistent with phenomenological estimates 
based on the bottomonium spectrum in NRQCD~\cite{Mateu:2017hlz}, \(\als(M_Z,N_f=5)=0.1178(51)\), 
or quarkonia spectra in pNQRCD~\cite{Peset:2018ria}, \(\als(M_Z,N_f=5)=0.1195(53)\), 
which have errors comparable to an earlier extraction from radiative 
bottomonium decays~\cite{Brambilla:2007cz}, \(\als(M_Z,N_f=5)=0.119^{+0.006}_{-0.005}\). 
Finally, a recent study of the static energy using a subset of our data in a 
much larger \(r\) window, with an assumption similar to the one of
\mbox{Ref.}~\cite{Takaura:2018vcy} and approximations to the weak-coupling 
result that prevent the quantification of the uncertainties, finds
\(\als(M_Z,N_f=5)=0.1168\)~\cite{Mateu:2018zym}, which agrees with our result.

\section{Conclusions}\label{sec:concl}

In this paper, we have calculated the static energy at zero temperature and the 
singlet free energy at finite temperature in 2+1 flavor QCD using the HISQ 
action, and determined the strong coupling. 
We improved and extended the previous 2+1 flavor HISQ analysis published
in \mbox{Ref.}~\cite{Bazavov:2014soa}. 
We included three finer lattice spacings, overhauled the analysis strategy, 
treated systematic uncertainties more conservatively, and restricted the analysis 
to much shorter distances that are deeper in the perturbative regime.  
Our main results are given in 
\mbox{Eqs.}~\eqref{eq:as t=0}~to~\eqref{eq:lq t=0 error}.
Moreover, for the first time we used the singlet free energy on finite 
temperature lattices to reach much smaller distances than ever used 
for the analysis of the static energy, obtaining an independent result 
with a complementary error budget given in 
\mbox{Eqs.}~\eqref{eq:as t>0}~to~\eqref{eq:lq t>0 error}. 
Both results agree without even having to invoke the systematic errors of the 
weak-coupling  calculation. 

Both of our results for the central value of $\als$ are lower than a few lattice 
QCD determinations as well as the PDG average. 
However, the result from the static energy is compatible within errors. 
Furthermore, our results agree with some other lattice determinations of
$\als$ based on moments of quarkonium correlators~\cite{Petreczky:2019ozv}.
Finally, we stress that the extraction of $\als$ from the static energy is 
the only lattice extraction based, in the continuum part, on an
observable known up to three loops and accurate at order $\als^4$. 
This is due to the fact that the static energy is sensitive to \(\als\) 
already at leading order.

\section*{Acknowledgments}

This work was supported by U.S. Department of Energy under 
Contract No. DE-SC0012704. 
\mbox{J.S.} has been supported by the FPA2016-76005-C2-1-P
and FPA2016-81114-P projects (Spain), and the 2017-SGR-929 grant
(Catalonia). 
The simulations have been carried out on the computing facilities of 
the Computational Center for Particle and Astrophysics (C2PAP) in the project 
``Static quark correlators in lattice QCD at non-zero temperature'' (pr83pu) 
as well as on the SuperMUC at the Leibniz-Rechenzentrum (LRZ) in the project 
``Static quark-antiquark energy at zero and finite temperature'' (pr48le). 
We thank the Institute for Nuclear Theory at the University of Washington 
for its kind hospitality and stimulating research environment. 
This research was supported in part by the INT's \mbox{U.S.} Department of Energy grant 
\mbox{No.} DE-FG02-00ER41132. 
This research was supported by the Munich Institute for Astro- and Particle 
Physics (MIAPP) of the DFG cluster of excellence ``Origins'' (www.origins-cluster.de). 
The lattice QCD calculations have been performed using the publicly available
MILC code. 
The data analysis was performed using the
\emph{R-base} and \emph{nlme} packages~\cite{Rpackage, nlme}. 

\appendix

\section{Gauge ensembles and correlators}\label{app:ensembles}

\begin{table} 
\parbox{.98\linewidth}{
  \begin{tabular}{|c|c|c|c|c|c|c|c|}
    \hline
    \multicolumn{8}{|c|}{\(m_l=m_s/20\):} \\
    \hline
    $ \beta $ & a\,(\rm{fm}) & \(N_\sigma,N_\tau\) & $am_s$ & \(m_\pi L\) & \#TUs & \#MEAS & Ref. \\
    \hline
    6.664 & 0.117 & \(32^4\) & 0.0514  & 3.0 & 4000 & 400 & \cite{Bazavov:2011nk}  \\
    6.740 & 0.109 & \(48^4\) & 0.0476  & 4.2 & 4000 & 800 & \cite{Bazavov:2011nk}  \\
    6.800 & 0.103 & \(32^4\) & 0.0448  & 2.7 & 4000 & 400 & \cite{Bazavov:2011nk}  \\
    6.880 & 0.095 & \(48^4\) & 0.0412  & 3.7 & 8200 & 820 & \cite{Bazavov:2011nk}  \\
    6.950 & 0.089 & \(32^4\) & 0.0386  & 2.3 & 19400 & 1940 & \cite{Bazavov:2011nk}  \\
    7.030 & 0.082 & \(48^4\) & 0.0356  & 3.2 & 7800 & 780 & \cite{Bazavov:2011nk}  \\
    7.150 & 0.070 & \(64^3\times 48\) & 0.0320 & 2.7 & 4047 & 667 & \cite{Bazavov:2011nk}  \\
    7.280 & 0.065 & \(64^3\times 48\) & 0.0284 & 2.5 & 3978 & 650 & \cite{Bazavov:2011nk}  \\
  \hline
    \hline
    \multicolumn{8}{|c|}{\(m_l=m_s/5\):} \\
    \hline
    $ \beta $ & a\,[\rm{fm}] & \(N_\sigma,N_\tau\) & $am_s$ & \(m_\pi L\) & \#TUs & \#MEAS & Ref. \\
    \hline
    7.030 & 0.082 & \(48^4\) & 0.0712  & 6.4 & 2990 & 598 & \cite{Bazavov:2017dsy} \\
    7.825 & 0.040 & \(64^4\) & 0.01542 & 4.0 & 2980 & 298 & \cite{Bazavov:2017dsy} \\
  \hline
  \end{tabular}
  \caption{\capstretch
\label{tab:t0 coarse}
  Parameters for the coarse $T=0$ ensembles. 
  In the seventh column we indicate the number of 
  correlator measurements performed.
  }
}
\end{table}

\begin{table}
\parbox{.98\linewidth}{
  \begin{tabular}{|c|c|c|c|c|}
    \hline
    \multicolumn{5}{|c|}{\(m_l=m_s/20\):} \\
    \hline
    $ \beta $ & $am_s$ & T (MeV) & \#TUs & Ref. \\
    \hline
    7.373 & 0.0250  &  273 &  85120 & \cite{Bazavov:2014pvz} \\
    7.596 & 0.0202  &  334 &  98010 & \cite{Bazavov:2014pvz} \\
    7.650 & 0.0202  &  350 &   3230 & \cite{Bazavov:2013uja} \\
    7.825 & 0.0164  &  408 & 134600 & \cite{Bazavov:2014pvz} \\
    8.570 & 0.008376 &  770 & 6320 & \cite{Bazavov:2018wmo} \\
    8.710 & 0.007394 &  866 & 6490 & \cite{Bazavov:2018wmo} \\
    8.850 & 0.006528 &  974 & 6340 & \cite{Bazavov:2018wmo} \\
    9.060 & 0.004834 & 1162 & 7430 & \cite{Bazavov:2018wmo} \\
    9.230 & 0.004148 & 1340 & 7280 & \cite{Bazavov:2018wmo} \\
    9.360 & 0.003691 & 1495 & 7910 & \cite{Bazavov:2018wmo} \\
    9.490 & 0.003285 & 1667 & 9780 & \cite{Bazavov:2018wmo} \\
    9.670 & 0.002798 & 1938 & 7650 & \cite{Bazavov:2018wmo} \\
    \hline
    \hline
    \multicolumn{5}{|c|}{\(m_l=m_s/5\):} \\
    \hline
    $ \beta $ & $am_s$ & T (MeV) & \#TUs & Ref. \\
    \hline
    8.000 & 0.01299  &  474 &   71670 & \cite{Bazavov:2018wmo} \\
    8.200 & 0.01071 &  562 &  71390 & \cite{Bazavov:2018wmo} \\
    8.400 & 0.00887 &  667 &  71170 & \cite{Bazavov:2018wmo} \\
    \hline
  \end{tabular}
  \caption{\capstretch
\label{tab:nt 12}
  Parameters of $N_{\tau}=12$ ensembles with aspect ratio $4$ and $m_l=m_s/20$. 
  Adjacent correlators are separated by $10$ TUs.
  }
}
\end{table}

\begin{table}[t]
\parbox{1.0\linewidth}{
  \begin{tabular}{|c|c|c|c|c|c|}
    \hline
    \multicolumn{5}{|c|}{\(m_l=m_s/20\):} \\
    \hline
    $ \beta $  & $am_s$ & T (MeV) & 
    \#TUs & $ L^{\mathrm{bare}} $, Ref.    \\
    \hline
    7.825 & 0.00164  & 306  & 67960 & 0.002962(06) \\
    8.570 & 0.008376 & 577  & 10400 & \cite{Bazavov:2018wmo} \\
    8.710 & 0.007394 & 650  & 10190 & \cite{Bazavov:2018wmo} \\
    8.850 & 0.006528 & 731  &  4480 & \cite{Bazavov:2018wmo} \\
    9.060 & 0.004834 & 872  & 41870 & 0.015993(20) \\
    9.230 & 0.004148 & 1005 &  3610 & \cite{Bazavov:2018wmo} \\
    9.360 & 0.003691 & 1121 &  3530 & \cite{Bazavov:2018wmo} \\
    9.490 & 0.003285 & 1250 &  6790 & \cite{Bazavov:2018wmo} \\
    9.670 & 0.002798 & 1454 & 42060 & 0.025530(29) \\
    \hline
    \hline
    \multicolumn{5}{|c|}{\(m_l=m_s/5\):} \\
    \hline
    $ \beta $  & $am_s$ & T (MeV) & 
    \#TUs & $ L^{\mathrm{bare}} $    \\
    \hline
    8.000 & 0.001299 & 356  & 11460 & \cite{Bazavov:2018wmo} \\
    8.200 & 0.001071 & 422  & 10660 & \cite{Bazavov:2018wmo} \\
    8.400 & 0.000887 & 500  & 64370 & 0.007819(10) \\
    \hline
  \end{tabular}
  \caption{\capstretch
\label{tab:nt 16}
  Parameters of the new $N_{\tau}=16$ ensembles with aspect ratio 
  \(4\) and expectation values of bare Polyakov loops. 
  Adjacent correlators are separated by $10$ TUs.
  }
}
\end{table}

In this appendix, we discuss additional ensembles that have not been 
summarized in \mbox{Tab.}~\ref{tab:t0 fine}, since they are used only 
for auxiliary calculations, and fits to the primary lattice correlators. 
Further zero temperature ensembles that have only been used for developing 
a procedure to treat the discretization artifacts are summarized in 
\mbox{Tab.}~\ref{tab:t0 coarse}. 
They also have been generated by the HotQCD collaboration~\cite{Bazavov:2011nk, 
Bazavov:2014pvz} using the RHMC algorithm and use 2+1 flavors with the strange 
quark mass at its physical value and the mass of the two degenerate light quarks 
at 5\% of the strange quark mass. 

We have also used the gauge configurations generated for the study of quark 
number susceptibilities at high temperatures~\cite{Bazavov:2013uja} and for 
the study of color screening~\cite{Bazavov:2018wmo} by the TUMQCD collaboration. 
The RHMC algorithm was used for 2+1 quark flavors with the strange quark mass 
at its physical value and the mass of the two degenerate light quarks at 5\% 
or 20\% of the strange quark mass, 
\mbox{i.e.}, corresponding to the zero temperature ensembles.  
We summarize the nonzero temperature gauge ensembles used in the determination 
of the strong coupling in \mbox{Tabs.}~\ref{tab:nt 12} 
and~\ref{tab:nt 16}.
We have increased the statistics of some \(N_\tau=16\) gauge ensembles specifically 
for this study and updated the Polyakov loop in \mbox{Tab.}~\ref{tab:nt 16}. 

In \mbox{Tab.}~\ref{tab:gsfit} we summarize the necessary cuts to the fit 
ranges for the extraction of the static energy from the lattice correlators. 

\begin{table}[t]
\parbox{1.0\linewidth}{
  \begin{tabular}{|c|c|}    
    \hline
    \multicolumn{2}{|c|}{\(\beta=7.373\):} \\
    \hline
    $\bm r/a$ & \(\max(\tau/a)\) \\
    \hline
    (1,0,0) & 10 \\
    (2,1,0) & 10 \\
    (2,2,2) & 10 \\
    (3,3,1) & 10 \\
    (4,0,0) & 10 \\
    (4,1,0) & 10 \\
    (4,2,0) & 10 \\
    \hline
  \end{tabular}
  \begin{tabular}{|c|c|}    
    \hline
    \multicolumn{2}{|c|}{\(\beta=7.596\):} \\
    \hline
    $\bm r/a$ & \(\max(\tau/a)\) \\
    \hline
    (1,0,0) & 11 \\
    (2,2,0) & 10 \\
    (2,2,1) & 10 \\
    (2,2,2) & 9 \\
    (3,1,0) & 10 \\
    (3,2,2) & 10 \\
    (3,3,1) & 11 \\
    (4,2,0) & 10 \\
    (4,3,0) & 10 \\
    \hline
  \end{tabular}
  \begin{tabular}{|c|c|}    
    \hline
    \multicolumn{2}{|c|}{\(\beta=7.825\):} \\
    \hline
    $\bm r/a$ & \(\max(\tau/a)\) \\
    \hline
    (1,1,0) & 10 \\
    (2,2,0) & 11 \\
    (3,0,0) & 9 \\
    (3,1,0) & 11 \\
    (4,0,0) & 9 \\
    (4,1,1) & 11 \\
    \hline
  \end{tabular}
  \begin{tabular}{|c|c|}    
    \hline
    \multicolumn{2}{|c|}{\(\beta=8.000\):} \\
    \hline
    $\bm r/a$ & \(\max(\tau/a)\) \\
    \hline
    (1,0,0) & 11 \\
    (3,1,1) & 11 \\
    (3,2,1) & 10 \\
    (3,2,2) & 10 \\
    (4,0,0) & 10 \\
    (4,3,0) & 10 \\
    \hline
  \end{tabular}
    \begin{tabular}{|c|c|}
    \hline
    \multicolumn{2}{|c|}{\(\beta=8.200\):} \\
    \hline
    $\bm r/a$ & \(\max(\tau/a)\) \\
    \hline
    (1,0,0) & 9 \\
    (1,1,1) & 9 \\
    (2,1,0) & 9 \\
    (2,1,1) & 11 \\
    (2,2,1) & 10 \\
    (3,1,0) & 10 \\
    (3,1,1) & 9 \\
    (3,2,0) & 10 \\
    (3,2,1) & 10 \\
    (3,2,2) & 10 \\
    (3,3,0) & 10 \\
    (3,3,1) & 10 \\
    (4,1,0) & 10 \\
    (4,1,1) & 10 \\
    (4,3,0) & 11 \\
    \hline
  \end{tabular}
  \begin{tabular}{|c|c|}
    \hline
    \multicolumn{2}{|c|}{\(\beta=8.400\):} \\
    \hline
    $\bm r/a$ & \(\max(\tau/a)\) \\
    \hline
    (2,0,0) & 9 \\
    (2,2,0) & 9 \\
    (2,2,2) & 9 \\
    (3,1,1) & 10 \\
    (3,2,2) & 10 \\
    (4,0,0) & 9 \\
    (5,5,1) & 9 \\
    \hline
  \end{tabular} 
  \caption{\capstretch
\label{tab:gsfit}
  For many data sets we have to restrict the fit windows in the ground 
  state extraction due to instabilities at large \(\tau/a\). 
  Otherwise, the fit window reaches up to \(\tau/a=12\).
  }
}
\end{table}

\section{Discretization artifacts}\label{app:discretization}

In this appendix, we briefly discuss the discretization artifacts of the static 
energy calculated on the lattice. 
We follow the treatment outlined in 
\mbox{Ref.}~\cite{Bazavov:2018wmo} with a few refinements. 

\begin{figure}
\centering
\includegraphics[width=8.6cm]{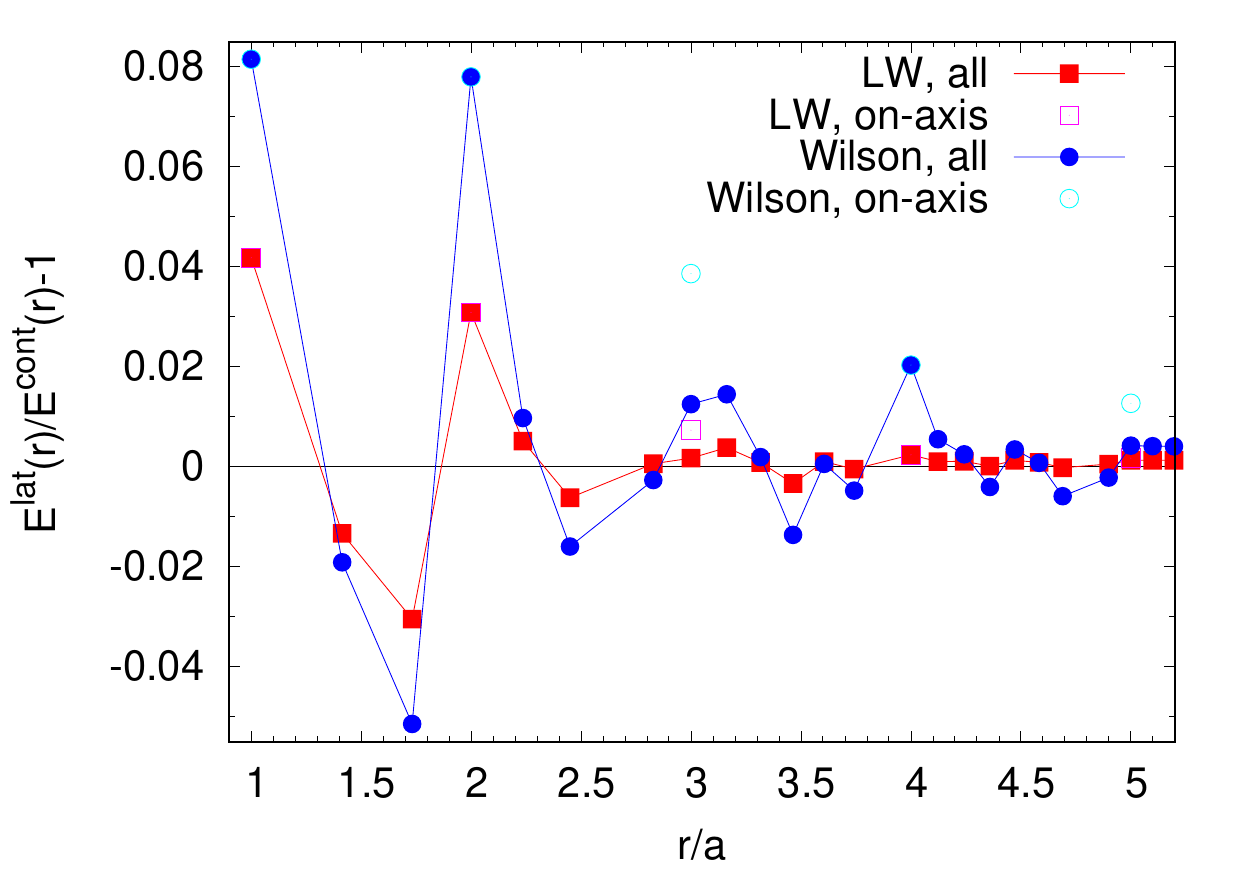}
\caption{\capstretch
\label{fig:tree-level}
Cutoff effects in the tree-level lattice static energy with L\"uscher--Weisz 
action are usually much smaller than for the Wilson gauge action. 
Both results are normalized by the corresponding continuum result.
}
\end{figure}

It has been known for a long time that the static energy is affected by 
discretization artifacts at distances that are only slightly larger than the 
lattice spacing, \mbox{i.e.}, \(r/a \gtrsim 1\). 
The ratio of the static energy on the lattice and in the continuum 
exhibits a distinct pattern of a few percent to permill level variation 
around 1, which dies out quickly for larger distances, 
see \mbox{Fig.}~\ref{fig:tree-level}.
This pattern is quite similar in the tree-level calculation and in the QCD 
result (with or without sea quarks).
Hence, one may partly account for these discretization artifacts by dividing 
the lattice QCD static energy by the ratio of the lattice and continuum static 
energies from a tree-level calculation, \mbox{i.e.}, assuming one-gluon 
exchange and no running coupling. 
This procedure is called tree-level improvement. 
At tree level the static energy is written as 
\al{
E^{\rm lat}_{\rm tree}(r)=-\frac{4}{3} g_0^2 \int \frac{d^3 k}{(2\pi)^3} e^{i\mathbf{k} \mathbf{r}} D_{00}(k),
}
with $D_{00}(k)$ being the temporal component of the lattice gluon propagator on the lattice.
The results on the lattice static energy at tree level calculated for Wilson 
and L\"uscher--Weisz action are shown in \mbox{Fig.}~\ref{fig:tree-level}. 
Writing 
\al{
E^{\rm lat}_{\rm tree}(r)=-\frac{4}{3} g_0^2 \frac{1}{4 \pi r_I}
}
defines the so-called improved distance $r_I$.
In practice, the tree-level improvement is included in the comparison between 
continuum and lattice results either as 
\al{
 E_{\rm imp}(r_I) = E^{\rm lat}(r), 
}
or -- by introducing a fit parameter \(\kappa_I(\bm r/a,\beta)\) for each distinctive 
path geometry \(\bm r/a\) between the quark and antiquark at each lattice spacing -- as 
\al{  
 E_{\rm imp}(r) = E^{\rm lat}(r) + 
 \kappa_{I}\left(\frac{\bm r}{a},\beta\right) \frac{r_I-r}{r r_I}.
}
Here we use the first approach. 

\begin{figure}
\centering
\includegraphics[width=8.6cm]{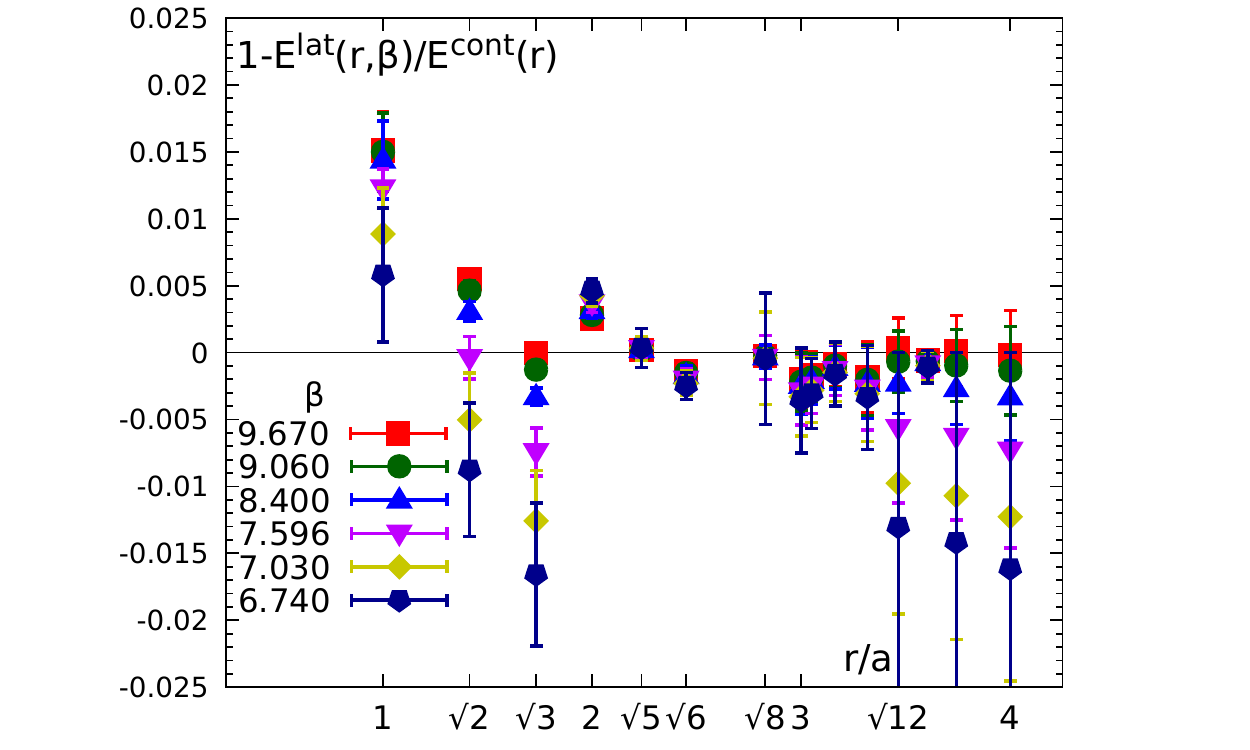}
\caption{\capstretch
\label{fig:hisq-imp}
Cutoff effects in the lattice static energy in (2+1)-flavor QCD
obtained in this work after using tree-level improvement.
}
\end{figure}

However, this is not sufficient to remove the discretization artifacts entirely. 
Ideally, one would replace the tree-level improvement by a one- or two-loop 
improvement, \mbox{i.e.}, calculating the correction factors using 
lattice perturbation theory at one- or two-loop level, and including the 
tadpole factors. 
We do not pursue this approach here as it requires extensive calculation 
beyond the scope of this study. 
As one can see from \mbox{Fig.}~\ref{fig:tree-level} discretization errors are at few percent
level for $r/a \le 2$. For $r/a=\sqrt{5}$ and $\sqrt{6}$ these errors are below a percent and are even smaller for $r/a\ge \sqrt{8}$. 
Based on previous studies we expect the same pattern of discretization 
effects in the interacting theory. 
We show in \mbox{Fig.}~\ref{fig:hisq-imp} that a similar pattern is 
observed for the HISQ action after the tree-level improvement. 
For distances $r/a\ge \sqrt{8}$
the statistical errors of the static energy are larger than the discretization errors. 

Hence, there are three options to cope with the situation:
The first option is to completely avoid the data with significant residual discretization artifacts. 
This corresponds to the first analysis strategy discussed in \mbox{Sec.}~\ref{sec:static}. 

The second option is to estimate the size of the residual discretization artifacts 
in the data at small \(r/a\) and combine a systematic error estimate of 
similar magnitude with the statistical errors to reduce the statistical 
weight of these data. 

The first quark-antiquark distance that we can access with distinctive 
path geometries (ignoring permutations) is \(r/a=3\), specifically, either 
through \(\bm{r}/a=(3,0,0)\), or \((2,2,1)\). 
We average these two results as \(E^{\rm lat}(r=3a)\), and use a Cornell 
fit to interpolate between the data for \(a = r_I(\sqrt{8}a),\ 3a,\) and 
\(r_I(\sqrt{10}a)\) 
to the distances \(r_I((3,0,0)\,a)=2.9788\,a\) and 
\(r_I((2,2,1)\,a)=3.0123\,a\). 
The difference 
\(\Delta^{\rm lat}(\bm r/a) = E^{\rm lat}(\bm r/a)-E^{\rm int}(\bm r/a)\) 
between the data and the interpolation at \(\bm{r}/a=(3,0,0)\), or 
\((2,2,1)\) already includes the tree-level improvement, \mbox{i.e.},  
\(\Delta^{\rm lat}( (3,0,0) ) - \Delta^{\rm lat}( (2,2,1) )\) is an 
estimate for the residual discretization artifacts after the tree-level 
improvement. 
On the other hand, \(E^{\rm lat}((3,0,0) ) - E^{\rm lat}( (2,2,1) )\) 
is an estimate for the discretization artifacts without the tree-level improvement. 
The latter is usually about 0.1\%, the former is usually one order of magnitude smaller. 
Since the latter certainly overestimates the residual discretization 
artifacts after tree-level improvement and the former may be accidentally 
small due to statistical fluctuations, we proceed using the geometric 
average of the two. 
We smooth out fluctuations between these geometric averages for different 
lattice spacings by demanding that the systematic error estimate has to 
be at least as large as the mean of the systematic error for the two 
neighboring lattice spacings.
We summarize the systematic error estimates for the different lattices in 
\mbox{Tab.}~\ref{tab:syst}. 
The absolute size of the estimate of the residual discretization error is 
typically less than 0.5\% of \(1/r_1\), namely, less than \(3\,{\rm MeV}\). 
This option is suitable for \(r/a \ge \sqrt{5}\) at reasonable accuracy and 
should be applied for all distances with at most two steps along any of the 
lattice axes.

\begin{table}[htb]
\begin{tabular}{|c|c|c|c|}
\hline
\(\beta\) & \(N_\tau\) & \(r_1\delta^{\rm syst}_{\rm abs}\) & 
\(\delta^{\rm syst}_{\rm rel}(\tfrac{r}{a}=3)\) \\
\hline
7.596 & 64 & 0.0041 & 0.00103 \\
7.825 & 64 & 0.0041 & 0.00093 \\
8.000 & 64 & 0.0038 & 0.00076 \\
8.200 & 64 & 0.0037 & 0.00066 \\
8.400 & 64 & 0.0034 & 0.00054 \\
\hline
7.825 & 16 & 0.0161 & 0.00360 \\
8.000 & 16 & 0.0179 & 0.00362 \\
8.200 & 16 & 0.0161 & 0.00290 \\
8.400 & 16 & 0.0108 & 0.00172 \\
8.570 & 16 & 0.0055 & 0.00079 \\
8.710 & 16 & 0.0057 & 0.00074 \\
8.850 & 16 & 0.0071 & 0.00086 \\
9.060 & 16 & 0.0078 & 0.00082 \\
9.230 & 16 & 0.0084 & 0.00079 \\
9.360 & 16 & 0.0098 & 0.00085 \\
9.490 & 16 & 0.0111 & 0.00089 \\
9.670 & 16 & 0.0054 & 0.00038 \\
\hline
7.596 & 12 & 0.0052 & 0.00128 \\
7.650 & 12 & 0.0065 & 0.00156 \\
7.825 & 12 & 0.0052 & 0.00116 \\
8.000 & 12 & 0.0038 & 0.00078 \\
8.200 & 12 & 0.0033 & 0.00060 \\
8.400 & 12 & 0.0028 & 0.00045 \\
8.570 & 12 & 0.0030 & 0.00044 \\
8.710 & 12 & 0.0041 & 0.00054 \\
8.850 & 12 & 0.0079 & 0.00095 \\
9.060 & 12 & 0.0097 & 0.00102 \\
9.230 & 12 & 0.0114 & 0.00108 \\
9.360 & 12 & 0.0066 & 0.00057 \\
9.490 & 12 & 0.0058 & 0.00046 \\
9.670 & 12 & 0.0049 & 0.00035 \\
\hline
\end{tabular}
\caption{\capstretch
Systematic uncertainty estimates for the discretization artifacts 
from the static energy at \(T=0\) or the singlet free energy with 
\(N_\tau=16\) or \(12\) at \(r/a=3\). 
The estimates have been calculated individually for each data set. 
The fluctuations within a factor 2 or 3 between the uncertainty estimates for 
the different data sets are due to statistical fluctuations of the data.  
}
\label{tab:syst}   
\end{table}
 
Finally, we may estimate the residual discretization artifacts in the data for each 
individual \(r/a\), and correct for these errors before comparing lattice
data with the weak coupling results.
This is the procedure that has been used in 
\mbox{Refs.}~\cite{Bazavov:2014soa, Bazavov:2018wmo}.
In this procedure, we define the correction factors 
\al{
 K\left(\frac{\bm r}{a},\beta\right)  = \frac{E^{\rm lat}\left(\bm r, a(\beta)\right)}{E^{\rm cont}(r_I)} - 1,
} 
and the nonperturbatively improved static energy 
\al{
 E^{\rm lat}_{\rm imp}(r,\beta) = \frac{E^{\rm lat}\left(\frac{\bm r}{a}, a(\beta)\right)}{1+K\left(\frac{\bm r}{a},\beta\right)}, 
}
which will be compared to the continuum static energy \(E^{\rm cont}(r_I)\). 
In the following, we spell out how we estimate the residual discretization 
artifacts, and how we correct for these. 
We use two different estimates for the continuum \(E^{\rm cont}(r_I)\).

\begin{enumerate}
  \item[a)] 
For the lattices under consideration and \(r/a \ge \sqrt{5}\) the relative 
discretization artifacts are less than 0.1\% after the tree-level improvement. 
We interpolate the static energy on the finest lattices at such distances to obtain 
a continuum estimate\footnote{We extrapolate to 10\% shorter distances, see 
\mbox{Ref.}~\cite{Bazavov:2018wmo}.} that can be used to take ratios of the 
QCD static energy on coarser lattices and in the continuum. 

Our updated interpolation procedure differs in three ways from the procedure 
outlined in \mbox{Ref.}~\cite{Bazavov:2018wmo}. 
Before the interpolation we add a systematic error estimate of 1.5\%, 0.5\%, 
0.5\%, 0.5\%, 0.04\%, 0.02\%, and 0.3\% to the data on fine lattices at 
\(r/a=1,\ \sqrt{2},\ \sqrt{3},\ 2, \sqrt{5}\), \(\sqrt{6}\), and \(\sqrt{12}\), 
respectively. 
Furthermore, we modify the data before the interpolation by -1.0\%, +0.5\%, 
-0.5\%, and -0.3\% for \(r/a=1,\ \sqrt{3},\ 2\), and \(\sqrt{12}\), respectively. 
These estimates of systematic errors are motivated by our previous analysis of 
the discretization artifacts at short distances~\cite{Bazavov:2018wmo}.
We split the interpolation with a modified Cornell potential into two 
intervals, a shorter distance (usually 7 points) and a larger distance 
interval, whose details are decided dynamically depending 
on the \(\cdf\) of the fit, and switch between the two interpolations in an 
overlap region. 
In each interval we require fewer higher order terms than for a single fit of 
all data. 
We use a spline interpolation to estimate the model dependence. 
In the next step, we calculate a weighted average \(\braket{K(\bm r/a,\beta)}\) 
of the ratios \(K(\bm r/a,\beta, \beta^{\rm ref})\) obtained with continuum 
estimates based on different reference lattices (indicated by 
\(\beta^{\rm ref}\)) for each \(\bm r/a\) and \(\beta\) as previously. 

Using the static energy at zero temperature restricted to \(r/a \le 8\) we 
vary the reference lattice down to \(\beta^{\rm ref} \ge 7.28\) and obtain 
correction factors for all but the finest lattice (\(\beta=8.4\)) down to 
\(\beta \ge 6.664\). 
Using the singlet free energy with \(N_\tau=12\) restricted to \(r/a \le \sqrt{12}\), 
namely, \(rT \le 1/\sqrt{12} \approx 0.289\), we obtain correction factors 
for all but the finest lattice (\(\beta=9.67\)) down to \(\beta \ge 7.373\). 
  \item[b)] 
We use the  \emph{three loop result with the leading ultra-soft resummation} as a fit form 
for a continuum estimate up to \(r \le 0.13\,{\rm fm}\). 
We restrict to this $r$- range to ascertain that the fit is performed in the perturbative 
regime, which limits this analysis to fine lattice spacings, \mbox{i.e.}, to 
\(\beta \ge 7.596\) in practical terms.
This fit form is physically more motivated but needs an input value for $\als$.
Before the interpolation we add a systematic error estimate of 1.5\%, 0.5\%, 
0.5\%, 0.5\%, 0.04\%, 0.02\%, and 0.3\% to the data at 
\(r/a=1,\ \sqrt{2},\ \sqrt{3},\ 2, \sqrt{5}\), \(\sqrt{6}\), and \(\sqrt{12}\), 
respectively. 
In order to minimize the bias towards a final determination of \(\als\), we vary the 
range of the input value \(\als\) between \(0.1155\) and \(0.1175\),  and determine 
for each input the normalization constant through a weighted average. 
Out of the set of inputs we accept either all of the fits with \(\cdf \le 1\) or the three fits with smallest \(\cdf\). 
We estimate the uncertainty of the correction factor by adding the statistical 
error and the variation of the continuum estimate with the input value of 
\(\als\) in quadrature. 

Using the static energy at zero temperature restricted to \(r/a \le 8\) we 
obtain correction factors for all lattices down to \(\beta \ge 7.596\). 
Using the singlet free energy with \(N_\tau=12\) restricted to \(r/a \le \sqrt{12}\), 
namely, \(rT \le 1/\sqrt{12} \approx 0.289\), we obtain correction factors 
for all lattices down to \(\beta \ge 7.825\). 
\end{enumerate}

\begin{figure*}
\includegraphics[width=0.32\textwidth]{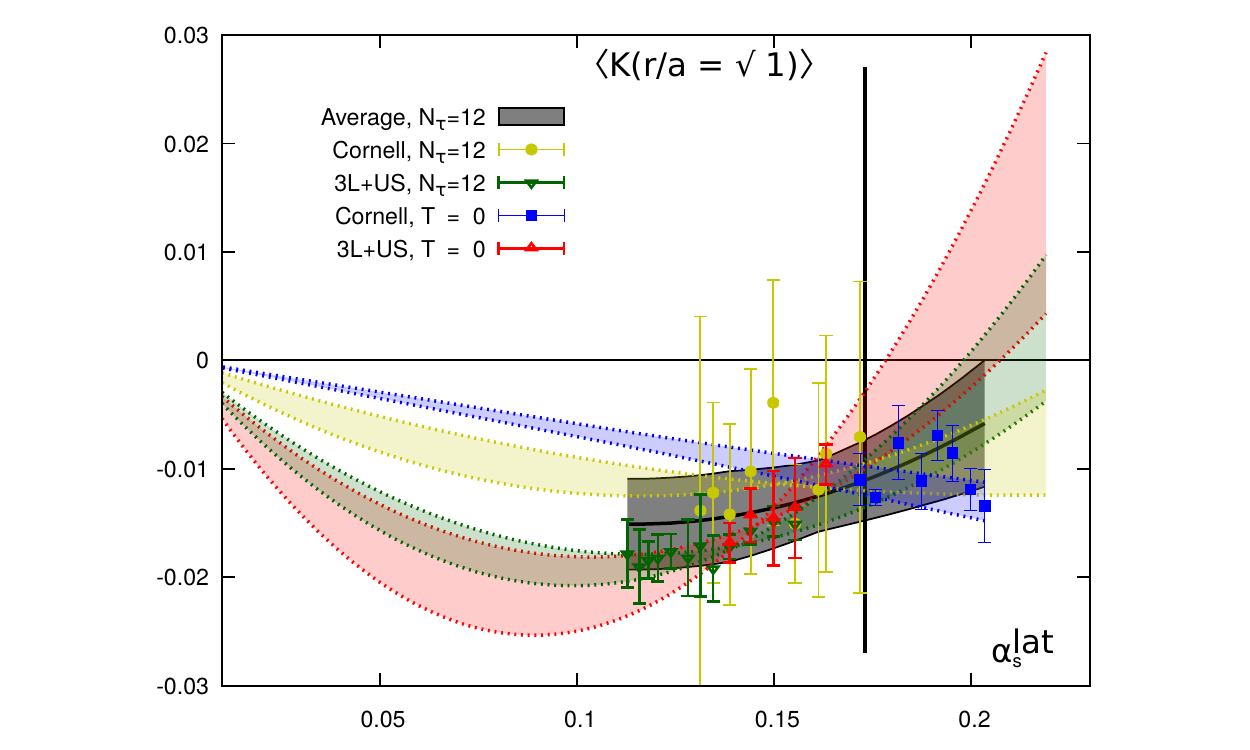}
\includegraphics[width=0.32\textwidth]{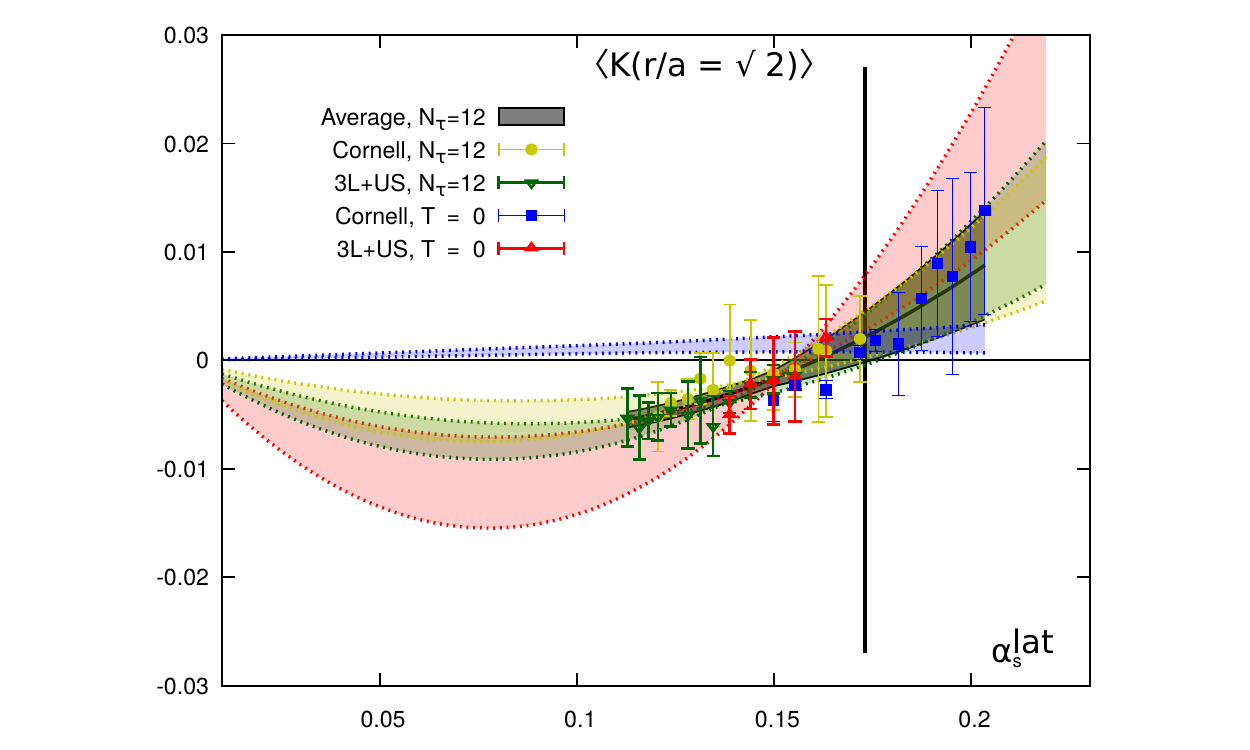}
\includegraphics[width=0.32\textwidth]{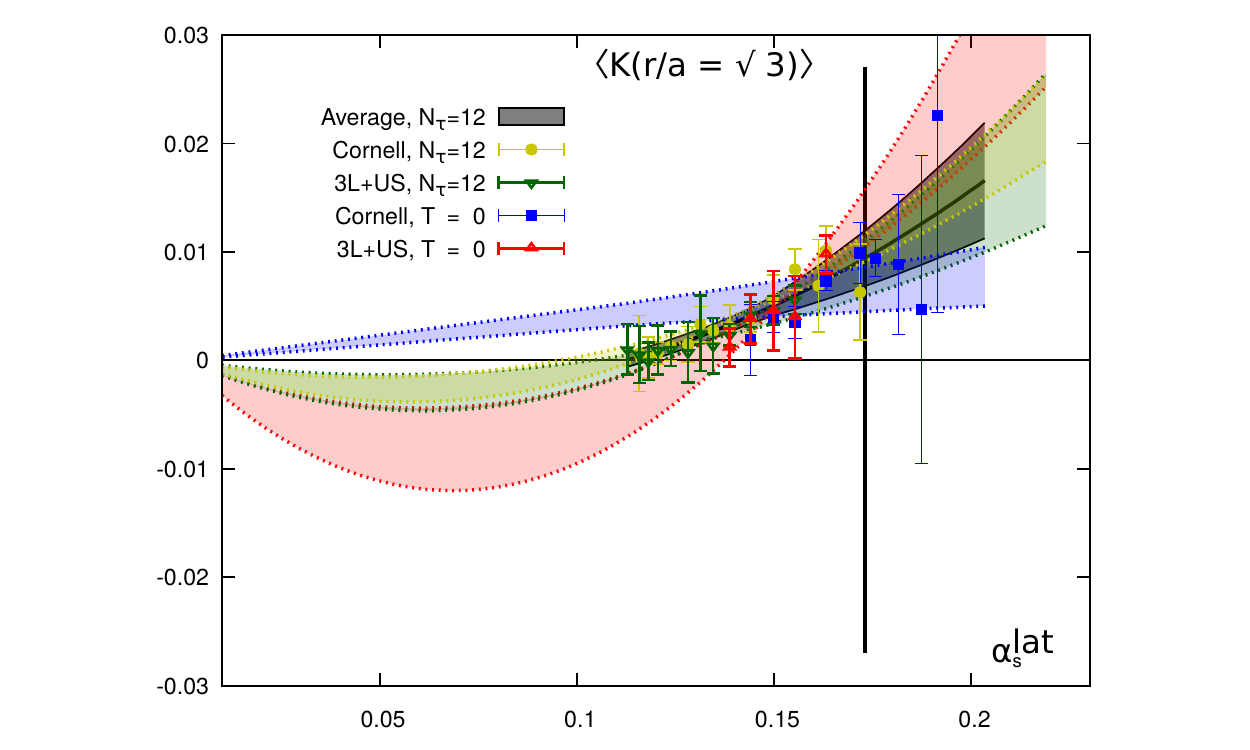}

\includegraphics[width=0.32\textwidth]{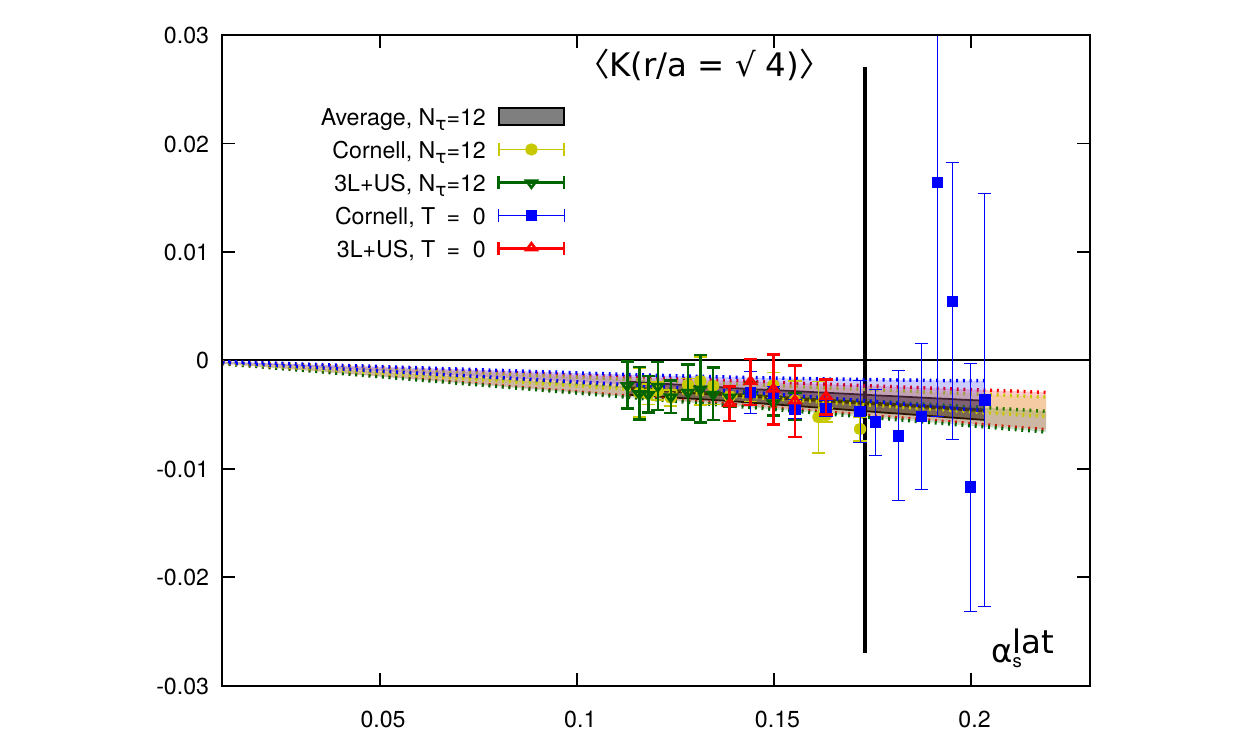}
\includegraphics[width=0.32\textwidth]{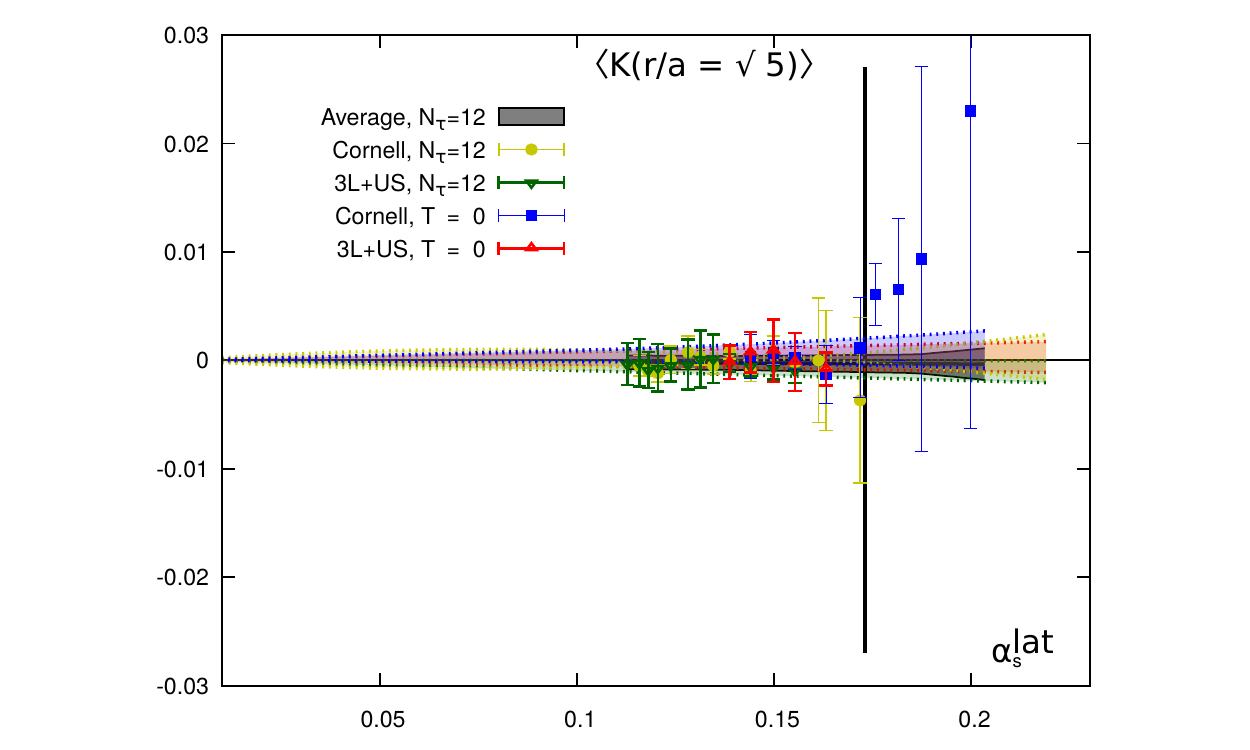}
\includegraphics[width=0.32\textwidth]{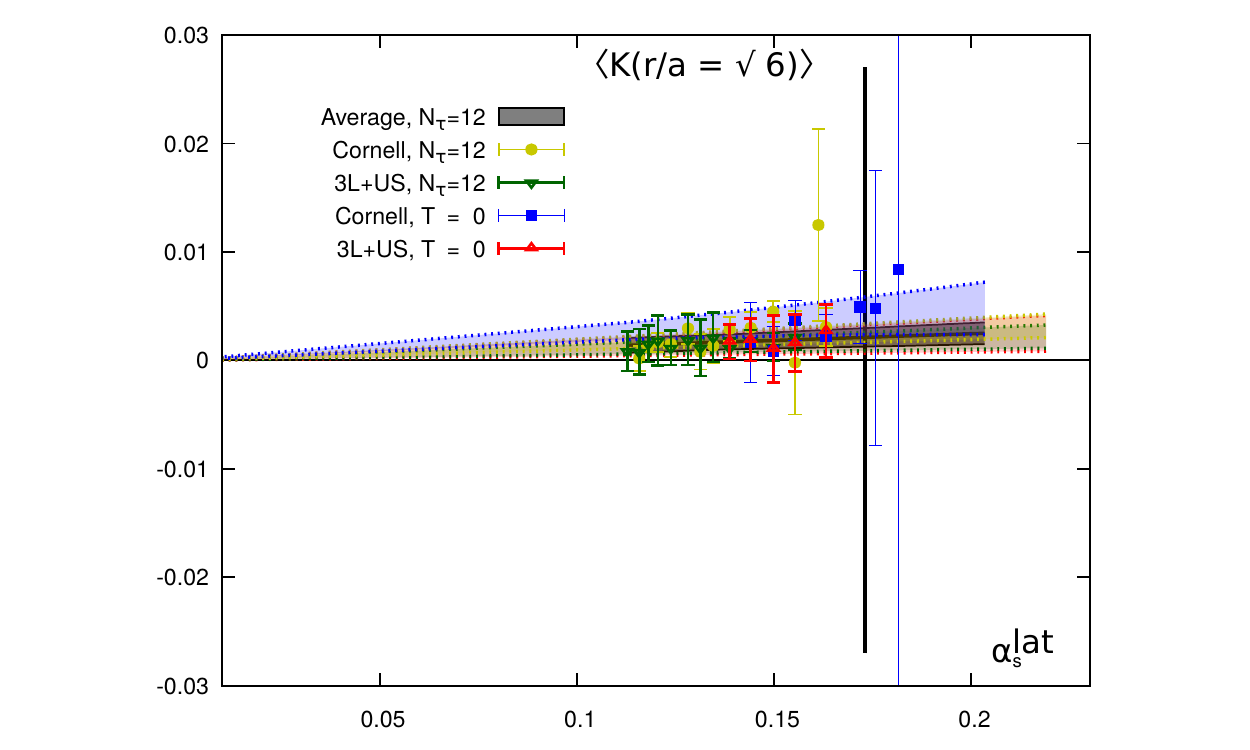}
\caption{\capstretch
The correction factors \(\braket{K}\) for the nonperturbative 
improvement of the static energy as function of the boosted coupling 
\(\als^{\rm lat} = 10/(4\pi\beta u_0^4)\) for 
\(r/a=\sqrt{1},~\sqrt{2}\), and~\(\sqrt{3}\) (top row), and
\(r/a=\sqrt{4},~\sqrt{5}\), and \(\sqrt{6}\) (bottom row).
The \(\Braket{K}\) are obtained for Cornell-type fits with \(\min(r/a)=\sqrt{5}\) 
and for weak-coupling type fits with \(\min(r/a)=\sqrt{1}\).
\(\Braket{K}\) obtained through extrapolation as function of \(\als^{\rm lat}\) 
is indicated through the colored bands, while the final result is shown as the 
gray band.  
Only data to the left of the vertical line corresponds to 
\(\beta \ge 7.373\) and has been used in this study. 
}
\label{fig:corr K}
\end{figure*}

All four estimates of the correction factors are complementary, since they have 
different systematic uncertainties. 
In \mbox{Fig.}~\ref{fig:corr K} we show the corresponding correction factors from 
Cornell fits using lattice data at \(r/a \ge \sqrt{5}\) on the reference lattice, or 
from fits with the weak-coupling result using \(r/a \ge \sqrt{1}\). 
All four results are in very good agreement. 
 
We extrapolate the \(\braket{K(\bm r/a,\beta)}\) towards the continuum limit using  
a boosted coupling \(\als^{\rm lat}= 10/({4\pi\beta u_0^4})\). 
We list the corresponding tadpole factors \(u_0\) in 
\mbox{Tab.}~\ref{tab:uzero}, which have been obtained from the plaquette 
at \(T=0\), or with \(N_\tau=12\) or \(16\).

\begin{table}[htb]
\begin{tabular}{|c|c|c|c|c|c|}
\hline
\(\beta\) & \(u_0\) & \(\beta\) & \(u_0\) & \(\beta\) & \(u_0\)  \\
\hline
6.608 & 0.86823 & 7.030 & 0.88173 & 8.200 & 0.90604 \\
6.664 & 0.87026 & 7.100 & 0.88363 & 8.400 & 0.90909 \\
6.700 & 0.87152 & 7.150 & 0.88493 & 8.570 & 0.91152 \\
6.740 & 0.87288 & 7.200 & 0.88621 & 8.710 & 0.91341 \\ 
6.770 & 0.87388 & 7.280 & 0.88817 & 8.850 & 0.91522 \\
6.800 & 0.87485 & 7.373 & 0.89035 & 9.060 & 0.91778 \\
6.840 & 0.87613 & 7.596 & 0.89517 & 9.230 & 0.91973 \\ 
6.880 & 0.87736 & 7.650 & 0.89627 & 9.360 & 0.92116 \\
6.910 & 0.87827 & 7.825 & 0.89962 & 9.490 & 0.92253 \\
6.950 & 0.87945 & 8.000 & 0.90274 & 9.670 & 0.92435 \\ 
\hline
\end{tabular}
\caption{\capstretch
Tadpole factors \(u_0\) for the lattices used in the 
improvement procedure. 
Tadpole factors have been obtained using zero temperature 
or finite temperature lattices. 
Temperature effects are always at most \(0.00001\). 
}
\label{tab:uzero}   
\end{table}

For the extrapolation we use
\al{
 \Braket{K\left(\frac{\bm r}{a}, \als^{\rm lat}\right)} = 
 \sum\limits_{i=1}^N \kappa_i\left(\frac{\bm r}{a}\right) \left(\als^{\rm lat}\right)^i. 
} 
We try extrapolations with up to \(N=3\) terms, but never require the third order term, 
\mbox{i.e.}, we cannot fix its coefficient with small uncertainties. 
While data are quite insensitive to the second order term for \(r/a \ge 2\), it cannot 
be avoided for \(r/a < 2\). 
The coefficients of the first and second order term are anti-correlated.
Only in the case of the Cornell fit to lattice data at zero temperature we cannot 
resolve the second order term successfully. 
We also tested a zeroth order term, which was never necessary.

For each estimate (Cornell vs weak-coupling fits, zero vs finite temperature) 
we take the median of the correction factors from fits including data starting at 
\(\min(r/a) = \sqrt{1}\) up to \(\min(r/a) = \sqrt{5}\).
We estimate the overall error from typical uncertainties, or the spread between results 
for different \(\min(r/a)\), whichever is larger. 
We show the median of the extrapolations for each fit type as function of 
\(\als^{\rm lat}\) as colored bands in \mbox{Fig.}~\ref{fig:corr K}. 
While the agreement between the median of Cornell fits and the median of 
weak-coupling fits is not satisfactory for fits to \(T=0\) data, the 
different fit results to \(T>0\) data agree very well. 
Moreover, for the correction factor \(\braket{K(r/a<2,\beta)}\), where the data are 
sensitive to the second order term in \(\als^{\rm lat}\), the extrapolation 
of the results obtained with \(T>0\) data also extends to the \(T=0\) results 
at larger values of the coupling, \mbox{i.e}, on coarse lattices. 
Therefore, in order to obtain our final results, we take the mean of the Cornell fit 
and the weak-coupling fit to the \(N_\tau=12\) data, and estimate the uncertainty by the 
largest among the errors of the two results, or half of the spread between the results. 
Since this uncertainty still underestimates the spread between \(\braket{K(r/a=1,\beta)}\) 
for different fit types and among different \(\beta\), we add a systematic error estimate 
of \(0.003\) in quadrature to the errors of this procedure. 
We summarize the coefficients corresponding to both fit types in 
\mbox{Tab.}~\ref{tab:corr K} and show the averaged result as a gray band 
in \mbox{Fig.}~\ref{fig:corr K}. 
While the corrections are significant and constrained quite well up to 
\(\bm r/a=(2,1,1)\), the corrections beyond tree-level are consistent with 
zero for larger distances. 

\begin{table}[htb]
\begin{adjustbox}{angle=00}
\begin{tabular}{|c|c|c|c|c|c|}
\hline
& tree level & \multicolumn{2}{|c|}{Cornell} & \multicolumn{2}{|c|}{three loop} \\
\hline
\(\bm{r}/a\) & \(\kappa_0 \!=\! \frac{r_I-r}{r_I}\) & \(\kappa_1\) & \(\kappa_2\) & \(\kappa_1\) & \(\kappa_2\) \\
\hline
(1,0,0) & -0.041777 & -0.166(49) & 0.60(33) & -0.364(56) & 1.73(39) \\
(1,1,0) & +0.013536 & -0.142(55) & 0.90(39) & -0.183(50) & 1.12(37) \\
(1,1,1) & +0.031486 & -0.099(34) & 0.92(24) & -0.102(51) & 0.87(38) \\
(2,0,0) & -0.029918 & -0.020(04) & 0.0      & -0.026(04) & 0.0      \\
(2,1,0) & -0.005012 & +0.001(23) & 0.0      & -0.004(05) & 0.0      \\
(2,1,1) & +0.006374 & +0.015(05) & 0.0      & +0.010(05) & 0.0      \\
(2,2,0) & -0.000432 & +0.009(47) & 0.0      & -0.004(05) & 0.0      \\
(2,2,1) & +0.004098 & +0.026(13) & 0.0      & +0.004(04) & 0.0      \\
(2,2,2) & +0.003694 & +0.016(50) & 0.0      & -0.171(08) & 1.47(58) \\
(3,0,0) & -0.007952 & +0.033(27) & 0.0      & +0.002(50) & 0.0      \\
(3,1,0) & -0.003564 & +0.020(08) & 0.0      & -0.004(05) & 0.0      \\
(3,1,1) & -0.000562 & +0.034(19) & 0.0      & -0.001(05) & 0.0      \\
(3,2,0) & -0.000657 & +0.010(53) & 0.0      & +0.010(06) & 0.0      \\
(3,2,1) & +0.000883 & +0.002(49) & 0.0      & -0.177(64) & 1.56(48) \\
(4,0,0) & -0.001854 & -0.022(45) & 0.0      & -0.191(80) & 1.72(63) \\
\hline
\end{tabular}
\end{adjustbox}
\caption{\capstretch
Coefficients of the nonperturbative improvement procedure 
up to order \((\als^{\rm lat})^2\). 
}
\label{tab:corr K}   
\end{table}

\section{Coefficients of the force at three loops}\label{app:pert}

In this appendix, we summarize the constants appearing in the weak-coupling 
result of the force at three-loop order in 
\mbox{Eqs.}~\eqref{eq:F3L} and \eqref{eq:FN3LL}.
The color factors read
\al{
&C_F=\frac{N_c^2-1}{2N_c},&& C_A=N_c,&& T_F=\frac{1}{2},
}
where $N_c$ is the number of colors.
The QCD beta function is
\al{
\frac{d\,\als(\nu)}{d\ln \nu}
&=
\als\beta(\als)
= -\frac{\als^2}{2\pi}\sum_{n=0}^\infty \left( \frac{\als}{4\pi} \right)^n \beta_n
\nonumber\\
&=-2\alpha_s\left[\beta_0\frac{\alpha_s}{4\pi}+\beta_1\left(\frac{\alpha_s}{4\pi}\right)^2+\cdots\right],
}
where the first three coefficients read~\cite{Herzog:2017ohr, Luthe:2017ttg}
\al{
\beta_0 & =  \frac{11}{3}C_A-\frac{4}{3}T_FN_f,\\
\beta_1 & =  \frac{34}{3}C_A^2-\frac{20}{3} C_A N_f T_F-4 C_F N_f T_F,\\
\beta_2 & =  \frac{2857}{54}C_A^3+
   \left(-\frac{1415}{27}C_A^2-\frac{205}{9}C_A C_F+2C_F^2\right) N_f
   T_F\nn\\
&+\left(\frac{158}{27}C_A+\frac{44}{9}C_F\right)N_f^2 T_F^2.
}
$N_f$ is the number of active massless quarks. 
Higher-order coefficients \(\beta_3\), and \(\beta_4\) are also 
known~\cite{Herzog:2017ohr, Luthe:2017ttg}, but are not reproduced for 
brevity's sake. 
We reproduce the coefficients \(\tilde{a}_i\) as given in \mbox{Ref.}~\cite{Tormo:2013tha} 
\al{ 
\tilde{a}_1 &= a_1+2\gamma_E\beta_0,
\\
\tilde{a}_2 &= a_2 +\left(\frac{\pi^2}{3}+4\gamma_E^2\right)\beta_0^2
  +\gamma_E\left(4a_1\beta_0+2\beta_1\right), \\
\tilde{a}_3 &= 
a_3+\Big(8\gamma_E^3+2\gamma_E\pi^2+16\zeta(3)\Big)
\beta_0^3+2\gamma_E\beta_2
\nonumber\\
&+\Big[\left(12\gamma_E^2+\pi^2\right)\beta_0^2+4\gamma_E\beta_1\Big]
a_1
\nonumber\\
&+\left[6a_2\gamma_E+\frac{5}{2}\left(4\gamma_E^2+\frac{\pi^2}{3}\right)\beta_1\right]\beta_0, 
}
where the coefficients \(a_1\) and \(a_2\) 
read~\cite{Fischler:1977yf, Billoire:1979ih, Schroder:1998vy} 
\al{
a_1 &= \frac{31}{9}C_A-\frac{20}{9}T_FN_f,
\\
a_2 &= \left({4343\over162}+4\pi^2-{\pi^4\over4}+{22\over3}\zeta(3)\right)C_A^2
\nonumber\\
&-\left({1798\over81}+{56\over3}\zeta(3)\right)C_AT_FN_f
\nonumber\\
& -\left({55\over3}-16\zeta(3)\right)C_FT_FN_f +\left({20\over9}T_FN_f\right)^2. 
} 
The coefficient \(a_3\) reads~\cite{Anzai:2009tm, Smirnov:2009fh} 
\al{ 
a_3 &=  a_3^{(3)}N_f^3+a_3^{(2)}N_f^2+a_3^{(1)}N_f+a_3^{(0)},\\
a_3^{(3)} &= - \left(\frac{20}{9}\right)^3 T_F^3
\nonumber\\
a_3^{(2)} &=
\left(\frac{12541}{243}
  + \frac{368}{3}\zeta(3)
  + \frac{64\pi^4}{135}
\right) C_A T_F^2
\nonumber\\
&+
\left(\frac{14002}{81}
  - \frac{416}{3}\zeta(3)
\right) C_F T_F^2
\\
a_3^{(1)} &=
\left(-709.717
\right) C_A^2 T_F
\nonumber\\
&+
\left(-\frac{71281}{162}
  + 264 \zeta(3)
  + 80 \zeta(5)
\right) C_AC_F T_F
\nonumber\\
&+
\left(\frac{286}{9}
  + \frac{296}{3}\zeta(3)
  - 160\zeta(5)
\right) C_F^2 T_F
\nonumber\\
&+
\left(-56.83(1)
\right) 
\frac{18-6N_c^2+N_c^4}{96N_c^2}
\\
a_3^{(0)} & =  502.24(1) C_A^3 -136.39(12)
\frac{N_c^3+6N_c}{48}
\nonumber\\
&+
\frac{8}{3}\pi^2C_A^3\left(-\frac{5}{3}+2\gamma_E+2\log2\right).\label{eq:a30}
}
The three numerical values given above are nowadays known analytically~\cite{Lee:2016cgz, Lee:2016lvq}. 
Lastly, we reproduce the coefficient \(a_{3}^L\) as given in \mbox{Ref.}~\cite{Brambilla:1999qa} 
\al{
 a_3^L & = \frac{16\pi^2}{3}C_A^3.
}
\bibliography{ref}

\begin{thebibliography}{46}%
\makeatletter
\providecommand \@ifxundefined [1]{%
 \@ifx{#1\undefined}
}%
\providecommand \@ifnum [1]{%
 \ifnum #1\expandafter \@firstoftwo
 \else \expandafter \@secondoftwo
 \fi
}%
\providecommand \@ifx [1]{%
 \ifx #1\expandafter \@firstoftwo
 \else \expandafter \@secondoftwo
 \fi
}%
\providecommand \natexlab [1]{#1}%
\providecommand \enquote  [1]{``#1''}%
\providecommand \bibnamefont  [1]{#1}%
\providecommand \bibfnamefont [1]{#1}%
\providecommand \citenamefont [1]{#1}%
\providecommand \href@noop [0]{\@secondoftwo}%
\providecommand \href [0]{\begingroup \@sanitize@url \@href}%
\providecommand \@href[1]{\@@startlink{#1}\@@href}%
\providecommand \@@href[1]{\endgroup#1\@@endlink}%
\providecommand \@sanitize@url [0]{\catcode `\\12\catcode `\$12\catcode
  `\&12\catcode `\#12\catcode `\^12\catcode `\_12\catcode `\%12\relax}%
\providecommand \@@startlink[1]{}%
\providecommand \@@endlink[0]{}%
\providecommand \url  [0]{\begingroup\@sanitize@url \@url }%
\providecommand \@url [1]{\endgroup\@href {#1}{\urlprefix }}%
\providecommand \urlprefix  [0]{URL }%
\providecommand \Eprint [0]{\href }%
\providecommand \doibase [0]{http://dx.doi.org/}%
\providecommand \selectlanguage [0]{\@gobble}%
\providecommand \bibinfo  [0]{\@secondoftwo}%
\providecommand \bibfield  [0]{\@secondoftwo}%
\providecommand \translation [1]{[#1]}%
\providecommand \BibitemOpen [0]{}%
\providecommand \bibitemStop [0]{}%
\providecommand \bibitemNoStop [0]{.\EOS\space}%
\providecommand \EOS [0]{\spacefactor3000\relax}%
\providecommand \BibitemShut  [1]{\csname bibitem#1\endcsname}%
\let\auto@bib@innerbib\@empty
\bibitem [{\citenamefont {Tanabashi}\ \emph {et~al.}(2018)\citenamefont
  {Tanabashi} \emph {et~al.}}]{Tanabashi:2018oca}%
  \BibitemOpen
  \bibfield  {author} {\bibinfo {author} {\bibfnamefont {M.}~\bibnamefont
  {Tanabashi}} \emph {et~al.} (\bibinfo {collaboration} {Particle Data
  Group}),\ }\href {\doibase 10.1103/PhysRevD.98.030001} {\bibfield  {journal}
  {\bibinfo  {journal} {Phys. Rev.}\ }\textbf {\bibinfo {volume} {D98}},\
  \bibinfo {pages} {030001} (\bibinfo {year} {2018})}\BibitemShut {NoStop}%
\bibitem [{\citenamefont {Bazavov}\ \emph
  {et~al.}(2018{\natexlab{a}})\citenamefont {Bazavov}, \citenamefont
  {Petreczky},\ and\ \citenamefont {Weber}}]{Bazavov:2017dsy}%
  \BibitemOpen
  \bibfield  {author} {\bibinfo {author} {\bibfnamefont {A.}~\bibnamefont
  {Bazavov}}, \bibinfo {author} {\bibfnamefont {P.}~\bibnamefont {Petreczky}},
  \ and\ \bibinfo {author} {\bibfnamefont {J.~H.}\ \bibnamefont {Weber}},\
  }\href {\doibase 10.1103/PhysRevD.97.014510} {\bibfield  {journal} {\bibinfo
  {journal} {Phys. Rev.}\ }\textbf {\bibinfo {volume} {D97}},\ \bibinfo {pages}
  {014510} (\bibinfo {year} {2018}{\natexlab{a}})},\ \Eprint
  {http://arxiv.org/abs/1710.05024} {arXiv:1710.05024 [hep-lat]} \BibitemShut
  {NoStop}%
\bibitem [{\citenamefont {Bazavov}\ \emph
  {et~al.}(2018{\natexlab{b}})\citenamefont {Bazavov}, \citenamefont
  {Brambilla}, \citenamefont {Petreczky}, \citenamefont {Vairo},\ and\
  \citenamefont {Weber}}]{Bazavov:2018wmo}%
  \BibitemOpen
  \bibfield  {author} {\bibinfo {author} {\bibfnamefont {A.}~\bibnamefont
  {Bazavov}}, \bibinfo {author} {\bibfnamefont {N.}~\bibnamefont {Brambilla}},
  \bibinfo {author} {\bibfnamefont {P.}~\bibnamefont {Petreczky}}, \bibinfo
  {author} {\bibfnamefont {A.}~\bibnamefont {Vairo}}, \ and\ \bibinfo {author}
  {\bibfnamefont {J.~H.}\ \bibnamefont {Weber}} (\bibinfo {collaboration}
  {TUMQCD}),\ }\href {\doibase 10.1103/PhysRevD.98.054511} {\bibfield
  {journal} {\bibinfo  {journal} {Phys. Rev.}\ }\textbf {\bibinfo {volume}
  {D98}},\ \bibinfo {pages} {054511} (\bibinfo {year} {2018}{\natexlab{b}})},\
  \Eprint {http://arxiv.org/abs/1804.10600} {arXiv:1804.10600 [hep-lat]}
  \BibitemShut {NoStop}%
\bibitem [{\citenamefont {Berwein}\ \emph {et~al.}(2017)\citenamefont
  {Berwein}, \citenamefont {Brambilla}, \citenamefont {Petreczky},\ and\
  \citenamefont {Vairo}}]{Berwein:2017thy}%
  \BibitemOpen
  \bibfield  {author} {\bibinfo {author} {\bibfnamefont {M.}~\bibnamefont
  {Berwein}}, \bibinfo {author} {\bibfnamefont {N.}~\bibnamefont {Brambilla}},
  \bibinfo {author} {\bibfnamefont {P.}~\bibnamefont {Petreczky}}, \ and\
  \bibinfo {author} {\bibfnamefont {A.}~\bibnamefont {Vairo}},\ }\href
  {\doibase 10.1103/PhysRevD.96.014025} {\bibfield  {journal} {\bibinfo
  {journal} {Phys. Rev.}\ }\textbf {\bibinfo {volume} {D96}},\ \bibinfo {pages}
  {014025} (\bibinfo {year} {2017})},\ \Eprint
  {http://arxiv.org/abs/1704.07266} {arXiv:1704.07266 [hep-ph]} \BibitemShut
  {NoStop}%
\bibitem [{\citenamefont {Bazavov}\ \emph
  {et~al.}(2014{\natexlab{a}})\citenamefont {Bazavov}, \citenamefont
  {Brambilla}, \citenamefont {Garcia~i Tormo}, \citenamefont {Petreczky},
  \citenamefont {Soto},\ and\ \citenamefont {Vairo}}]{Bazavov:2014soa}%
  \BibitemOpen
  \bibfield  {author} {\bibinfo {author} {\bibfnamefont {A.}~\bibnamefont
  {Bazavov}}, \bibinfo {author} {\bibfnamefont {N.}~\bibnamefont {Brambilla}},
  \bibinfo {author} {\bibfnamefont {X.}~\bibnamefont {Garcia~i Tormo}},
  \bibinfo {author} {\bibfnamefont {P.}~\bibnamefont {Petreczky}}, \bibinfo
  {author} {\bibfnamefont {J.}~\bibnamefont {Soto}}, \ and\ \bibinfo {author}
  {\bibfnamefont {A.}~\bibnamefont {Vairo}},\ }\href {\doibase
  10.1103/PhysRevD.90.074038} {\bibfield  {journal} {\bibinfo  {journal} {Phys.
  Rev.}\ }\textbf {\bibinfo {volume} {D90}},\ \bibinfo {pages} {074038}
  (\bibinfo {year} {2014}{\natexlab{a}})},\ \Eprint
  {http://arxiv.org/abs/1407.8437} {arXiv:1407.8437 [hep-ph]} \BibitemShut
  {NoStop}%
\bibitem [{\citenamefont {Bazavov}\ \emph
  {et~al.}(2014{\natexlab{b}})\citenamefont {Bazavov} \emph
  {et~al.}}]{Bazavov:2014pvz}%
  \BibitemOpen
  \bibfield  {author} {\bibinfo {author} {\bibfnamefont {A.}~\bibnamefont
  {Bazavov}} \emph {et~al.} (\bibinfo {collaboration} {HotQCD}),\ }\href
  {\doibase 10.1103/PhysRevD.90.094503} {\bibfield  {journal} {\bibinfo
  {journal} {Phys. Rev.}\ }\textbf {\bibinfo {volume} {D90}},\ \bibinfo {pages}
  {094503} (\bibinfo {year} {2014}{\natexlab{b}})},\ \Eprint
  {http://arxiv.org/abs/1407.6387} {arXiv:1407.6387 [hep-lat]} \BibitemShut
  {NoStop}%
\bibitem [{\citenamefont {Follana}\ \emph {et~al.}(2007)\citenamefont
  {Follana}, \citenamefont {Mason}, \citenamefont {Davies}, \citenamefont
  {Hornbostel}, \citenamefont {Lepage}, \citenamefont {Shigemitsu},
  \citenamefont {Trottier},\ and\ \citenamefont {Wong}}]{Follana:2006rc}%
  \BibitemOpen
  \bibfield  {author} {\bibinfo {author} {\bibfnamefont {E.}~\bibnamefont
  {Follana}}, \bibinfo {author} {\bibfnamefont {Q.}~\bibnamefont {Mason}},
  \bibinfo {author} {\bibfnamefont {C.}~\bibnamefont {Davies}}, \bibinfo
  {author} {\bibfnamefont {K.}~\bibnamefont {Hornbostel}}, \bibinfo {author}
  {\bibfnamefont {G.~P.}\ \bibnamefont {Lepage}}, \bibinfo {author}
  {\bibfnamefont {J.}~\bibnamefont {Shigemitsu}}, \bibinfo {author}
  {\bibfnamefont {H.}~\bibnamefont {Trottier}}, \ and\ \bibinfo {author}
  {\bibfnamefont {K.}~\bibnamefont {Wong}} (\bibinfo {collaboration} {HPQCD,
  UKQCD}),\ }\href {\doibase 10.1103/PhysRevD.75.054502} {\bibfield  {journal}
  {\bibinfo  {journal} {Phys. Rev.}\ }\textbf {\bibinfo {volume} {D75}},\
  \bibinfo {pages} {054502} (\bibinfo {year} {2007})},\ \Eprint
  {http://arxiv.org/abs/hep-lat/0610092} {arXiv:hep-lat/0610092 [hep-lat]}
  \BibitemShut {NoStop}%
\bibitem [{\citenamefont {Weber}\ \emph {et~al.}(2018)\citenamefont {Weber},
  \citenamefont {Bazavov},\ and\ \citenamefont {Petreczky}}]{Weber:2018bam}%
  \BibitemOpen
  \bibfield  {author} {\bibinfo {author} {\bibfnamefont {J.~H.}\ \bibnamefont
  {Weber}}, \bibinfo {author} {\bibfnamefont {A.}~\bibnamefont {Bazavov}}, \
  and\ \bibinfo {author} {\bibfnamefont {P.}~\bibnamefont {Petreczky}},\ }in\
  \href@noop {} {\emph {\bibinfo {booktitle} {{13th Conference on Quark
  Confinement and the Hadron Spectrum (Confinement XIII) Maynooth, Ireland,
  July 31-August 6, 2018}}}}\ (\bibinfo {year} {2018})\ \Eprint
  {http://arxiv.org/abs/1811.12902} {arXiv:1811.12902 [hep-lat]} \BibitemShut
  {NoStop}%
\bibitem [{\citenamefont {Hasenfratz}\ and\ \citenamefont
  {Knechtli}(2001)}]{Hasenfratz:2001hp}%
  \BibitemOpen
  \bibfield  {author} {\bibinfo {author} {\bibfnamefont {A.}~\bibnamefont
  {Hasenfratz}}\ and\ \bibinfo {author} {\bibfnamefont {F.}~\bibnamefont
  {Knechtli}},\ }\href {\doibase 10.1103/PhysRevD.64.034504} {\bibfield
  {journal} {\bibinfo  {journal} {Phys. Rev.}\ }\textbf {\bibinfo {volume}
  {D64}},\ \bibinfo {pages} {034504} (\bibinfo {year} {2001})},\ \Eprint
  {http://arxiv.org/abs/hep-lat/0103029} {arXiv:hep-lat/0103029 [hep-lat]}
  \BibitemShut {NoStop}%
\bibitem [{\citenamefont {Bazavov}\ \emph
  {et~al.}(2012{\natexlab{a}})\citenamefont {Bazavov}, \citenamefont
  {Brambilla}, \citenamefont {Garcia~i Tormo}, \citenamefont {Petreczky},
  \citenamefont {Soto},\ and\ \citenamefont {Vairo}}]{Bazavov:2012ka}%
  \BibitemOpen
  \bibfield  {author} {\bibinfo {author} {\bibfnamefont {A.}~\bibnamefont
  {Bazavov}}, \bibinfo {author} {\bibfnamefont {N.}~\bibnamefont {Brambilla}},
  \bibinfo {author} {\bibfnamefont {X.}~\bibnamefont {Garcia~i Tormo}},
  \bibinfo {author} {\bibfnamefont {P.}~\bibnamefont {Petreczky}}, \bibinfo
  {author} {\bibfnamefont {J.}~\bibnamefont {Soto}}, \ and\ \bibinfo {author}
  {\bibfnamefont {A.}~\bibnamefont {Vairo}},\ }\href {\doibase
  10.1103/PhysRevD.86.114031} {\bibfield  {journal} {\bibinfo  {journal} {Phys.
  Rev.}\ }\textbf {\bibinfo {volume} {D86}},\ \bibinfo {pages} {114031}
  (\bibinfo {year} {2012}{\natexlab{a}})},\ \Eprint
  {http://arxiv.org/abs/1205.6155} {arXiv:1205.6155 [hep-ph]} \BibitemShut
  {NoStop}%
\bibitem [{\citenamefont {Takaura}\ \emph
  {et~al.}(2019{\natexlab{a}})\citenamefont {Takaura}, \citenamefont {Kaneko},
  \citenamefont {Kiyo},\ and\ \citenamefont {Sumino}}]{Takaura:2018lpw}%
  \BibitemOpen
  \bibfield  {author} {\bibinfo {author} {\bibfnamefont {H.}~\bibnamefont
  {Takaura}}, \bibinfo {author} {\bibfnamefont {T.}~\bibnamefont {Kaneko}},
  \bibinfo {author} {\bibfnamefont {Y.}~\bibnamefont {Kiyo}}, \ and\ \bibinfo
  {author} {\bibfnamefont {Y.}~\bibnamefont {Sumino}},\ }\href {\doibase
  10.1016/j.physletb.2018.12.060} {\bibfield  {journal} {\bibinfo  {journal}
  {Phys. Lett.}\ }\textbf {\bibinfo {volume} {B789}},\ \bibinfo {pages} {598}
  (\bibinfo {year} {2019}{\natexlab{a}})},\ \Eprint
  {http://arxiv.org/abs/1808.01632} {arXiv:1808.01632 [hep-ph]} \BibitemShut
  {NoStop}%
\bibitem [{\citenamefont {Takaura}\ \emph
  {et~al.}(2019{\natexlab{b}})\citenamefont {Takaura}, \citenamefont {Kaneko},
  \citenamefont {Kiyo},\ and\ \citenamefont {Sumino}}]{Takaura:2018vcy}%
  \BibitemOpen
  \bibfield  {author} {\bibinfo {author} {\bibfnamefont {H.}~\bibnamefont
  {Takaura}}, \bibinfo {author} {\bibfnamefont {T.}~\bibnamefont {Kaneko}},
  \bibinfo {author} {\bibfnamefont {Y.}~\bibnamefont {Kiyo}}, \ and\ \bibinfo
  {author} {\bibfnamefont {Y.}~\bibnamefont {Sumino}},\ }\href {\doibase
  10.1007/JHEP04(2019)155} {\bibfield  {journal} {\bibinfo  {journal} {JHEP}\
  }\textbf {\bibinfo {volume} {04}},\ \bibinfo {pages} {155} (\bibinfo {year}
  {2019}{\natexlab{b}})},\ \Eprint {http://arxiv.org/abs/1808.01643}
  {arXiv:1808.01643 [hep-ph]} \BibitemShut {NoStop}%
\bibitem [{\citenamefont {Garcia~i Tormo}(2013)}]{Tormo:2013tha}%
  \BibitemOpen
  \bibfield  {author} {\bibinfo {author} {\bibfnamefont {X.}~\bibnamefont
  {Garcia~i Tormo}},\ }\href {\doibase 10.1142/S0217732313300280} {\bibfield
  {journal} {\bibinfo  {journal} {Mod. Phys. Lett.}\ }\textbf {\bibinfo
  {volume} {A28}},\ \bibinfo {pages} {1330028} (\bibinfo {year} {2013})},\
  \Eprint {http://arxiv.org/abs/1307.2238} {arXiv:1307.2238 [hep-ph]}
  \BibitemShut {NoStop}%
\bibitem [{\citenamefont {Brambilla}\ \emph {et~al.}(2000)\citenamefont
  {Brambilla}, \citenamefont {Pineda}, \citenamefont {Soto},\ and\
  \citenamefont {Vairo}}]{Brambilla:1999xf}%
  \BibitemOpen
  \bibfield  {author} {\bibinfo {author} {\bibfnamefont {N.}~\bibnamefont
  {Brambilla}}, \bibinfo {author} {\bibfnamefont {A.}~\bibnamefont {Pineda}},
  \bibinfo {author} {\bibfnamefont {J.}~\bibnamefont {Soto}}, \ and\ \bibinfo
  {author} {\bibfnamefont {A.}~\bibnamefont {Vairo}},\ }\href {\doibase
  10.1016/S0550-3213(99)00693-8} {\bibfield  {journal} {\bibinfo  {journal}
  {Nucl. Phys.}\ }\textbf {\bibinfo {volume} {B566}},\ \bibinfo {pages} {275}
  (\bibinfo {year} {2000})},\ \Eprint {http://arxiv.org/abs/hep-ph/9907240}
  {arXiv:hep-ph/9907240 [hep-ph]} \BibitemShut {NoStop}%
\bibitem [{\citenamefont {Brambilla}\ \emph {et~al.}(1999)\citenamefont
  {Brambilla}, \citenamefont {Pineda}, \citenamefont {Soto},\ and\
  \citenamefont {Vairo}}]{Brambilla:1999qa}%
  \BibitemOpen
  \bibfield  {author} {\bibinfo {author} {\bibfnamefont {N.}~\bibnamefont
  {Brambilla}}, \bibinfo {author} {\bibfnamefont {A.}~\bibnamefont {Pineda}},
  \bibinfo {author} {\bibfnamefont {J.}~\bibnamefont {Soto}}, \ and\ \bibinfo
  {author} {\bibfnamefont {A.}~\bibnamefont {Vairo}},\ }\href {\doibase
  10.1103/PhysRevD.60.091502} {\bibfield  {journal} {\bibinfo  {journal} {Phys.
  Rev.}\ }\textbf {\bibinfo {volume} {D60}},\ \bibinfo {pages} {091502}
  (\bibinfo {year} {1999})},\ \Eprint {http://arxiv.org/abs/hep-ph/9903355}
  {arXiv:hep-ph/9903355 [hep-ph]} \BibitemShut {NoStop}%
\bibitem [{\citenamefont {Appelquist}\ \emph {et~al.}(1977)\citenamefont
  {Appelquist}, \citenamefont {Dine},\ and\ \citenamefont
  {Muzinich}}]{Appelquist:1977tw}%
  \BibitemOpen
  \bibfield  {author} {\bibinfo {author} {\bibfnamefont {T.}~\bibnamefont
  {Appelquist}}, \bibinfo {author} {\bibfnamefont {M.}~\bibnamefont {Dine}}, \
  and\ \bibinfo {author} {\bibfnamefont {I.~J.}\ \bibnamefont {Muzinich}},\
  }\href {\doibase 10.1016/0370-2693(77)90651-7} {\bibfield  {journal}
  {\bibinfo  {journal} {Phys. Lett.}\ }\textbf {\bibinfo {volume} {69B}},\
  \bibinfo {pages} {231} (\bibinfo {year} {1977})}\BibitemShut {NoStop}%
\bibitem [{\citenamefont {Pineda}\ and\ \citenamefont
  {Soto}(2000)}]{Pineda:2000gza}%
  \BibitemOpen
  \bibfield  {author} {\bibinfo {author} {\bibfnamefont {A.}~\bibnamefont
  {Pineda}}\ and\ \bibinfo {author} {\bibfnamefont {J.}~\bibnamefont {Soto}},\
  }\href {\doibase 10.1016/S0370-2693(00)01261-2} {\bibfield  {journal}
  {\bibinfo  {journal} {Phys. Lett.}\ }\textbf {\bibinfo {volume} {B495}},\
  \bibinfo {pages} {323} (\bibinfo {year} {2000})},\ \Eprint
  {http://arxiv.org/abs/hep-ph/0007197} {arXiv:hep-ph/0007197 [hep-ph]}
  \BibitemShut {NoStop}%
\bibitem [{\citenamefont {Brambilla}\ \emph {et~al.}(2005)\citenamefont
  {Brambilla}, \citenamefont {Pineda}, \citenamefont {Soto},\ and\
  \citenamefont {Vairo}}]{Brambilla:2004jw}%
  \BibitemOpen
  \bibfield  {author} {\bibinfo {author} {\bibfnamefont {N.}~\bibnamefont
  {Brambilla}}, \bibinfo {author} {\bibfnamefont {A.}~\bibnamefont {Pineda}},
  \bibinfo {author} {\bibfnamefont {J.}~\bibnamefont {Soto}}, \ and\ \bibinfo
  {author} {\bibfnamefont {A.}~\bibnamefont {Vairo}},\ }\href {\doibase
  10.1103/RevModPhys.77.1423} {\bibfield  {journal} {\bibinfo  {journal} {Rev.
  Mod. Phys.}\ }\textbf {\bibinfo {volume} {77}},\ \bibinfo {pages} {1423}
  (\bibinfo {year} {2005})},\ \Eprint {http://arxiv.org/abs/hep-ph/0410047}
  {arXiv:hep-ph/0410047 [hep-ph]} \BibitemShut {NoStop}%
\bibitem [{\citenamefont {Brambilla}\ \emph {et~al.}(2009)\citenamefont
  {Brambilla}, \citenamefont {Vairo}, \citenamefont {Garcia~i Tormo},\ and\
  \citenamefont {Soto}}]{Brambilla:2009bi}%
  \BibitemOpen
  \bibfield  {author} {\bibinfo {author} {\bibfnamefont {N.}~\bibnamefont
  {Brambilla}}, \bibinfo {author} {\bibfnamefont {A.}~\bibnamefont {Vairo}},
  \bibinfo {author} {\bibfnamefont {X.}~\bibnamefont {Garcia~i Tormo}}, \ and\
  \bibinfo {author} {\bibfnamefont {J.}~\bibnamefont {Soto}},\ }\href {\doibase
  10.1103/PhysRevD.80.034016} {\bibfield  {journal} {\bibinfo  {journal} {Phys.
  Rev.}\ }\textbf {\bibinfo {volume} {D80}},\ \bibinfo {pages} {034016}
  (\bibinfo {year} {2009})},\ \Eprint {http://arxiv.org/abs/0906.1390}
  {arXiv:0906.1390 [hep-ph]} \BibitemShut {NoStop}%
\bibitem [{\citenamefont {Husung}\ \emph {et~al.}(2018)\citenamefont {Husung},
  \citenamefont {Koren}, \citenamefont {Krah},\ and\ \citenamefont
  {Sommer}}]{Husung:2017qjz}%
  \BibitemOpen
  \bibfield  {author} {\bibinfo {author} {\bibfnamefont {N.}~\bibnamefont
  {Husung}}, \bibinfo {author} {\bibfnamefont {M.}~\bibnamefont {Koren}},
  \bibinfo {author} {\bibfnamefont {P.}~\bibnamefont {Krah}}, \ and\ \bibinfo
  {author} {\bibfnamefont {R.}~\bibnamefont {Sommer}},\ }\bibfield  {booktitle}
  {\emph {\bibinfo {booktitle} {{Proceedings, 35th International Symposium on
  Lattice Field Theory (Lattice 2017): Granada, Spain, June 18-24, 2017}}},\
  }\href {\doibase 10.1051/epjconf/201817514024} {\bibfield  {journal}
  {\bibinfo  {journal} {EPJ Web Conf.}\ }\textbf {\bibinfo {volume} {175}},\
  \bibinfo {pages} {14024} (\bibinfo {year} {2018})},\ \Eprint
  {http://arxiv.org/abs/1711.01860} {arXiv:1711.01860 [hep-lat]} \BibitemShut
  {NoStop}%
\bibitem [{\citenamefont {Aoki}\ \emph {et~al.}(2019)\citenamefont {Aoki} \emph
  {et~al.}}]{Aoki:2019cca}%
  \BibitemOpen
  \bibfield  {author} {\bibinfo {author} {\bibfnamefont {S.}~\bibnamefont
  {Aoki}} \emph {et~al.} (\bibinfo {collaboration} {Flavour Lattice Averaging
  Group}),\ }\href@noop {} {\  (\bibinfo {year} {2019})},\ \Eprint
  {http://arxiv.org/abs/1902.08191} {arXiv:1902.08191 [hep-lat]} \BibitemShut
  {NoStop}%
\bibitem [{\citenamefont {Bruno}\ \emph {et~al.}(2017)\citenamefont {Bruno},
  \citenamefont {Dalla~Brida}, \citenamefont {Fritzsch}, \citenamefont
  {Korzec}, \citenamefont {Ramos}, \citenamefont {Schaefer}, \citenamefont
  {Simma}, \citenamefont {Sint},\ and\ \citenamefont {Sommer}}]{Bruno:2017gxd}%
  \BibitemOpen
  \bibfield  {author} {\bibinfo {author} {\bibfnamefont {M.}~\bibnamefont
  {Bruno}}, \bibinfo {author} {\bibfnamefont {M.}~\bibnamefont {Dalla~Brida}},
  \bibinfo {author} {\bibfnamefont {P.}~\bibnamefont {Fritzsch}}, \bibinfo
  {author} {\bibfnamefont {T.}~\bibnamefont {Korzec}}, \bibinfo {author}
  {\bibfnamefont {A.}~\bibnamefont {Ramos}}, \bibinfo {author} {\bibfnamefont
  {S.}~\bibnamefont {Schaefer}}, \bibinfo {author} {\bibfnamefont
  {H.}~\bibnamefont {Simma}}, \bibinfo {author} {\bibfnamefont
  {S.}~\bibnamefont {Sint}}, \ and\ \bibinfo {author} {\bibfnamefont
  {R.}~\bibnamefont {Sommer}} (\bibinfo {collaboration} {ALPHA}),\ }\href
  {\doibase 10.1103/PhysRevLett.119.102001} {\bibfield  {journal} {\bibinfo
  {journal} {Phys. Rev. Lett.}\ }\textbf {\bibinfo {volume} {119}},\ \bibinfo
  {pages} {102001} (\bibinfo {year} {2017})},\ \Eprint
  {http://arxiv.org/abs/1706.03821} {arXiv:1706.03821 [hep-lat]} \BibitemShut
  {NoStop}%
\bibitem [{\citenamefont {McNeile}\ \emph {et~al.}(2010)\citenamefont
  {McNeile}, \citenamefont {Davies}, \citenamefont {Follana}, \citenamefont
  {Hornbostel},\ and\ \citenamefont {Lepage}}]{McNeile:2010ji}%
  \BibitemOpen
  \bibfield  {author} {\bibinfo {author} {\bibfnamefont {C.}~\bibnamefont
  {McNeile}}, \bibinfo {author} {\bibfnamefont {C.~T.~H.}\ \bibnamefont
  {Davies}}, \bibinfo {author} {\bibfnamefont {E.}~\bibnamefont {Follana}},
  \bibinfo {author} {\bibfnamefont {K.}~\bibnamefont {Hornbostel}}, \ and\
  \bibinfo {author} {\bibfnamefont {G.~P.}\ \bibnamefont {Lepage}},\ }\href
  {\doibase 10.1103/PhysRevD.82.034512} {\bibfield  {journal} {\bibinfo
  {journal} {Phys. Rev.}\ }\textbf {\bibinfo {volume} {D82}},\ \bibinfo {pages}
  {034512} (\bibinfo {year} {2010})},\ \Eprint {http://arxiv.org/abs/1004.4285}
  {arXiv:1004.4285 [hep-lat]} \BibitemShut {NoStop}%
\bibitem [{\citenamefont {Chakraborty}\ \emph {et~al.}(2015)\citenamefont
  {Chakraborty}, \citenamefont {Davies}, \citenamefont {Galloway},
  \citenamefont {Knecht}, \citenamefont {Koponen}, \citenamefont {Donald},
  \citenamefont {Dowdall}, \citenamefont {Lepage},\ and\ \citenamefont
  {McNeile}}]{Chakraborty:2014aca}%
  \BibitemOpen
  \bibfield  {author} {\bibinfo {author} {\bibfnamefont {B.}~\bibnamefont
  {Chakraborty}}, \bibinfo {author} {\bibfnamefont {C.~T.~H.}\ \bibnamefont
  {Davies}}, \bibinfo {author} {\bibfnamefont {B.}~\bibnamefont {Galloway}},
  \bibinfo {author} {\bibfnamefont {P.}~\bibnamefont {Knecht}}, \bibinfo
  {author} {\bibfnamefont {J.}~\bibnamefont {Koponen}}, \bibinfo {author}
  {\bibfnamefont {G.~C.}\ \bibnamefont {Donald}}, \bibinfo {author}
  {\bibfnamefont {R.~J.}\ \bibnamefont {Dowdall}}, \bibinfo {author}
  {\bibfnamefont {G.~P.}\ \bibnamefont {Lepage}}, \ and\ \bibinfo {author}
  {\bibfnamefont {C.}~\bibnamefont {McNeile}},\ }\href {\doibase
  10.1103/PhysRevD.91.054508} {\bibfield  {journal} {\bibinfo  {journal} {Phys.
  Rev.}\ }\textbf {\bibinfo {volume} {D91}},\ \bibinfo {pages} {054508}
  (\bibinfo {year} {2015})},\ \Eprint {http://arxiv.org/abs/1408.4169}
  {arXiv:1408.4169 [hep-lat]} \BibitemShut {NoStop}%
\bibitem [{\citenamefont {Petreczky}\ and\ \citenamefont
  {Weber}(2019)}]{Petreczky:2019ozv}%
  \BibitemOpen
  \bibfield  {author} {\bibinfo {author} {\bibfnamefont {P.}~\bibnamefont
  {Petreczky}}\ and\ \bibinfo {author} {\bibfnamefont {J.~H.}\ \bibnamefont
  {Weber}},\ }\href@noop {} {\  (\bibinfo {year} {2019})},\ \Eprint
  {http://arxiv.org/abs/1901.06424} {arXiv:1901.06424 [hep-lat]} \BibitemShut
  {NoStop}%
\bibitem [{\citenamefont {Maezawa}\ and\ \citenamefont
  {Petreczky}(2016)}]{Maezawa:2016vgv}%
  \BibitemOpen
  \bibfield  {author} {\bibinfo {author} {\bibfnamefont {Y.}~\bibnamefont
  {Maezawa}}\ and\ \bibinfo {author} {\bibfnamefont {P.}~\bibnamefont
  {Petreczky}},\ }\href {\doibase 10.1103/PhysRevD.94.034507} {\bibfield
  {journal} {\bibinfo  {journal} {Phys. Rev.}\ }\textbf {\bibinfo {volume}
  {D94}},\ \bibinfo {pages} {034507} (\bibinfo {year} {2016})},\ \Eprint
  {http://arxiv.org/abs/1606.08798} {arXiv:1606.08798 [hep-lat]} \BibitemShut
  {NoStop}%
\bibitem [{\citenamefont {Nakayama}\ \emph {et~al.}(2016)\citenamefont
  {Nakayama}, \citenamefont {Fahy},\ and\ \citenamefont
  {Hashimoto}}]{Nakayama:2016atf}%
  \BibitemOpen
  \bibfield  {author} {\bibinfo {author} {\bibfnamefont {K.}~\bibnamefont
  {Nakayama}}, \bibinfo {author} {\bibfnamefont {B.}~\bibnamefont {Fahy}}, \
  and\ \bibinfo {author} {\bibfnamefont {S.}~\bibnamefont {Hashimoto}},\ }\href
  {\doibase 10.1103/PhysRevD.94.054507} {\bibfield  {journal} {\bibinfo
  {journal} {Phys. Rev.}\ }\textbf {\bibinfo {volume} {D94}},\ \bibinfo {pages}
  {054507} (\bibinfo {year} {2016})},\ \Eprint
  {http://arxiv.org/abs/1606.01002} {arXiv:1606.01002 [hep-lat]} \BibitemShut
  {NoStop}%
\bibitem [{\citenamefont {Hudspith}\ \emph {et~al.}(2018)\citenamefont
  {Hudspith}, \citenamefont {Lewis}, \citenamefont {Maltman},\ and\
  \citenamefont {Shintani}}]{Hudspith:2018bpz}%
  \BibitemOpen
  \bibfield  {author} {\bibinfo {author} {\bibfnamefont {R.~J.}\ \bibnamefont
  {Hudspith}}, \bibinfo {author} {\bibfnamefont {R.}~\bibnamefont {Lewis}},
  \bibinfo {author} {\bibfnamefont {K.}~\bibnamefont {Maltman}}, \ and\
  \bibinfo {author} {\bibfnamefont {E.}~\bibnamefont {Shintani}},\ }\href@noop
  {} {\  (\bibinfo {year} {2018})},\ \Eprint {http://arxiv.org/abs/1804.10286}
  {arXiv:1804.10286 [hep-lat]} \BibitemShut {NoStop}%
\bibitem [{\citenamefont {Zafeiropoulos}\ \emph {et~al.}(2019)\citenamefont
  {Zafeiropoulos}, \citenamefont {Boucaud}, \citenamefont {De~Soto},
  \citenamefont {Rodríguez-Quintero},\ and\ \citenamefont
  {Segovia}}]{Zafeiropoulos:2019flq}%
  \BibitemOpen
  \bibfield  {author} {\bibinfo {author} {\bibfnamefont {S.}~\bibnamefont
  {Zafeiropoulos}}, \bibinfo {author} {\bibfnamefont {P.}~\bibnamefont
  {Boucaud}}, \bibinfo {author} {\bibfnamefont {F.}~\bibnamefont {De~Soto}},
  \bibinfo {author} {\bibfnamefont {J.}~\bibnamefont {Rodríguez-Quintero}}, \
  and\ \bibinfo {author} {\bibfnamefont {J.}~\bibnamefont {Segovia}},\ }\href
  {\doibase 10.1103/PhysRevLett.122.162002} {\bibfield  {journal} {\bibinfo
  {journal} {Phys. Rev. Lett.}\ }\textbf {\bibinfo {volume} {122}},\ \bibinfo
  {pages} {162002} (\bibinfo {year} {2019})},\ \Eprint
  {http://arxiv.org/abs/1902.08148} {arXiv:1902.08148 [hep-ph]} \BibitemShut
  {NoStop}%
\bibitem [{\citenamefont {Mateu}\ and\ \citenamefont
  {Ortega}(2018)}]{Mateu:2017hlz}%
  \BibitemOpen
  \bibfield  {author} {\bibinfo {author} {\bibfnamefont {V.}~\bibnamefont
  {Mateu}}\ and\ \bibinfo {author} {\bibfnamefont {P.~G.}\ \bibnamefont
  {Ortega}},\ }\href {\doibase 10.1007/JHEP01(2018)122} {\bibfield  {journal}
  {\bibinfo  {journal} {JHEP}\ }\textbf {\bibinfo {volume} {01}},\ \bibinfo
  {pages} {122} (\bibinfo {year} {2018})},\ \Eprint
  {http://arxiv.org/abs/1711.05755} {arXiv:1711.05755 [hep-ph]} \BibitemShut
  {NoStop}%
\bibitem [{\citenamefont {Peset}\ \emph {et~al.}(2018)\citenamefont {Peset},
  \citenamefont {Pineda},\ and\ \citenamefont {Segovia}}]{Peset:2018ria}%
  \BibitemOpen
  \bibfield  {author} {\bibinfo {author} {\bibfnamefont {C.}~\bibnamefont
  {Peset}}, \bibinfo {author} {\bibfnamefont {A.}~\bibnamefont {Pineda}}, \
  and\ \bibinfo {author} {\bibfnamefont {J.}~\bibnamefont {Segovia}},\ }\href
  {\doibase 10.1007/JHEP09(2018)167} {\bibfield  {journal} {\bibinfo  {journal}
  {JHEP}\ }\textbf {\bibinfo {volume} {09}},\ \bibinfo {pages} {167} (\bibinfo
  {year} {2018})},\ \Eprint {http://arxiv.org/abs/1806.05197} {arXiv:1806.05197
  [hep-ph]} \BibitemShut {NoStop}%
\bibitem [{\citenamefont {Brambilla}\ \emph {et~al.}(2007)\citenamefont
  {Brambilla}, \citenamefont {Garcia~i Tormo}, \citenamefont {Soto},\ and\
  \citenamefont {Vairo}}]{Brambilla:2007cz}%
  \BibitemOpen
  \bibfield  {author} {\bibinfo {author} {\bibfnamefont {N.}~\bibnamefont
  {Brambilla}}, \bibinfo {author} {\bibfnamefont {X.}~\bibnamefont {Garcia~i
  Tormo}}, \bibinfo {author} {\bibfnamefont {J.}~\bibnamefont {Soto}}, \ and\
  \bibinfo {author} {\bibfnamefont {A.}~\bibnamefont {Vairo}},\ }\href
  {\doibase 10.1103/PhysRevD.75.074014} {\bibfield  {journal} {\bibinfo
  {journal} {Phys. Rev.}\ }\textbf {\bibinfo {volume} {D75}},\ \bibinfo {pages}
  {074014} (\bibinfo {year} {2007})},\ \Eprint
  {http://arxiv.org/abs/hep-ph/0702079} {arXiv:hep-ph/0702079 [hep-ph]}
  \BibitemShut {NoStop}%
\bibitem [{\citenamefont {Mateu}\ \emph {et~al.}(2019)\citenamefont {Mateu},
  \citenamefont {Ortega}, \citenamefont {Entem},\ and\ \citenamefont
  {Fernández}}]{Mateu:2018zym}%
  \BibitemOpen
  \bibfield  {author} {\bibinfo {author} {\bibfnamefont {V.}~\bibnamefont
  {Mateu}}, \bibinfo {author} {\bibfnamefont {P.~G.}\ \bibnamefont {Ortega}},
  \bibinfo {author} {\bibfnamefont {D.~R.}\ \bibnamefont {Entem}}, \ and\
  \bibinfo {author} {\bibfnamefont {F.}~\bibnamefont {Fernández}},\ }\href
  {\doibase 10.1140/epjc/s10052-019-6808-2} {\bibfield  {journal} {\bibinfo
  {journal} {Eur. Phys. J.}\ }\textbf {\bibinfo {volume} {C79}},\ \bibinfo
  {pages} {323} (\bibinfo {year} {2019})},\ \Eprint
  {http://arxiv.org/abs/1811.01982} {arXiv:1811.01982 [hep-ph]} \BibitemShut
  {NoStop}%
\bibitem [{\citenamefont {{R Core Team}}(2018)}]{Rpackage}%
  \BibitemOpen
  \bibfield  {author} {\bibinfo {author} {\bibnamefont {{R Core Team}}},\
  }\href {https://www.R-project.org} {\emph {\bibinfo {title} {R: A Language
  and Environment for Statistical Computing}}},\ \bibinfo {organization} {R
  Foundation for Statistical Computing},\ \bibinfo {address} {Vienna, Austria}
  (\bibinfo {year} {2018})\BibitemShut {NoStop}%
\bibitem [{\citenamefont {Pinheiro}\ \emph {et~al.}(2018)\citenamefont
  {Pinheiro}, \citenamefont {Bates}, \citenamefont {DebRoy}, \citenamefont
  {Sarkar},\ and\ \citenamefont {{R Core Team}}}]{nlme}%
  \BibitemOpen
  \bibfield  {author} {\bibinfo {author} {\bibfnamefont {J.}~\bibnamefont
  {Pinheiro}}, \bibinfo {author} {\bibfnamefont {D.}~\bibnamefont {Bates}},
  \bibinfo {author} {\bibfnamefont {S.}~\bibnamefont {DebRoy}}, \bibinfo
  {author} {\bibfnamefont {D.}~\bibnamefont {Sarkar}}, \ and\ \bibinfo {author}
  {\bibnamefont {{R Core Team}}},\ }\href
  {https://CRAN.R-project.org/package=nlme} {\emph {\bibinfo {title} {{nlme}:
  Linear and Nonlinear Mixed Effects Models}}} (\bibinfo {year} {2018}),\
  \bibinfo {note} {r package version 3.1-137}\BibitemShut {NoStop}%
\bibitem [{\citenamefont {Bazavov}\ \emph
  {et~al.}(2012{\natexlab{b}})\citenamefont {Bazavov} \emph
  {et~al.}}]{Bazavov:2011nk}%
  \BibitemOpen
  \bibfield  {author} {\bibinfo {author} {\bibfnamefont {A.}~\bibnamefont
  {Bazavov}} \emph {et~al.},\ }\href {\doibase 10.1103/PhysRevD.85.054503}
  {\bibfield  {journal} {\bibinfo  {journal} {Phys. Rev.}\ }\textbf {\bibinfo
  {volume} {D85}},\ \bibinfo {pages} {054503} (\bibinfo {year}
  {2012}{\natexlab{b}})},\ \Eprint {http://arxiv.org/abs/1111.1710}
  {arXiv:1111.1710 [hep-lat]} \BibitemShut {NoStop}%
\bibitem [{\citenamefont {Bazavov}\ \emph {et~al.}(2013)\citenamefont
  {Bazavov}, \citenamefont {Ding}, \citenamefont {Hegde}, \citenamefont
  {Karsch}, \citenamefont {Miao}, \citenamefont {Mukherjee}, \citenamefont
  {Petreczky}, \citenamefont {Schmidt},\ and\ \citenamefont
  {Velytsky}}]{Bazavov:2013uja}%
  \BibitemOpen
  \bibfield  {author} {\bibinfo {author} {\bibfnamefont {A.}~\bibnamefont
  {Bazavov}}, \bibinfo {author} {\bibfnamefont {H.~T.}\ \bibnamefont {Ding}},
  \bibinfo {author} {\bibfnamefont {P.}~\bibnamefont {Hegde}}, \bibinfo
  {author} {\bibfnamefont {F.}~\bibnamefont {Karsch}}, \bibinfo {author}
  {\bibfnamefont {C.}~\bibnamefont {Miao}}, \bibinfo {author} {\bibfnamefont
  {S.}~\bibnamefont {Mukherjee}}, \bibinfo {author} {\bibfnamefont
  {P.}~\bibnamefont {Petreczky}}, \bibinfo {author} {\bibfnamefont
  {C.}~\bibnamefont {Schmidt}}, \ and\ \bibinfo {author} {\bibfnamefont
  {A.}~\bibnamefont {Velytsky}},\ }\href {\doibase 10.1103/PhysRevD.88.094021}
  {\bibfield  {journal} {\bibinfo  {journal} {Phys. Rev.}\ }\textbf {\bibinfo
  {volume} {D88}},\ \bibinfo {pages} {094021} (\bibinfo {year} {2013})},\
  \Eprint {http://arxiv.org/abs/1309.2317} {arXiv:1309.2317 [hep-lat]}
  \BibitemShut {NoStop}%
\bibitem [{\citenamefont {Herzog}\ \emph {et~al.}(2017)\citenamefont {Herzog},
  \citenamefont {Ruijl}, \citenamefont {Ueda}, \citenamefont {Vermaseren},\
  and\ \citenamefont {Vogt}}]{Herzog:2017ohr}%
  \BibitemOpen
  \bibfield  {author} {\bibinfo {author} {\bibfnamefont {F.}~\bibnamefont
  {Herzog}}, \bibinfo {author} {\bibfnamefont {B.}~\bibnamefont {Ruijl}},
  \bibinfo {author} {\bibfnamefont {T.}~\bibnamefont {Ueda}}, \bibinfo {author}
  {\bibfnamefont {J.~A.~M.}\ \bibnamefont {Vermaseren}}, \ and\ \bibinfo
  {author} {\bibfnamefont {A.}~\bibnamefont {Vogt}},\ }\href {\doibase
  10.1007/JHEP02(2017)090} {\bibfield  {journal} {\bibinfo  {journal} {JHEP}\
  }\textbf {\bibinfo {volume} {02}},\ \bibinfo {pages} {090} (\bibinfo {year}
  {2017})},\ \Eprint {http://arxiv.org/abs/1701.01404} {arXiv:1701.01404
  [hep-ph]} \BibitemShut {NoStop}%
\bibitem [{\citenamefont {Luthe}\ \emph {et~al.}(2017)\citenamefont {Luthe},
  \citenamefont {Maier}, \citenamefont {Marquard},\ and\ \citenamefont
  {Schr\mbox{\"o}der}}]{Luthe:2017ttg}%
  \BibitemOpen
  \bibfield  {author} {\bibinfo {author} {\bibfnamefont {T.}~\bibnamefont
  {Luthe}}, \bibinfo {author} {\bibfnamefont {A.}~\bibnamefont {Maier}},
  \bibinfo {author} {\bibfnamefont {P.}~\bibnamefont {Marquard}}, \ and\
  \bibinfo {author} {\bibfnamefont {Y.}~\bibnamefont {Schr\mbox{\"o}der}},\
  }\href {\doibase 10.1007/JHEP10(2017)166} {\bibfield  {journal} {\bibinfo
  {journal} {JHEP}\ }\textbf {\bibinfo {volume} {10}},\ \bibinfo {pages} {166}
  (\bibinfo {year} {2017})},\ \Eprint {http://arxiv.org/abs/1709.07718}
  {arXiv:1709.07718 [hep-ph]} \BibitemShut {NoStop}%
\bibitem [{\citenamefont {Fischler}(1977)}]{Fischler:1977yf}%
  \BibitemOpen
  \bibfield  {author} {\bibinfo {author} {\bibfnamefont {W.}~\bibnamefont
  {Fischler}},\ }\href {\doibase 10.1016/0550-3213(77)90026-8} {\bibfield
  {journal} {\bibinfo  {journal} {Nucl. Phys.}\ }\textbf {\bibinfo {volume}
  {B129}},\ \bibinfo {pages} {157} (\bibinfo {year} {1977})}\BibitemShut
  {NoStop}%
\bibitem [{\citenamefont {Billoire}(1980)}]{Billoire:1979ih}%
  \BibitemOpen
  \bibfield  {author} {\bibinfo {author} {\bibfnamefont {A.}~\bibnamefont
  {Billoire}},\ }\href {\doibase 10.1016/0370-2693(80)90279-8} {\bibfield
  {journal} {\bibinfo  {journal} {Phys. Lett.}\ }\textbf {\bibinfo {volume}
  {92B}},\ \bibinfo {pages} {343} (\bibinfo {year} {1980})}\BibitemShut
  {NoStop}%
\bibitem [{\citenamefont {Schr\mbox{\"o}der}(1999)}]{Schroder:1998vy}%
  \BibitemOpen
  \bibfield  {author} {\bibinfo {author} {\bibfnamefont {Y.}~\bibnamefont
  {Schr\mbox{\"o}der}},\ }\href {\doibase 10.1016/S0370-2693(99)00010-6}
  {\bibfield  {journal} {\bibinfo  {journal} {Phys. Lett.}\ }\textbf {\bibinfo
  {volume} {B447}},\ \bibinfo {pages} {321} (\bibinfo {year} {1999})},\ \Eprint
  {http://arxiv.org/abs/hep-ph/9812205} {arXiv:hep-ph/9812205 [hep-ph]}
  \BibitemShut {NoStop}%
\bibitem [{\citenamefont {Anzai}\ \emph {et~al.}(2010)\citenamefont {Anzai},
  \citenamefont {Kiyo},\ and\ \citenamefont {Sumino}}]{Anzai:2009tm}%
  \BibitemOpen
  \bibfield  {author} {\bibinfo {author} {\bibfnamefont {C.}~\bibnamefont
  {Anzai}}, \bibinfo {author} {\bibfnamefont {Y.}~\bibnamefont {Kiyo}}, \ and\
  \bibinfo {author} {\bibfnamefont {Y.}~\bibnamefont {Sumino}},\ }\href
  {\doibase 10.1103/PhysRevLett.104.112003} {\bibfield  {journal} {\bibinfo
  {journal} {Phys. Rev. Lett.}\ }\textbf {\bibinfo {volume} {104}},\ \bibinfo
  {pages} {112003} (\bibinfo {year} {2010})},\ \Eprint
  {http://arxiv.org/abs/0911.4335} {arXiv:0911.4335 [hep-ph]} \BibitemShut
  {NoStop}%
\bibitem [{\citenamefont {Smirnov}\ \emph {et~al.}(2010)\citenamefont
  {Smirnov}, \citenamefont {Smirnov},\ and\ \citenamefont
  {Steinhauser}}]{Smirnov:2009fh}%
  \BibitemOpen
  \bibfield  {author} {\bibinfo {author} {\bibfnamefont {A.~V.}\ \bibnamefont
  {Smirnov}}, \bibinfo {author} {\bibfnamefont {V.~A.}\ \bibnamefont
  {Smirnov}}, \ and\ \bibinfo {author} {\bibfnamefont {M.}~\bibnamefont
  {Steinhauser}},\ }\href {\doibase 10.1103/PhysRevLett.104.112002} {\bibfield
  {journal} {\bibinfo  {journal} {Phys. Rev. Lett.}\ }\textbf {\bibinfo
  {volume} {104}},\ \bibinfo {pages} {112002} (\bibinfo {year} {2010})},\
  \Eprint {http://arxiv.org/abs/0911.4742} {arXiv:0911.4742 [hep-ph]}
  \BibitemShut {NoStop}%
\bibitem [{\citenamefont {Lee}\ \emph {et~al.}(2016)\citenamefont {Lee},
  \citenamefont {Smirnov}, \citenamefont {Smirnov},\ and\ \citenamefont
  {Steinhauser}}]{Lee:2016cgz}%
  \BibitemOpen
  \bibfield  {author} {\bibinfo {author} {\bibfnamefont {R.~N.}\ \bibnamefont
  {Lee}}, \bibinfo {author} {\bibfnamefont {A.~V.}\ \bibnamefont {Smirnov}},
  \bibinfo {author} {\bibfnamefont {V.~A.}\ \bibnamefont {Smirnov}}, \ and\
  \bibinfo {author} {\bibfnamefont {M.}~\bibnamefont {Steinhauser}},\ }\href
  {\doibase 10.1103/PhysRevD.94.054029} {\bibfield  {journal} {\bibinfo
  {journal} {Phys. Rev.}\ }\textbf {\bibinfo {volume} {D94}},\ \bibinfo {pages}
  {054029} (\bibinfo {year} {2016})},\ \Eprint
  {http://arxiv.org/abs/1608.02603} {arXiv:1608.02603 [hep-ph]} \BibitemShut
  {NoStop}%
\bibitem [{\citenamefont {Lee}\ and\ \citenamefont
  {Smirnov}(2016)}]{Lee:2016lvq}%
  \BibitemOpen
  \bibfield  {author} {\bibinfo {author} {\bibfnamefont {R.~N.}\ \bibnamefont
  {Lee}}\ and\ \bibinfo {author} {\bibfnamefont {V.~A.}\ \bibnamefont
  {Smirnov}},\ }\href {\doibase 10.1007/JHEP10(2016)089} {\bibfield  {journal}
  {\bibinfo  {journal} {JHEP}\ }\textbf {\bibinfo {volume} {10}},\ \bibinfo
  {pages} {089} (\bibinfo {year} {2016})},\ \Eprint
  {http://arxiv.org/abs/1608.02605} {arXiv:1608.02605 [hep-ph]} \BibitemShut
  {NoStop}%
\end{thebibliography}%

\end{document}